\documentclass[aps,showpacs,twocolumn,superscriptaddress]{revtex4-2}
\usepackage[utf8]{inputenc}

\usepackage[T1]{fontenc}
\usepackage{lmodern}
\usepackage{braket}
\usepackage[version=3]{mhchem} 
\usepackage{here}
\usepackage{multirow}
\usepackage{longtable}
\usepackage{ascmac}
\usepackage{color}
\usepackage[all]{xy}
\usepackage{algorithm2e}
\usepackage{array}
\usepackage{qcircuit}
\usepackage{hyperref}

\usepackage{natbib}
\usepackage{mathtools} 
\usepackage{graphicx}
\usepackage{amsfonts,amsmath,amssymb,amsthm}
\usepackage{bbm}
\usepackage{multirow}

\theoremstyle{plain}
\newtheorem{thm}{Theorem}
\newtheorem{lemma}{Lemma}
\newtheorem{remark}{Remark}
\newtheorem*{thm*}{Theorem}
\newtheorem*{lemma*}{Lemma}
\newtheorem*{remark*}{Remark}
\newtheorem{corollary}{Corollary}[thm]

\usepackage{bm}
\usepackage{comment}

\hypersetup{
           breaklinks=true,   
           colorlinks=true,   
           pdfusetitle=true,  
           citecolor=blue, 
           urlcolor=blue
           }

\begin{document}

\title{Doubly optimal parallel wire cutting without ancilla qubits}

\author{Hiroyuki Harada}
\email{hiro.041o_i7@keio.jp}
\affiliation{Department of Applied Physics and Physico-Informatics, Keio University, Hiyoshi 3-14-1, Kohoku, Yokohama 223-8522, Japan}

\author{Kaito Wada}
\email{wkai1013keio840@keio.jp}
\thanks{\\H.H. and K.W. contributed equally to this work.}
\affiliation{Department of Applied Physics and Physico-Informatics, Keio University, Hiyoshi 3-14-1, Kohoku, Yokohama 223-8522, Japan}

\author{Naoki Yamamoto}
\email{yamamoto@appi.keio.ac.jp}
\affiliation{Department of Applied Physics and Physico-Informatics, Keio University, Hiyoshi 3-14-1, Kohoku, Yokohama 223-8522, Japan}
\affiliation{Quantum Computing Center, Keio University, Hiyoshi 3-14-1, Kohoku, Yokohama 223-8522, Japan}


\begin{abstract}

A restriction in the quality and quantity of available qubits presents a substantial obstacle to the application of near-term and early fault-tolerant quantum computers in practical tasks. 
To confront this challenge, some techniques for effectively augmenting the system size through classical processing have been proposed; one promising approach is quantum circuit cutting. 
The main idea of quantum circuit cutting is to decompose an original circuit into smaller sub-circuits and combine outputs from these sub-circuits to recover the original output. 
Although this approach enables us to simulate larger quantum circuits beyond physically available circuits, it needs classical overheads quantified by the two metrics: the sampling overhead in the number of measurements to reconstruct the original output, and the number of channels in the decomposition. 
Thus, it is crucial to devise a decomposition method that minimizes both of these metrics, thereby reducing the overall execution time. 
This paper studies the problem of decomposing the parallel $n$-qubit identity channel, i.e., $n$-parallel wire cutting, into a set of local operations and classical communication; 
then we give an optimal wire-cutting method comprised of channels based on mutually unbiased bases, that achieves minimal overheads in both the sampling overhead and the number of channels, without ancilla qubits. 
This is in stark contrast to the existing method that achieves the optimal sampling overhead yet with ancilla qubits. 
Moreover, we derive a tight lower bound of the number of channels in parallel wire cutting without ancilla systems and show that only our method achieves this lower bound among the existing methods. 
Notably, our method shows an exponential improvement in the number of channels, compared to the aforementioned ancilla-assisted method that achieves optimal sampling overhead. 
\end{abstract}

\maketitle

\RestyleAlgo{ruled}


\section{Introduction}\label{sec:intro}

\begin{figure*}[htbp]
\centering
\begin{center}
 \includegraphics[width=170mm]{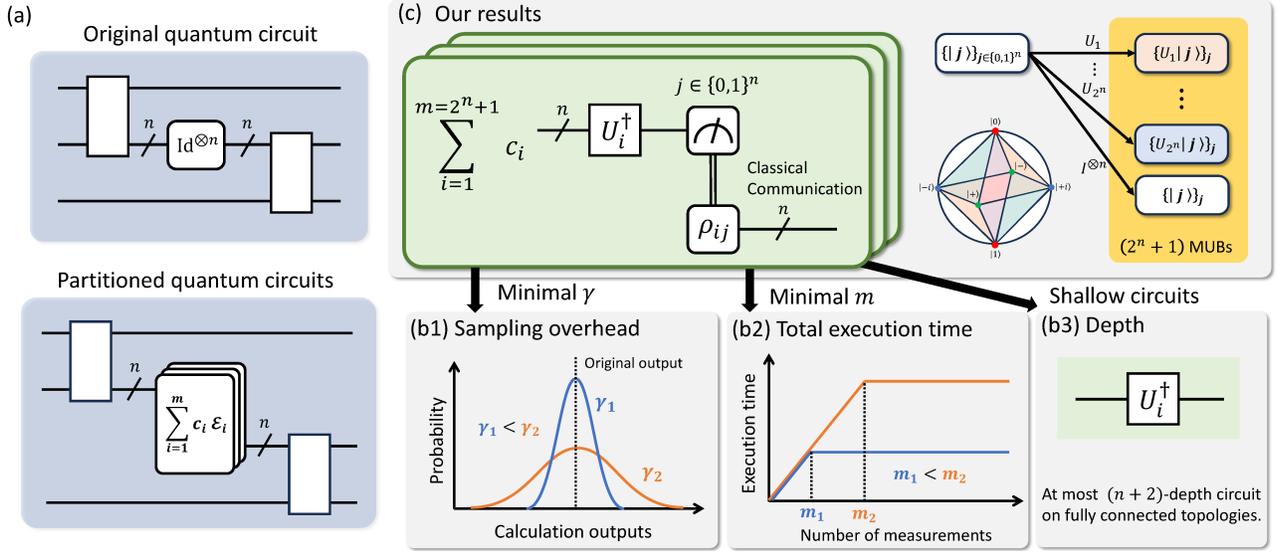}    
\end{center}
\caption{Summary of the proposed method. 
(a) The idea of the parallel $n$-wire cutting, where the $n$-qubit identity channel ${\rm Id}^{\otimes n}$ is quasiprobabilistically decomposed into the form of Eq.~(\ref{eq:quasiprobability}).
(b) Several metrices for the performance of wire-cutting methods.
(c) The proposed decomposition of ${\rm Id}^{\otimes n}$ consists of $m=2^n+1$ m-p channels based on MUBs. 
Importantly, our decomposition achieves the minimal overheads in terms of both (b1) the sampling overheads $\gamma({\rm Id}^{\otimes n})$ and (b2) the number of channels $m$, without ancilla qubits; Tables~\ref{intro_table} and \ref{intro_table_2} give detailed information on $\gamma$ and $m$ with comparison to other methods. 
$\gamma$ and $m$ are closely related to the number of measurements required to achieve a given precision and the total execution time in implementing the wire-cutting method, respectively. 
Furthermore, as shown in (b3), the circuit $U_i$ in the m-p channels of our decomposition can be implemented on a circuit with at most $(n+2)$-depth and fully connected qubits.}
\label{intro:MUBs_cut}
\end{figure*}

\subsection{Background}

Quantum computers are expected to have a significant advantage in certain tasks over classical ones~\cite{10.1145/237814.237866,doi:10.1137/S0097539795293172}. 
{\color{black}In recent years, with the rapid progress in the development of quantum devices, 
extensive research has been conducted to explore algorithms that may demonstrate the usefulness of near-term and early fault-tolerant quantum computing devices, e.g., quantum machine learning~\cite{Moll_2018,Havlíček2019,Amaro2022,fuller2021approximate}, quantum simulation~\cite{Yuan2019theoryofvariational,PhysRevResearch.3.033083,PhysRevA.105.062421}, and quantum chemical computation~\cite{peruzzo2014variational,kandala2017hardware,gao2021applications}. 
However, those devices have fundamental limitations in both quality and quantity of available qubits, which are major obstacles to their practical use.}

To mitigate this problem, various approaches have been proposed, which effectively  augment the size of quantum systems with the help of classical processing~\cite{PhysRevX.6.021043,PhysRevLett.125.150504,Perlin2021,10.1145/3445814.3446758,Lowe2023fastquantumcircuit,10025537,brenner2023optimal,Mitarai_2021,Mitarai2021overheadsimulating,10236453,Ufrecht2023cuttingmulticontrol,PhysRevLett.127.040501,PRXQuantum.3.010346,PRXQuantum.3.010309}.
In particular, the methods of partitioning a quantum circuit~\cite{PhysRevLett.125.150504,Perlin2021,10.1145/3445814.3446758,Mitarai_2021,Mitarai2021overheadsimulating,10236453,Lowe2023fastquantumcircuit,10025537,Ufrecht2023cuttingmulticontrol,brenner2023optimal} are useful.
The idea is to decompose a large quantum circuit into a set of smaller quantum circuits and recover the original output by appropriately combining the output results from those smaller circuits.

One approach to realize the perfect partitioning
is the quasiprobability simulation, which is also widely used in classical simulation of quantum systems~\cite{PhysRevLett.115.070501, PhysRevLett.118.090501, doi:10.1098/rspa.2019.0251, PRXQuantum.2.010345} and quantum error mitigation techniques~\cite{PhysRevLett.119.180509,PhysRevX.8.031027,Piveteau2022}.
More specifically, we replace a target quantum channel $\Gamma$ in a circuit with implementable channels $\mathcal{E}_i$ via the following decomposition:
\begin{equation}\label{eq:quasiprobability}
    \Gamma(\bullet) = \sum_{i=1}^m c_i \mathcal{E}_i(\bullet),
\end{equation}
where $m$ is the number of the principle channels and $c_i$ are real coefficients. 
Because the coefficients are not necessarily positive, this type of decomposition is referred to as a {\it quasiprobability decomposition}.
Depending on the target channel $\Gamma$, such a quasiprobability decomposition is mainly categorized into two techniques. One is the method to decompose the identity channel (or the {\it wires}) into a linear combination of channels that contain measurement followed by preparation of a quantum state, which is called the {\it wire cutting} (or time-like cut)~\cite{PhysRevLett.125.150504,Perlin2021,10.1145/3445814.3446758,Lowe2023fastquantumcircuit,10025537,brenner2023optimal}; see Fig.~\ref{intro:MUBs_cut}(a).
The other one is called the {\it gate cutting} (or space-like cut), which decomposes a non-local channel into a linear combination of tensor products of local channels~\cite{Mitarai_2021,Mitarai2021overheadsimulating,10236453, Ufrecht2023cuttingmulticontrol}.

Although the quasiprobability decomposition enables us to simulate a large quantum circuit with fewer quantum resources, the overhead for classical processing may become a critical issue. 
Specifically, there are two important metrics to quantify this overhead.
The first metric is the {\it sampling overhead} resulting from the normalization of the quasiprobability distribution, which quantifies the number of measurements required to achieve a given precision in recovering the original output, as shown in Fig.~\ref{intro:MUBs_cut}(b1). 
The sampling overhead $\gamma(\Gamma)^2$ of Eq.~(\ref{eq:quasiprobability}) is characterized by
\begin{equation}
    \label{intro:sampling_overhead}
    \mbox{Sampling overhead}:=\gamma(\Gamma)^2,~~ 
    \gamma(\Gamma):=\sum_{i=1}^{m}|c_{i}|.
\end{equation}
Note that applying the decomposition (\ref{eq:quasiprobability}) to $k$ channels $\{\Gamma_{l}\}_{l=1}^{k}$ in a given circuit, the overall sampling overhead results in the product of each sampling overhead as $\prod_{l=1}^{k}\gamma(\Gamma_{l})^2$.
This means that the variance of the target quantity calculated with the subsystems exponentially increases with respect to the number of quantum channels to be decomposed.
Clearly, this sampling issue will become serious in practice, and several studies have been conducted to pursue a less-costly decomposition~\cite{Mitarai2021overheadsimulating,10236453,10025537,Lowe2023fastquantumcircuit,brenner2023optimal}.
The other crucial metric is the (classical) execution time during the implementation of a quasiprobability decomposition, which directly relates to the number of channels $m$ in the decomposition.
Actually, as the compilation of many quantum circuits accounts for a large part of the execution time in the current quantum computing, the number $m$ is an important factor in a decomposition. 
In the (single) application of Eq.~(\ref{eq:quasiprobability}), the worst-case scaling of the number of quantum circuits to be compiled exhibits a transition from a linear scaling to a constant scaling with respect to the number of measurements, as illustrated in Fig.~\ref{intro:MUBs_cut}(b2), where the critical point of this transition is determined by the number of channels $m$ in a decomposition.
Therefore, $m$ should be as small as possible in reducing the execution time, among decompositions that need similar values in the sampling overhead.
\begin{table*}[ht]
\renewcommand{\arraystretch}{1.3}
\begin{tabular}{|c||c|c|c|c|}
\hline
&~No classical communication~~&\multicolumn{2}{c|}{~LOCC without ancilla qubits~~}&~LOCC with ancilla qubits~~\\
\cline{2-5}
& Ref.~\cite{PhysRevLett.125.150504} & Ref.~\cite{Lowe2023fastquantumcircuit} & Our work & Ref.~\cite{brenner2023optimal} \\
\hline\hline
1 wire & 16 & 25 & 9 & 9 \\
Parallel $n$ wires & $16^{n}$ &~~$(2^{n+1}+1)^2$~~& $(2^{n+1}-1)^2$ & $(2^{n+1}-1)^2$ \\
~~Non-parallel $n$ wires~~& $16^{n}$ & $25^n$ & $9^n$ & $(2^{n+1}-1)^2$ \\
\hline
\end{tabular}
\caption{Comparison of the sampling overheads $\gamma({\rm Id}^{\otimes n})^2$ for some wire-cutting methods. 
The ``1 wire'', ``Parallel $n$ wires'', and ``Non-parallel $n$ wires'' indicate the sampling overhead for the 1-qubit, parallel $n$-qubit, and non-parallel $n$-qubit identity channel decomposition, respectively. 
It is proven in~\cite{brenner2023optimal} that, if an arbitrary LOCC can be used, the optimal sampling overhead for $n$-wire cutting is $(2^{n+1}-1)^2$, regardless of ancilla assistance.} 
\label{intro_table}
\end{table*}
\begin{table*}[ht]
\renewcommand{\arraystretch}{1.3}
\begin{tabular}{|c||c|c|c|c|}
\hline
&~No classical communication~~&\multicolumn{2}{c|}{~LOCC without ancilla qubits~~}&~LOCC with ancilla qubits~~\\
\cline{2-5}
& ~Ref.~\cite{PhysRevLett.125.150504}~ & ~~~Ref.~\cite{Lowe2023fastquantumcircuit}~~~ & ~Our work~ & ~Ref.~\cite{brenner2023optimal}~ \\
\hline\hline
\multirow{2}{*}{~Parallel $n$ wires~}& \multirow{2}{*}{$8^n$} & ~$16^n-2\cdot4^n+3$~ & \multirow{2}{*}{$2^n+1$} & \multirow{2}{*}{$2^{2^n}+4^n-2^n-1$} \\
&& (At least) & &\\
\hline
\end{tabular}
\caption{Comparison of the number of m-p channels $m$ required for the $n$-qubit  parallel wire-cutting. Note that this table does not include the case of non-parallel wire-cutting because the m-p channels do not provide the basis for arbitrary non-parallel cuts, making it non-trivial to discuss $m$ in this case.
Corollary~\ref{corollary:smallest_m} proves $m\geq 2^n+1$ when no ancilla qubits are allowed to use, hence only our work achieves this lower bound.
}
\label{intro_table_2}
\end{table*}

\subsection{Contribution}

In this paper, we develop a new decomposition for the $n$-qubit identity channel ${\rm Id}^{\otimes n}$ with ${\rm Id}$ the single-qubit identity channel, i.e., the $n$-parallel wire cutting, as summarized in Fig.~\ref{intro:MUBs_cut}(c). 
In the wire-cutting problem, we need \textit{measure-and-prepare (m-p) channels} (or \textit{entanglement breaking channels}~\cite{doi:10.1142/S0129055X03001709}), each of which is comprised of a positive operator-valued measure (POVM) followed by the preparation of an $n$-qubit density matrix. 
Then, among all the decomposition methods that use m-p channels (which are slightly extended version from the original definition~\cite{doi:10.1142/S0129055X03001709}), our decomposition achieves the minimal overheads in both the sampling overhead $\gamma({\rm Id}^{\otimes n})^2$ and the number of channels $m$, due to the usage of mutually unbiased bases (MUBs); this result will be formally presented in Theorem~\ref{thm 2}, together with Theorem~\ref{thm:number_mp_channels}. 
This double optimality is attributed to the nature of MUBs that form a minimal set of measurements for the optimal quantum state estimation~\cite{I_D_Ivonovic_1981,Wootters1986,WOOTTERS1989363}.
Below we give a detailed explanation on this main result.

As for the sampling overhead, our decomposition achieves $\gamma({\rm Id}^{\otimes n})^2=(2^{n+1}-1)^2$ without any ancilla qubits. 
This is the minimum sampling overhead in the setup where arbitrary local operations and classical communication (LOCC) can be included in the principle channels $\mathcal{E}_i$ for all the $n$-wire cutting problems (not limited to the parallel wires), regardless of using ancilla assistance~\cite{brenner2023optimal}. 
Our result is in stark contrast to the previous $n$ wire-cutting method~\cite{brenner2023optimal} that achieves the same minimal sampling overhead using the quantum teleportations {\it with} $n$ ancilla qubits. 
In the absence of any ancilla assistance, the best sampling overhead of $\gamma({\rm Id}^{\otimes n})^2$ was known to be $(2^{n+1}+1)^2$ by using the randomized measurements~\cite{Lowe2023fastquantumcircuit} with the help of classical communication, which is slightly worse than our case. 
These results, together with those of the original method \cite{PhysRevLett.125.150504} and our method, are summarized in Table~\ref{intro_table}.

Next, as for the number of channels necessary for decomposing ${\rm Id}^{\otimes n}$, we prove $m\geq 2^{n}+1$ among all decomposition method using m-p channels without ancilla qubits (Theorem \ref{thm:number_mp_channels} and Corollary \ref{corollary:smallest_m}); then we show that our method achieves $m=2^{n}+1$, as proven in Theorem~\ref{thm 2}. 
It is notable that the number of channels $m=2^{n}+1$ in our decomposition is exponentially smaller than that of the method proposed in \cite{brenner2023optimal}, as shown in Table~\ref{intro_table_2}. 
Also, in our method, the m-p process can be efficiently performed on a device with fully connected qubits and at most $O(n)$-depth circuit, 
while the randomized measurement~\cite{Lowe2023fastquantumcircuit} uses at most $O(n\log n)$-depth circuits.

In addition, we discuss the problem of cutting non-parallel $n$ wires, i.e., the target to be decomposed is the set of single-qubit identity channels Id that cannot be collected to ${\rm Id}^{\otimes n}$.
Remarkably, the teleportation-based scheme~\cite{brenner2023optimal} assisted by ancilla qubits can achieve the minimum sampling overhead $(2^{n+1}-1)^2$, which is the same as the parallel $n$ wires cutting. 
This is thanks to the ability of collective quasiprobability decomposition on the entangling gates over the ancilla qubits and the output qubits for quantum teleportation. 
However, in the absence of ancilla, such global operations are not allowed, and there has been no proposal to reach the minimum sampling overhead $(2^{n+1}-1)^2$. 
A simple strategy is to independently apply the 1-wire cutting method for all wires.
Then, the sampling overhead for the original method~\cite{PhysRevLett.125.150504}, the method with randomized measurements~\cite{Lowe2023fastquantumcircuit}, and our proposed method are $16^n$, $25^n$, and $9^n$, respectively. 
Hence, though limited to the situation taking such a simple strategy, our method surely improves the overhead reached by the original one.
{\color{red}}


\subsection{Related works}

\begin{figure*}[ht]
\centering
\begin{center}
\includegraphics[width=170mm]{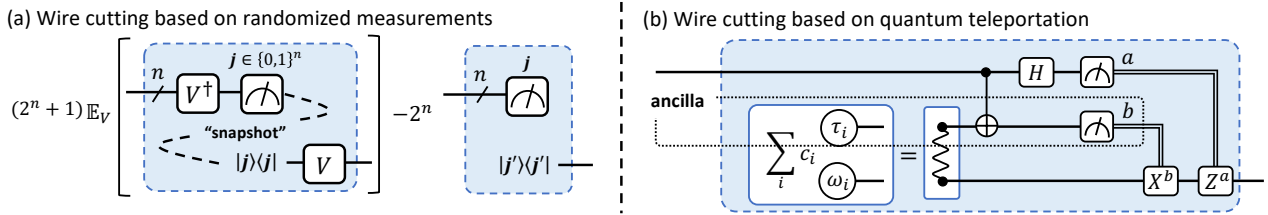}  
\end{center}
\caption{Previously proposed methods for the parallel $n$-wire cutting, i.e., the decomposition of $\mathrm{Id}^{\otimes n}$. (a) Wire cutting based on randomized measurements~\cite{Lowe2023fastquantumcircuit}. 
(b) Wire cutting based on quantum teleportation~\cite{brenner2023optimal}. For simplicity, the figure shows the case of the 1-wire cutting ($n=1$). 
Note that this method can be applied to non-parallel wire cutting at the cost of additional qubits. Both methods (a) and (b) can be written as a linear combination of m-p channels. 
A more detailed description is given in Appendix~\ref{sec:correspondence}.}
\label{intro_fig}
\end{figure*}

Finally, we briefly describe the previous methods~\cite{Lowe2023fastquantumcircuit,brenner2023optimal}. 
Actually, their techniques for improving the original sampling overhead \cite{PhysRevLett.125.150504} 
would offer better understanding on our method and thus it is worth being presented here in advance; a more detailed description, especially for $\gamma$ and $m$, is given in Appendix.~\ref{sec:correspondence}. 
As illustrated in Fig.~\ref{intro:MUBs_cut}(a), both methods replace ${\rm Id}^{\otimes n}$ with a linear combination of channels $\mathcal{E}_{i}$, which functions as an identity channel from the top $n$ wires to the bottom $n$ wires; 
additionally, to realize the wire cutting, the top and bottom wires in each channel $\mathcal{E}_{i}$ are not connected via any quantum resource, but LOCC are allowed.
Ref.~\cite{Lowe2023fastquantumcircuit} employed the idea of randomized measurements for wire cutting, and the series of operations can be seen as the implementation of classical shadow~\cite{Huang2020} in the middle of a quantum circuit, as illustrated in Fig.~\ref{intro_fig}(a).
Specifically, a random unitary $V^\dagger$ followed by the computational basis measurement on the top wire yields the output $\bm{j}\in\{0,1\}^{n}$, from which we input the state $V\ket{\bm{j}}\bra{\bm{j}}V^\dagger$ (called the classical snapshot in~\cite{Huang2020}) to the bottom wire.
If $V$ forms a unitary $t$-design ($t\geq 2$) such as the Clifford random gate, then their ensemble average over $V$ and $\bm{j}$ together with an additional input $\bm{j'}$ (which is a uniformly random vector in $\{0,1\}^n$) recovers the input state at the bottom wire. Clearly, the top and bottom wires in $\mathcal{E}_i$ in this case are separated, and thus the entire circuit depicted in Fig.~\ref{intro_fig}(a) is decomposed into two smaller sub-circuits.
Note that this scheme does not use any ancilla qubit other than the top and bottom wires. 

Next, the idea of Ref.~\cite{brenner2023optimal} is to use the quantum teleportation to realize the identity channel from the top to bottom wires, as shown in Fig.~\ref{intro_fig}(b) in the case of $n=1$.
As is well known, the teleportation needs to introduce an ancilla qubit and a pre-shared entangled state (i.e., a Bell pair) between the ancilla and the bottom wire, followed by a LOCC.
To realize 1-wire cutting between the top and bottom wires, the authors of~\cite{brenner2023optimal} employ a Bell pair quasiprobabilistically generated by a linear combination of separable states $\sum_{i}c_{i} \tau_{i}\otimes \omega_{i}$ on the ancilla and bottom qubits,
where $c_{i}$ are real numbers, and $\tau_{i}$ and $\omega_{i}$ are 1-qubit states, derived from the decomposition of Ref.~\cite{PhysRevA.59.141}.
As a result, the Bell pair becomes virtual and thereby the wire cutting is achieved.
Notably, this approach is extended to cutting $n$ wires by the quasiprobabilistic preparation of $n$ Bell pairs at the same time and the following replacement of the $n$ identity channels at $n$ places with the $n$ teleportation protocols.

\section{Preliminaries}\label{sec:preliminaries}

In this section, we first introduce the m-p channel as a building block for (parallel) wire cutting. 
Then, we review the basic idea of wire cutting, with the focus on the original work by~\textcite{PhysRevLett.125.150504}. 
Finally, we discuss two important metrics to evaluate the classical cost for decomposing a (large) quantum system into smaller subsystems; that is, the sampling overhead $\gamma^2$ in Sec.~\ref{sec:sampling_overhead} and the number of decomposed channels $m$, which is directly connected to the total execution time, in Sec.~\ref{sec:total execution time}.

%
\begin{figure}[ht]
\centering
\begin{center}
 \includegraphics[width=70mm]{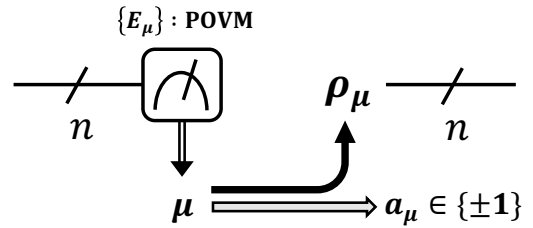}    
\end{center}
\caption{Schematic diagram of the $n$-qubit m-p channel defined in Eq.~(\ref{measure-and-prepare channel}).
This channel consists of three processes: (i) measurement with a POVM $\{E_{\bm{i}}\}$, (ii) multiplication of $a_{\bm{i}}$ on the output, and (iii) preparation of the new input state $\rho_{\bm{i}}$ depending on the measurement result $\bm{i}$.}
\label{fig of m-and-p channel}
\end{figure}

\subsection{Measure-and-prepare (m-p) channel}\label{mp_channel}

\begin{figure*}[t]
\centering
\begin{center}
 \includegraphics[width=150mm]{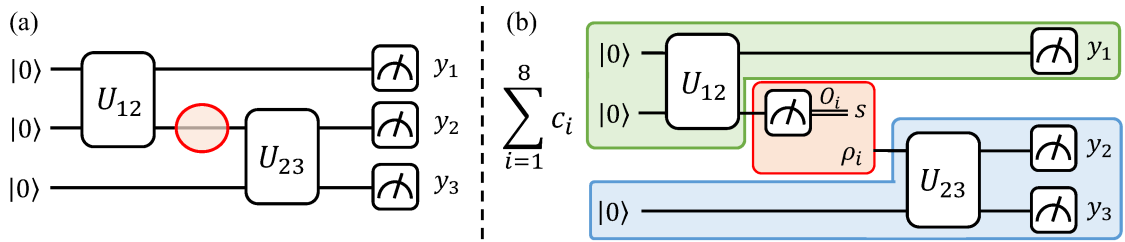}    
\end{center}
\caption{Schematic diagram of the wire cutting procedure, for the case of Ref.~\cite{PhysRevLett.125.150504}. 
(a) A quantum circuit before the decomposition. 
The circuit is composed of the 3-qubit initial state $\ket{0}^{\otimes 3}$, 
two 2-qubit unitary gates $U_{12}$, $U_{23}$, and the computational basis measurements producing $\bm{y}=(y_{1},y_{2},y_{3})$. 
(b) Quantum circuits after the decomposition. 
By applying Eq.~(\ref{quantum circuit cutting}) to the identity channel indicated by 
the red circle in (a), we obtain two sub-circuits indicated by the green and blue areas. 
Here, $s$ denotes the result of the eigenbasis measurement on $O_i$ in the red box in (b). 
This wire-cut enables us to simulate the 3-qubit quantum system by running the 2-qubit 
quantum systems with some post-processing.} 
\label{fig of qcc}
\end{figure*}

To execute wire-cutting, the principle channel $\mathcal{E}_i$ should combine the following operations: (i) measuring the input state, (ii) (if necessary) classically sending the measurement outcomes, and (iii) preparing a quantum state upon the received classical data. 
The channel satisfying these operations is called the m-p channel (or entanglement breaking channel), 
and it takes the following form: 
\begin{equation}
\label{measure-and-prepare channel}
    \mathcal{E}(\bullet):=\sum_{\mu} a_{\mu} \mathrm{Tr}[E_{\mu}(\bullet)] \rho_{\mu},
\end{equation}
where $a_{\mu}=\pm 1$. 
This definition of m-p channel is a slight extension of the original one given in Ref.~\cite{doi:10.1142/S0129055X03001709}, which restricts $a_\mu=1~\forall \mu$. 
If $a_\mu=-1$ for some $\mu$, our m-p channel \eqref{measure-and-prepare channel} may not be a quantum channel (i.e., a completely positive and trace-preserving channel). 
The operator set $\{E_{\mu}\}$ is a POVM which satisfies $\sum_{\mu} E_{\mu}=I^{\otimes n}$ and $E_{\mu}\ge 0$ for all $\mu$. 
Also, $\rho_{\mu}$ is an $n$-qubit density operator, which is assumed to be efficiently prepared in quantum circuits.
Note that this m-p channel is contained in the class of channels considered by Ref.~\cite{brenner2023optimal}, in which any LOCC between the top and bottom wires in Fig.~\ref{intro:MUBs_cut}(a) is allowed to be used. 
Actually, all of the existing (parallel) wire-cutting methods can be captured in the formalism introduced here; see the next subsection and Appendix~\ref{sec:correspondence}. 
We also remark that such channels expressed by the difference of two completely positive and trace-nonincreasing (CPTN) maps can be simulated efficiently~\cite{Mitarai_2021,Mitarai2021overheadsimulating,brenner2023optimal}, likewise the quantum channels represented by Eq.~(\ref{measure-and-prepare channel}).


An example of the channel \eqref{measure-and-prepare channel} is depicted in Fig.~\ref{fig of m-and-p channel}; in this figure, the bottom wires are lifted to the same line as that of the top wires. 
In this paper, we employ this $n$-qubit m-p channel as a channel $\mathcal{E}_i$ for the $n$-parallel wire cutting, i.e., the decomposition \eqref{eq:quasiprobability} with $\Gamma={\rm Id}^{\otimes n}$. 
In what follows, for simplicity, we denote the sampling overhead $\gamma({\rm Id}^{\otimes n})$ defined in Eq.~(\ref{intro:sampling_overhead}) by $\gamma_{n}^{\rm (mp)}$, reflecting the setup where we employ the $n$-qubit m-p channel as the principle channel $\mathcal{E}_i$ in Eq.~(\ref{eq:quasiprobability}).

\subsection{Wire cutting}
\label{sec:quantum circuit cutting}

Here we review the original wire cutting method \cite{PhysRevLett.125.150504} that 
enables simulating a large quantum circuit by classically processing the measurement result produced from smaller quantum circuits. 

Let $\bm{y}\in\{0,1\}^{L}$ be the result of the computational basis measurement on 
a given $L$-qubit quantum circuit with the input state $\ket{0}^{\otimes L}$. 
Now, with $f:\{0,1\}^{L}\to [-1,1]$ a classical postprocessing function, 
the expectation $\mathbb{E}_{\bm{y}}[f(\bm{y})]$ over the outputs of the $L$-qubit 
circuit is the target quantity to be estimated. 
Note that this setting includes the estimation of the expectation of Pauli strings.

The method splits a large circuit into smaller ones by decomposing the single-qubit identity channel ${\rm Id}(\bullet)$ as follows; 
\begin{equation}\label{quantum circuit cutting}
    {\rm Id}(\bullet) 
     = \sum_{i=1}^{8} c_{i} \mathrm{Tr}\left[{O_{i}(\bullet)}\right] \rho_{i},
\end{equation}
where the observables $O_i$, quantum states $\rho_i$, and real coefficients $c_i$ 
are given by 
\begin{alignat}{8}
    O_{1}&=I, &\quad \rho_{1}&=\ket{0}\bra{0}, &\quad c_{1} = +1/2,\notag\\
    O_{2}&=I, & \rho_{2}&=\ket{1}\bra{1}, & c_{2} = +1/2,\notag\\
    O_{3}&=X, & \rho_{3}&=\ket{+}\bra{+}, & c_{3} = +1/2,\notag\\
    O_{4}&=X, & \rho_{4}&=\ket{-}\bra{-}, & c_{4} = -1/2,\notag\\
    O_{5}&=Y, & \rho_{5}&=\ket{+i}\bra{+i}, & c_{5} = +1/2,\notag\\
    O_{6}&=Y, & \rho_{6}&=\ket{-i}\bra{-i}, & c_{6} = -1/2,\notag\\
    O_{7}&=Z, & \rho_{7}&=\ket{0}\bra{0}, & c_{7} = +1/2,\notag\\
    O_{8}&=Z, & \rho_{8}&=\ket{1}\bra{1}, & c_{8} = -1/2.\notag
\end{alignat}
Here, $I,X,Y,Z$ denote the single-qubit identity and Pauli matrices; 
$\ket{\pm}$, $\ket{\pm i}$, and $\{\ket{0},\ket{1}\}$ are eigenstates of $X, Y$, 
and $Z$, respectively. 
Note that the above set $\{O_i,\rho_i,c_i\}$ is just one of decompositions 
satisfying Eq.~(\ref{quantum circuit cutting}).
In this decomposition, $\mathrm{Tr}[{O_{i}(\bullet)}] \rho_{i}$ represents a channel 
that measures the expectation value of the observable $O_{i}$ on the quantum state 
entering the identity channel and prepares the new quantum state $\rho_i$. 
That is, this channel is the m-p channel (\ref{measure-and-prepare channel}) that has no classical communications, which can be easily seen by the spectral decomposition of the observables ${O}_i$; see Appendix~\ref{sec:correspondence}. 
Therefore, the decomposition (\ref{quantum circuit cutting}) separates the prepared state from the entering state by local operations (LO).

The decomposition of ${\rm Id}$ given by Eq.~\eqref{quantum circuit cutting} allows 
us to simulate a large circuit by the weighted summation of smaller circuits.
To see the procedure, let us consider the decomposition of a 3-qubit quantum circuit shown in Fig.~\ref{fig of qcc} as an example.
This quantum circuit consists of the 3-qubit initial state $\ket{0}^{\otimes 3}$ and 
two unitary operators $U_{12}$ and $U_{23}$ where $U_{ab}$ acts on the $a$-th and $b$-th qubits, followed by the computational basis measurements. Here, we decompose ${\rm Id}$ on the 2nd qubit wire between $U_{12}$ and $U_{23}$, indicated by the red circle in Fig.~\ref{fig of qcc}(a).
In this case, for each quantum circuit equipped with the $i$-th channel in Eq.~\eqref{quantum circuit cutting}, there are 2 types of measurement results: the computational-basis measurement at the terminal of the circuit, $\bm{y}=(y_1,y_2,y_3)$, and the eigenbasis measurement on $O_i$ in the middle of the circuit, $s\in\{-1,1\}$; see Fig.~\ref{fig of qcc}(b). 
These measurement results follow the probability distribution conditioned on the index $i$ as
\begin{eqnarray}
    P\left[y_1,s|i\right]&:=&{\rm Tr}\left[\left(\ket{y_1}\bra{y_1}\otimes E_{is}\right)U_{12}\ket{0}\bra{0}^{\otimes 2} U_{12}^\dagger\right],\notag\\[6pt]
    P\left[y_2,y_3|i\right]&:=&{\rm Tr}\left[\ket{y_2,y_3}\bra{y_2,y_3}U_{23}\left(\rho_i\otimes\ket{0}\bra{0} \right)U_{23}^\dagger\right],\notag
\end{eqnarray}
where $E_{is}$ denotes the projector on the eigenspace with eigenvalue $s$ of $O_i$.
Using these probabilities, the target expectation value can be written as
\begin{align}
\label{eq:harrow_decomposition}
    &\mathbb{E}_{\bm{y}}\left[f(\bm{y})\right]\notag\\
    &=\gamma_1^{(\rm mp)} \sum_{i=1}^8 p_{i}\sum_{\bm{y},s} {\rm sgn}(c_i) s f(\bm{y}) { P}\left[y_1,s|i\right]{P}\left[y_2,y_3|i\right],
\end{align}
where ${\rm sgn}(c_i)$ is the sign of $c_{i}$. 
Also, $p_{i}:=|c_{i}|/\gamma_{1}^{(\mathrm{mp})}$, where 
\begin{align}\label{eq:gamma_factor_harrowsdecomposition}
    \gamma_1^{(\rm mp)}~\mbox{of Eq.~(\ref{quantum circuit cutting})} = \sum_{i=1}^8 |c_i|=4.
\end{align}
The subscript ``1'' means the 1-qubit wire cutting.
Because $p_{i}$ represents the probability of the $i$-th channel, 
Eq.~(\ref{eq:harrow_decomposition}) enables us to estimate the target quantity using the Monte-Carlo method with samples obtained from the smaller quantum circuits. 
More precisely, for each shot, we first choose the index $i$ with probability $p_{i}$ and then run the small sub-circuits obtained from the quantum circuit equipped with the $i$-th channel as shown in Fig.~\ref{fig of qcc}(b). 
Then, using the measurement outcomes $\bm{y}$ and $s$, we calculate the value of ${\rm sgn}(c_i) s f(\bm{y})$. 
By repeating the above process enough times and computing their arithmetic mean with the multiplication of $\gamma_1^{(\rm mp)}$, we finally obtain the target expectation value. 
Importantly, $(y_1,s)$ and $(y_2,y_3)$ can be sampled from each small circuit in Fig.~\ref{fig of qcc}(b) respectively, and as a result we can estimate $\mathbb{E}_{\bm{y}}[f(\bm{y})]$ over the 3-qubit system with the use of only 2-qubit circuits.


\subsection{Sampling overhead}\label{sec:sampling_overhead}

The above example of ${\rm Id}$ decomposition indicates that the total number of measurements $N$ in the Monte-Carlo method for $n$-parallel wire cutting with $\gamma_n^{\rm (mp)}$ is given by 
\begin{align}\label{eq:old_samplingcost}
     N=O\left({\left[\gamma_n^{\rm (mp)}\right]^2}\times\frac{1}{ {\varepsilon^2}}\right),
\end{align}
where $\varepsilon$ is a given estimation error for the target quantity $\mathbb{E}_{\bm{y}}[f(\bm{y})]$. 
This equation shows that $[\gamma_n^{\rm (mp)}]^2$ defined by Eq.~(\ref{intro:sampling_overhead}) has a meaning of the overhead for the circuit cutting in the number of samples; thus, $[\gamma_n^{\rm (mp)}]^2$ is called the sampling overhead.
Furthermore, if we apply the decomposition with $\gamma_n^{\rm (mp)}$ at $k$ places in a quantum circuit, the total number of measurements with the error $\varepsilon$ results in 
$O([\gamma_n^{\rm (mp)}]^{2k}/\varepsilon^2)$. 
This exponential increase in the number of required measurements, with respect to the number of cut locations, is a serious problem for practical use.
Hence, it is important to find a decomposition that achieves the smallest $\gamma_n^{\rm (mp)}$.

\subsection{Total execution time}\label{sec:total execution time}

In the implementation of the circuit-cutting scheme, the number of channels $m$ in a decomposition should be minimized to reduce the total execution time $T$.
To clarify this point, we here give an expression of $T$ in terms of $m$, by specifying the classical processing part $T_{\rm C}$ and the quantum processing part $T_{\rm Q}$ of $T$, which thus satisfy $T:=T_{\rm C}+T_{\rm Q}$.

\begin{figure}[ht]
\centering
\begin{center}
 \includegraphics[width=85mm]{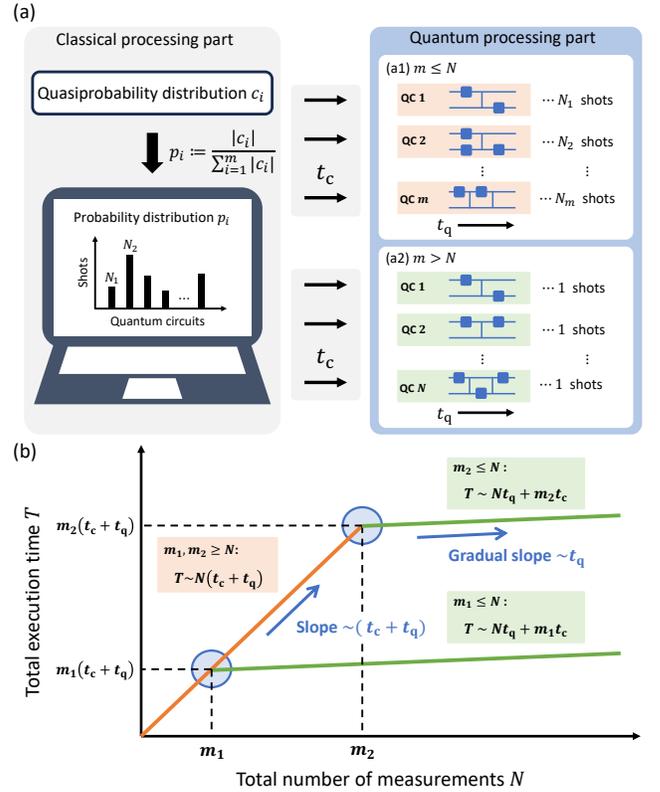}    
\end{center}
\caption{
(a) Illustration of the quasiprobability sampling scheme.
In the classical processing part, we assign the total number of shots $N$ to each quantum circuit based on the probability distribution $p_{i}:=|c_{i}|/\sum_{i=1}^{m}|c_{i}|$ and compile each quantum circuit with the unit time $t_{\rm c}$. Then, we run each of the compiled circuits the assigned number of times, 
with the unit time $t_{\rm q}$ per shot.
(a1) When the number of channels $m$ in a decomposition is larger than $N$, then $N$ different types of channels are employed for the Monte-Carlo procedure in the worst case. 
(a2) When $m \leq N$, all $m$ types of channels in a decomposition are realized in the worst case. 
(b) The relation between the upper bound of the total execution time $T$ and the total number of measurements $N$. 
At the threshold $N=m$, the slope gets smaller. 
The two distinct curves correspond to the cutting schemes with different $m$, being $m_{1}\leq m_{2}$. 
In view of the execution time, the cutting scheme with $m_{1}$ is more desirable than the scheme with $m_{2}$.}
\label{fig:total_execution_time}
\end{figure}

According to the Monte-Carlo procedure mentioned in the previous subsection, it can be reasonably assumed that the compilation time of the quantum circuits equipped with sampled channels is dominant in the classical processing part $T_{\rm C}$.
For quantitative analysis on $T_{\rm C}$, we here count the number of types of sampled channels, or equivalently the number of quantum circuits to be compiled, in the Monte-Carlo procedure with $N$ samples, where $N$ is specified by $\gamma$ and $\epsilon$ in advance.
First, let us consider the case where $N$ is larger than $m$.
While the total number of quantum circuits to be compiled stochastically varies depending on $N$ and the probability distribution $p_i=|c_i|/\gamma_1^{(\rm mp)}$, 
it is equal to $m$ in the worst case; 
see Fig.~\ref{fig:total_execution_time}(a1).
For instance, if we have an $n$-parallel application of the wire cut (\ref{quantum circuit cutting}) on the circuit, all $8^n$ types of m-p channels would be realized in the entire Monte-Carlo procedure with $N$ $(\geq m=8^n)$ samples, bacause all the channels have an equal probability in this decomposition.
Thus, we can evaluate $T_{\rm C}$ (in the worst case) as
\begin{equation}\label{eq:T_classical}
    T_{\rm C}\sim m\times t_{\rm c},~~~(m\leq N),
\end{equation}
where $t_{\rm c}$ is the unit time for compiling any quantum circuit.
As for the case of $m> N$, some channels are not sampled even in the worst case; 
actually, the most right method of Table~\ref{intro_table_2} falls into this case in a wide range of $N$.
For such a decomposition of the identity channel, 
$T_{\rm C}$ should be taken as $N\times t_{\rm c}$ instead of Eq.~(\ref{eq:T_classical}), as shown in Fig.~\ref{fig:total_execution_time}(a2).

On the other hand, the quantum processing part $T_{\rm Q}$ is directly specified by the total number of measurements $N$.
That is, assuming that the running time of all sub-circuits obtained from the original circuit equipped with a sampled channel is a constant value $t_{\rm q}$ regardless of the principle channels, $T_{\rm Q}$ is calculated as
\begin{equation}
    T_{\rm Q}\sim N\times t_{\rm q}.
\end{equation}

Therefore, in our model, the total execution time of the circuit-cutting method with the use of an identity channel decomposition comprised of $m$ channels is given by
\begin{align}\label{eq:execution_time}
    T=T_{\rm C}+T_{\rm Q}\sim \begin{dcases} 
    mt_{\rm c} + Nt_{\rm q},~~\mbox{if}~~~m\leq N,\\[6pt]
    Nt_{\rm c} + Nt_{\rm q},~~\mbox{if}~~~m > N,
    \end{dcases}
\end{align}
in the worst case. 
Because 
$t_{\rm q}$ is considered to be much smaller than 
$t_{\rm c}$ in the current quantum computing devices, the transition time in Eq.~(\ref{eq:execution_time}) has a significant impact on the total execution time as depicted in Fig.~\ref{fig:total_execution_time}(b). 
Thus, the number $m$ corresponding to the transition time ($m_1$ or $m_2$ in Fig.~\ref{fig:total_execution_time}(b)) is the main factor that may reduce the total execution time $T$.
More specifically, suppose we have some parallel $n$-wire cutting methods that give similar sampling overheads (and have similar values in $t_{\rm q}$ and $t_{\rm c}$), 
the decomposition with the smallest number of channels $m$ is the best because the number of measurements $N$ is comparable among those decompositions.
Therefore, it is important to devise a decomposing method for an identity channel, with a smaller value of channels $m$.

\section{Main Results}\label{sec:main_result}

\subsection{Lower bound of the number of decomposed channels $m$
}
\label{sec:measure-and-prepare channel}

As detailed in the previous subsection, the number $m$ of m-p channels in a decomposition of ${\rm Id}^{\otimes n}$ directly affects the total execution time, and thus it should be as small as possible. 
To discuss a desirable decomposition, here we first identify the lower bound of $m$ as follows.

\begin{thm}\label{thm:number_mp_channels}
    Suppose that an $n$-qubit quantum channel $\Gamma(\bullet)$ can be decomposed by $m$ (extended) m-p channels $\{\mathcal{E}_i(\bullet)\}_{i=1}^m$ 
    defined in Eq.~(\ref{measure-and-prepare channel}) and some weights $c_i\in \mathbb{R}$, in the form of Eq.~(\ref{eq:quasiprobability}).
    If the channels $\{\mathcal{E}_i(\bullet)\}_{i=1}^m$ do not use any ancilla qubits to implement each POVM, then $m$ 
    is lower bounded as
    \begin{align}
        \frac{{\rm Rank}\left(\Gamma\right)-1}{2^n-1}\leq m,
    \end{align}
    where ${\rm Rank}\left(\Gamma\right)$ is the rank of the transfer matrix of $\Gamma(\bullet)$.
\end{thm}
\noindent
The proof of Theorem~\ref{thm:number_mp_channels} is provided in Appendix~\ref{sec:proof of lower bound_basis}, together with a brief review of quantum channels and their transfer matrix representation. 
Although we assume that each m-p channel does not use any ancilla qubits for practical importance, if we are allowed to use ancilla qubits and thereby realize over-complete measurements, the lower bound of $m$ may be improved. 
In particular, if $\Gamma={\rm Id}^{\otimes n}$, 
Theorem~\ref{thm:number_mp_channels} immediately leads to the following result, because the transfer matrix of ${\rm Id}^{\otimes n}$ is full-rank, i.e., $4^n$.
\begin{corollary}
\label{corollary:smallest_m}
For the problem of parallel $n$-wire cutting without any ancilla assistance, the number $m_{{\rm Id}^{\otimes n}}$ of m-p channels~(\ref{measure-and-prepare channel}) in the ${\rm Id}^{\otimes n}$ decomposition is lower bounded as 
\begin{align}\label{eq:lower_bound_of_m}
     2^n+1\leq m_{{\rm Id}^{\otimes n}}.
\end{align}
\end{corollary}
\noindent
Importantly, there exists a decomposition that achieves this lower bound and, at the same time, the smallest sampling overhead for any $n$, as shown in the following subsections. 


\subsection{Doubly optimal single-qubit wire cutting}
\label{1_qubit_cut}

In this subsection, we study the problem of 1 wire cutting without ancilla, with particular focus on the classical overhead $\gamma$ and $m$; that is, we aim to develop the optimal decomposition of the single-qubit identity channel ${\rm Id}$ via a linear combination of single-qubit m-p channels. 
The results shown here provide a basis for the general multi-qubit case. 
Also, as will be shown later, the 1 wire cutting method itself is useful due to its applicability to arbitrary places on a quantum circuit.

To discuss the optimality, let us recall the lower bounds of $\gamma$ and $m$. 
First, Collorary~\ref{corollary:smallest_m} states that $m_{\rm Id}$ is bounded by $3\leq m_{\rm Id}$, when no ancilla qubits are allowed. 
We again note that there have not been wire-cutting methods that achieve the lower bound on $m_{\rm Id}$. 
Second, $\gamma^{(\rm mp)}_1$ is lower bounded as
\begin{align}
\label{eq:lower_bound}
    3\leq \gamma^{\mathrm{(mp)}}_{1}.
\end{align}
The proof is provided in Appendix~\ref{sec:proof of lower bound}.
Note that \textcite{brenner2023optimal} proved that, when an arbitrary LOCC 
(including the m-p channels) is allowed to use for a decomposition of ${\rm Id}^{\otimes n}$, the lower bound of $\gamma_{n}^{\rm (LOCC)}$ is given by
\begin{align}
\label{general lower bound}
    2^{n+1}-1 \leq \gamma_{n}^{\rm (LOCC)},
\end{align}
where the superscript means a decomposition of ${\rm Id}^{\otimes n}$ based on LOCC channels. 
Because the m-p channel (\ref{measure-and-prepare channel}) is an LOCC operation, we readily have $\gamma_{n}^{\rm (LOCC)}\leq\gamma_{n}^{\rm (mp)}$, and thus Eq.~\eqref{eq:lower_bound} is not new.
However, the achievability of the lower bound in Eq.~\eqref{general lower bound} was proven only for the teleportation-based wire-cutting method that uses ancilla qubits; that is, the achievability of \eqref{eq:lower_bound} has been unknown. 

Now, recall that the decomposition of ${\rm Id}$ via the original method \cite{PhysRevLett.125.150504} realizes $m=8$ and $\gamma_1^{\rm (mp)}=4$, as shown in the previous section; thus, this method is not optimal in both metrics. 
In contrast, here we explicitly construct a set of ancilla-free m-p channels that simultaneously achieves both the lower bounds $m_{\rm Id}=3$ and $\gamma^{\mathrm{(mp)}}_{1}=3$, as follows. 
\begin{thm}\label{thm 1}
The single-qubit identity channel ${\rm Id}(\bullet)$ can be decomposed as
\begin{align}\label{1-qubit cut in main text}
    {\rm Id}(\bullet)
    &=\sum_{i=1,2} \sum_{j\in\{0,1\}} \mathrm{Tr}\left[ U_{i} \ket{j}\bra{j}U_{i}^{\dagger}(\bullet) \right]U_{i} \ket{j}\bra{j}U_{i}^{\dagger}\notag\\
    &~~~-\sum_{j\in\{0,1\}} \mathrm{Tr}\left[ \ket{j}\bra{j}(\bullet) \right]X \ket{j}\bra{j}X,
\end{align}
where $U_1=H$ and $U_2=SH$, with $H$ the Hadamard gate and $S$ the phase gate. 
This decomposition achieves both of the lower bounds in Eqs.~(\ref{eq:lower_bound_of_m}) and (\ref{eq:lower_bound}); that is, 
\begin{equation}\label{smallest gamma}
    m~\mbox{of Eq.~(\ref{1-qubit cut in main text})}=3,~\mbox{and}~~\gamma^{\mathrm{(mp)}}_1~\mbox{of Eq.~(\ref{1-qubit cut in main text})}=3.
\end{equation}
\end{thm}

The proof of Theorem~\ref{thm 1} is given in Appendix~\ref{sec:proof of thm1}.
Fig.~\ref{fig:graphicalrepresentations}(a) shows the graphical representation of the decomposition (\ref{1-qubit cut in main text}), which consists of $m=3$ m-p channels in total. 
The idea of proof is as follows. 
The original decomposition in Eq.~(\ref{quantum circuit cutting}) has redundancy in the measurement and the input state in terms of the basis set that simultaneously diagonalize commuting observables.
This redundancy can be eliminated by the grouping technique~\cite{9248636,Crawford2021efficientquantum} via simultaneous diagonalization for the commuting observables such as $I$ and $Z$.
Noting that the input states and the coefficients $c_i$ in Eq.~(\ref{quantum circuit cutting}) can be rewritten as observables (see Appendix~\ref{sec:proof of thm1}), we can apply the grouping technique for both the measurement and the input state.
As a result, we obtain the m-p channels that have one-to-one correspondence between each POVM element $U_i\ket{j}\bra{j}U_i^\dagger$ and the prepared input state $U_i\ket{j}$ with respect to the measurement outcome $j$; hence, we arrive at the decomposition (\ref{1-qubit cut in main text}). 

%
\begin{figure*}[ht]
\centering
\begin{center}
 \includegraphics[width=170mm]{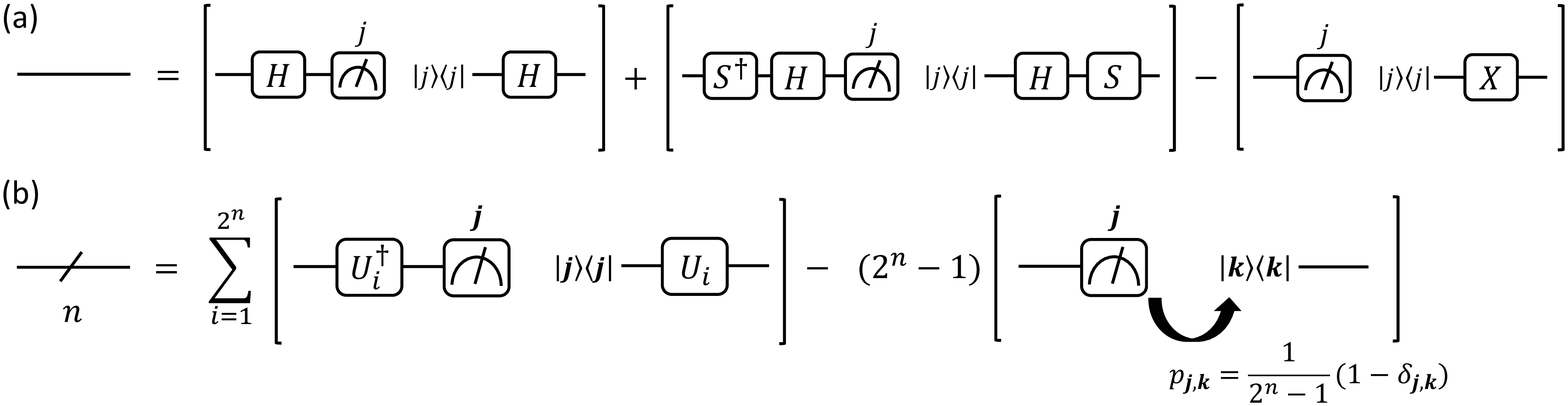}
\end{center}
\caption{Graphical representation for the optimal decomposition of (a) the 1 wire 
(single-qubit identity channel) and (b) the parallel $n$ wires ($n$-qubit parallel identity channel). 
Both decompositions are sum of m-p channels composed of unitary operation followed by the computational basis measurement, and thus they can be implemented without any ancilla qubit. 
In the second channel with coefficient $2^n-1$ in the decomposition (b), the input 
state is a mixed state that mixes the computational basis $\ket{\bm{k}}$ with the 
classical probability distribution $p_{\bm{j,k}}=(1-\delta_{\bm{j,k}})/(2^n-1)$ 
conditioned by the measurement result $\bm{j}$. 
The unitary operators $\{U_i\}$ change the computational base into MUBs as described in Sec.~\ref{sec:n-qubit_extension}.}
\label{fig:graphicalrepresentations}
\end{figure*}

Based on the result of 1-wire cutting obtained above, we can have a simple discussion on how to decompose non-parallel $n$ wires, i.e., the set of single-qubit identity channel Id that cannot be collected to the parallel one, ${\rm Id}^{\otimes n}$. 
Even in this case, our decomposition \eqref{1-qubit cut in main text} improves the base of the exponential scaling of the sampling overhead for the non-parallel $n$-wire cutting, from $16^n$ obtained by the original 1-wire cutting (\ref{quantum circuit cutting}) to $9^n$, when no ancilla qubit can be used. 
In other words, the decomposition (\ref{1-qubit cut in main text}) allows us to cut arbitrary $n$ wires with the sampling overhead $9^n$ even in the worst case (i.e., the case of non-parallel $n$-wire cutting). 
Note that the multiple applications of the decomposition (\ref{1-qubit cut in main text}) for each wire cannot achieve the optimal sampling overhead $(\gamma_{n}^{\rm (LOCC)})^2=(2^{n+1}-1)^2$ \cite{brenner2023optimal} except for the 1-wire cutting. 
However, for the parallel $n$-wire cutting, there exists an extension of the optimal 1-wire cutting that gives the optimal sampling overhead 
without the help of ancilla qubits, as shown in the next subsection.

Another important feature of our decomposition is that it contains only 3 m-p channels, which matches the lower bound of Eq.~(\ref{eq:lower_bound_of_m}).
Note that, when we apply the decomposition (\ref{1-qubit cut in main text}) to arbitrary $n$ wires in a circuit, the total number of channels to be simulated is $3^n$ in the worst case, which increases exponentially with respect to the number of cuts.
In comparison with $n$ applications of the original decomposition (\ref{quantum circuit cutting}), our method improves the base of the exponential scaling from $8^n$ to $3^n$.
Here, the number of channels in a decomposition of non-parallel identity channels should also be as small as possible to reduce the total execution time, as described in Sec.~\ref{sec:total execution time}. Then, the noticeable reduction on $m$ can decrease the overall execution time compared to the conventional method as long as the unit time for quantum processing $t_{\rm q}$ is comparable regardless of the existence of classical communications between the sub-circuits, although this assumption is slightly favorable for our method.

Finally, to clearly see the function of classical communication between the top and 
bottom wires, let us reconsider the 3-qubits example studied in Sec.~\ref{sec:quantum circuit cutting}. 
As before, we replace $\rm{Id}$ indicated by the red circle in Fig.~\ref{fig of qcc}(a) with the m-p channels in Eq.~(\ref{1-qubit cut in main text}); then we have 2 types 
of measurement results; $\bm{y}=(y_1,y_2,y_3)$ obtained at the terminal of the circuit and $j\in \{0,1\}$ in the middle of the circuit (instead of $s$ in Eq.~(\ref{eq:harrow_decomposition})). 
Unlike the original decomposition (\ref{quantum circuit cutting}), each of the three m-p channels (\ref{1-qubit cut in main text}) requires us to input the quantum state $U_i\ket{j}$ or $X\ket{j}$, which depends on the measurement result $j$, into the subsequent quantum circuit to sample $(y_2,y_3)$. 
Thus, the probability distribution of these measurement results are given by 
\begin{widetext}
\begin{eqnarray}
    P\left[y_1,j|i\right]&:=&{\rm Tr}\left[\left(\ket{y_1}\bra{y_1}\otimes U_i\ket{j}\bra{j}U_i^\dagger\right)U_{12}\ket{0}\bra{0}^{\otimes 2} U_{12}^\dagger\right],\notag\\[6pt]
    P\left[y_2,y_3|i,j\right]&:=&{\rm Tr}\left[\ket{y_2,y_3}\bra{y_2,y_3}U_{23}\left( U_i\ket{j}\bra{j}U^\dagger_i\otimes\ket{0}\bra{0} \right)U_{23}^\dagger\right],\notag
\end{eqnarray}
\end{widetext}
where the probability distribution of $(y_2,y_3)$ is conditioned by both the index of channels $i$ and the measurement result $j$; note that these are the case of $i=1,2$, and the case of $i=3$ is not shown due to its lengthy expression. 
Comparing with Eq.~(\ref{eq:harrow_decomposition}), it is clear that our decomposition employs classical communications between the splitted quantum circuits.

\subsection{Extension to parallel multi-qubit wire cutting}\label{sec:n-qubit_extension}

In this subsection, we show an extension of Eq.~(\ref{1-qubit cut in main text}) to the doubly optimal decomposition of the $n$-qubit identity channel ${\rm Id}^{\otimes n}$ via 
$n$-qubit m-p channels. 
First, we provide an intuitive discussion on how to achieve this goal. 
Similar to Theorem~\ref{thm 1}, our starting point is the $n$-qubit version of the original decomposition~(\ref{quantum circuit cutting}).
More precisely, the application of Eq.~(\ref{quantum circuit cutting}) to the parallel $n$ wires leads to the following expression:
\begin{equation}
\label{Pauli_cut}
    {\rm Id}^{\otimes n}(\bullet)= \frac{1}{2^n} \sum_{P\in\{I,X,Y,Z\}^{\otimes n}} {\rm Tr}[P(\bullet)]P,
\end{equation}
where $P$ is an $n$-qubit Pauli string such as $X\otimes Y \otimes \cdots \otimes Z$. 
Each term $\mathrm{Tr}[P(\bullet)]P$ can be viewed as a process of measuring the expectation value of $P$ followed by the input of eigenstates of $P$ into the subsequent circuit. 
That is, letting the spectral decomposition of $P$ as $\sum_{i=1}^{2^n} \alpha_{i} \ket{\psi_{i}}\bra{\psi_{i}}$ ($\alpha_{i}\in\{\pm1\}$), the process independently performs the projective measurement by the POVM $\{\ket{\psi_{i}}\bra{\psi_{i}}\}_{i=1}^{2^n}$ consisting of rank-1 operators and the state preparation $\ket{\psi_{j}}$ with a uniform-randomly selected $j\in\{1,2,\ldots,2^n\}$, followed by the classical multiplication of $\alpha_i\alpha_j$ on output $f(\bm{y})$.
Importantly, if the Pauli string $Q$ commutes with $P$ (i.e., they share the same set of eigenstates), then the processes for $Q$ and $P$ can be implemented using the same quantum operations. 
This fact implies that they should be performed simultaneously for more efficient decomposition of ${\rm Id}^{\otimes n}$.

Hence, our idea is to eliminate the redundancy among channels in the original decomposition \eqref{Pauli_cut} by the simultaneous diagonalization of the Pauli strings $P$. 
In particular, it is required to simultaneously diagonalize as many Pauli strings as possible. 
Such diagonalizations are known to be possible with the use of MUBs~\cite{WOOTTERS1989363,PhysRevA.65.032320,seyfarth2019cyclic,9248636}. 
Here, two orthonormal bases $\{\ket{\phi_i}\}$ and $\{\ket{\psi_j}\}$ in the $n$-qubit system are said to be MUBs if and only if 
\begin{align}
    |\braket{\phi_i|\psi_j}|^2=\frac{1}{2^{n}}
\end{align}
holds for all $i,j=0,1,\ldots,2^n-1$.
In the $n$-qubit system, there exist $2^n+1$ distinct orthonormal bases that are mutually unbiased~\cite{WOOTTERS1989363}, and it is known that they describe a grouping of the $4^n-1$ $n$-qubit Pauli strings $\{I,X,Y,Z\}^{\otimes n} \setminus \{I^{\otimes n}\}$ (up to unitarily equivalence)~\cite{PhysRevA.65.032320,seyfarth2019cyclic,9248636}.
That is, MUBs define $2^n+1$ disjoint sets of mutually commuting Pauli strings, each set of which is of the maximal size i.e., $2^n-1$ (excluding the identity).
Applying such a grouping of commuting Pauli strings based on MUBs to the original decomposition (\ref{Pauli_cut}), we derive the following decomposition of ${\rm Id}^{\otimes n}$.

\begin{thm}\label{thm 2}
The $n$-qubit identity channel ${\rm Id}^{\otimes n}(\bullet)$ can be decomposed as
\begin{align}
\label{n-qubit cut in main text}
    {\rm Id}^{\otimes n}(\bullet) &=\sum_{i=1}^{2^{n}} \sum_{\bm{j}\in\{0,1\}^{ n}} \mathrm{Tr}\left[ U_{i} \ket{\bm{j}}\bra{\bm{j}}U_{i}^{\dagger}(\bullet) \right]U_{i} \ket{\bm{j}}\bra{\bm{j}}U_{i}^{\dagger}\nonumber\\
    &\quad-(2^{n}-1)\sum_{\bm{j}\in\{0,1\}^{ n}} \mathrm{Tr}\left[ \ket{\bm{j}}\bra{\bm{j}}(\bullet) \right] \rho_{\bm{j}},
\end{align}
where $\ket{\bm{j}}$ denotes the $n$-qubit computational basis. 
The new input state $\rho_{\bm{j}}$, which depends on the measurement result $\bm{j}$, is defined as
\begin{equation}
    \rho_{\bm{j}}=\sum_{\bm{k}\in\{0,1\}^{n}}\frac{1}{2^n-1}(1-\delta_{\bm{j},\bm{k}})\ket{\bm{k}}\bra{\bm{k}},
\end{equation}
where $\delta_{\bm{j},\bm{k}}$ is the Kronecker delta. 
Also, $\{U_i\}_{i=1}^{2^n}\cup \{I^{\otimes n}\}$ denotes a set of unitary operators that transform the computational base into the $2^n+1$ MUBs. 
In particular, each unitary $U_i$ can be implemented by a Clifford circuit with the maximal depth of $n+2$ on a device with fully-connected qubits. 
\end{thm}
\noindent
The proof of Theorem~\ref{thm 2} is given in Appendix~\ref{sec:proof of thm2}.
Figure~\ref{fig:graphicalrepresentations}(b) shows the graphical representation of the decomposition \eqref{n-qubit cut in main text}. 

\subsubsection{Optimalities in two aspects}

Importantly, the decomposition \eqref{n-qubit cut in main text} achieves the lowest overhead in two aspects: the sampling overhead $\gamma$ and the number of m-p channels $m$.

First, the sampling overhead \eqref{intro:sampling_overhead} of the decomposition is calculated by
\begin{align}\label{eq:gamma_nmp_ours}
    [\gamma^{\mathrm{(mp)}}_n]^2~\mbox{of Eq.~(\ref{n-qubit cut in main text})}=(2^{n+1}-1)^2.
\end{align}
From Eq.~\eqref{general lower bound}, the sampling overhead of our decomposition achieves the lower bound of the sampling overhead. 
That is, the decomposition \eqref{n-qubit cut in main text} is one of the optimal parallel $n$-wire cutting methods among all the LOCC-based schemes.
Surely the most notable feature of our method is that it does not use any ancilla qubits, while the existing optimal $n$-wire cutting method~\cite{brenner2023optimal} uses $n$ ancilla qubits for quantum teleportations.
Moreover, when comparing our decomposition to the best decomposition without ancilla qubits proposed by \textcite{Lowe2023fastquantumcircuit}, we can see an improvement in terms of the sampling overhead from the previous method $(2^{n+1}+1)^2$ to our method $(2^{n+1}-1)^2$.
While the difference is asymptotically negligible as the number of parallel wires $n$ increases, it is crucial when $n$ is relatively small such as $n=1,2,3$.
For instance, for parallel $n~(=1,2,3)$-wire cut, the sampling overhead of Ref.~\cite{Lowe2023fastquantumcircuit} scale as $25,81,289$, and those of our methods scale as $9,49,225$.
In particular, for an original circuit with many clusters of entangling gates, the circuit partitioning of such circuits may require only small number of cuts at multiple locations to split the original circuit into smaller subcircuits~\footnote{For instance, the decomposition of a 12-qubit linear-cluster state demonstrated in Ref.~\cite{PhysRevLett.130.110601} requires the application of 1-wire cuts at three locations.}; as mentioned above, the improvement of sampling overhead in the case of such small $n$ is practically large. 

Second, the number of m-p channels in Eq.~(\ref{n-qubit cut in main text}) is 
\begin{align}\label{eq:channels_ours}
    m_{{\rm Id}^{\otimes n}}~\mbox{of Eq.~(\ref{n-qubit cut in main text})}=2^{n}+1.
\end{align}
Thus, our decomposition saturates the lower bound of $m_{{\rm Id}^{\otimes n}}$ proved in Corollary~\ref{corollary:smallest_m}.
Similar to the case of 1-wire cutting, there are no previous methods that achieve this lower bound on $m_{{\rm Id}^{\otimes n}}$.
Note that $2^n+1$ is equal to the maximal number of MUBs in the $n$-qubit system, and also this number matches the minimum number of distinct orthogonal measurements that in principle are required for the $n$-qubit quantum state tomography~\cite{Bandyopadhyay2002}.
Here, we can see notable differences in the number of m-p channels between our method and the previous parallel $n$-wire cutting methods~\cite{Lowe2023fastquantumcircuit,brenner2023optimal} that have optimal (or nearly optimal) sampling overhead.
Recall that these methods for parallel $n$-wire cut can be written in the unified form (\ref{eq:quasiprobability}) with the m-p channels defined by~Eq.~\eqref{measure-and-prepare channel} (see Appendix~\ref{sec:correspondence}).
In comparison with the optimal parallel $n$-wire cut with quantum teleportations~\cite{brenner2023optimal}, our method makes an exponential improvement on $m$.
That is, the previous decomposition consists of $2^{2^n}+4^n-2^n-1$ m-p channels in total, while our decomposition~\eqref{n-qubit cut in main text} has only $2^n+1$ channels.
This super-exponential scaling is attributed to the quasiprobabilistic state preparation for $n$ Bell pairs (see Appendix~\ref{apdxsub:teleportationcutting}).
Since the number of channels $m$ is directly related to the total execution time $T=T_{\rm C}+T_{\rm Q}$ as described in Sec.~\ref{sec:total execution time}, the exponential improvement on $m$ result in the significant reduction for the classical processing time $T_{\rm C}$.
As for the ${\rm Id}^{\otimes n}$ decomposition with randomized measurements given in \cite{Lowe2023fastquantumcircuit}, which does not reach the optimal sampling overhead, the number of m-p channels $m$ is calculated as (at least) $2^{4n}-2\cdot2^{2n}+3$.
This is because a finite set of random unitary operators that forms a unitary 2-design has at least $2^{4n}-2\cdot2^{2n}+2$ elements~\cite{10.1063/1.2716992,Roy2009} (see Appendix~\ref{sec:proof of lower bound_basis}).
Thus, our method improves the wire-cutting method with randomized measurement in terms of $m$ in addition to $\gamma$.

\subsubsection{The set of quantum circuits $\{U_{i}\}_{i=1}^{2^n}$ in Theorem~\ref{thm 2}}

For completeness, we show a detailed algorithm to explicitly construct the quantum circuits corresponding to $\{U_{i}\}_{i=1}^{2^n}$ in Theorem~\ref{thm 2}.
Theorem~\ref{thm 2} is derived by the application of the MUBs-based grouping to the Pauli strings in Eq.~(\ref{Pauli_cut}). 
In particular, among possible MUBs-based groupings, Theorem~\ref{thm 2} is formulated with respect to the groupings that contain the group $\{I,Z\}^{\otimes n}\setminus\{I^{\otimes n}\}$.
This grouping fulfills the following conditions: 
\begin{itemize}
\item[(i)] $G_i$ and $G_j$ are disjoint for all $i\neq j$.
\item[(ii)] All elements of $G_i$ are $n$-qubit Pauli strings in $\{I,X,Y,Z\}^{\otimes n} \setminus I^{\otimes n}$ which commute with each other, and the cardinality of $G_i$ is $2^n-1$ \cite{9248636,PhysRevA.65.032320}. 
\item[(iii)] $G_{2^n+1}=\{I,Z\}^{\otimes n} \setminus I^{\otimes n}$.
\end{itemize}
For instance, a set $\{G_{i}\}_{i=1}^{5}$ with $G_{1}=\{XI,IX,XX\}$, $G_{2}=\{YZ,ZX,XY\}$, $G_{3}=\{XZ,ZY,YX\}$, $G_{4}=\{YI,IY,YY\}$, and $G_{5}=\{ZI,IZ,ZZ\}$ is a disjoint set of 2-qubit Pauli groups that fulfills the above conditions. 
Note that the existence of such a set satisfying all these conditions is guaranteed by the existing construction~\cite{Bandyopadhyay2002, seyfarth2019cyclic, reggio2023fast}.
%
Then, a unitary operator that simultaneously diagonalizes all elements of $G_i$ corresponds to $U_{i}$ appearing in Theorem~\ref{thm 2}; 
in particular, $U_{2^n+1}=I^{\otimes n}$ for $G_{2^n+1}=\{I,Z\}^{\otimes n}\setminus\{I^{\otimes n}\}$. 
As an approach to generate a unitary circuit that simultaneously diagonalizes mutually commuting Pauli strings, we take the method via the stabilizer formalism \cite{9248636,Crawford2021efficientquantum,KAWASE2023108720}. 
Adapting this circuit construction methods to conform Theorem~\ref{thm 2}, we here provide Algorithm~\ref{alg:circuit_construction} to generate a set 
$\{U_{i}\}_{i=1}^{2^n}$ of quantum circuits in Theorem~\ref{thm 2} from a set of Pauli groups $\{G_{i}\}_{i=1}^{2^n}$ 
that fulfills the conditions (i)--(iii) (see Appendix~\ref{sec:construct_qc} for more detail).
Importantly, Algorithm~\ref{alg:circuit_construction} yields 
the set of Clifford circuits $\{U_i\}$ 
that has at most $n+2$ depth circuit on a device with fully connected qubits; see Lemma~\ref{apdx:maximal_depth_proof} in Appendix~\ref{sec:construct_qc}.
Moreover, as shown in the previous analysis~\cite{Crawford2021efficientquantum}, each Clifford circuit $U_{i}$ produced by this algorithm is composed of $H$, $S$, and $CZ$ gates; the total number of these elementary gates satisfy 
\begin{align}
    N_{H} = n, ~~ N_{S} \leq n, ~~ N_{CZ} \leq \frac{n(n-1)}{2}. 
\end{align}

Here, we supplement the necessary information to execute Algorithm~\ref{alg:circuit_construction}. 
Firstly, this algorithm requires a set $\{G_{i}\}_{i=1}^{2^n+1}$ of Pauli groups that fulfill the above conditions (i)--(iii). 
Secondly, as mentioned above, the algorithm is established via the stabilizer formalism. In this formalism, we use the $2n$-bit binary representation of the $n$-qubit Pauli strings; to clarify this relation, let us define the following function $\phi:\{0,1\}^{2n} \rightarrow \{I,X,Y,Z\}^{\otimes n}$, which maps a $2n$-dimensional binary vector to a single Pauli string with $+1$ phase as
\begin{equation}\label{eq:definition of phi_main}
    \phi(\bm{b}) :=(-i)^{(\bm{b}^{z})^T\bm{b}^{x}} Z^{b^{z}_1}X^{b^{x}_1} \otimes Z^{b^{z}_2}X^{b^{x}_2} \otimes ... \otimes Z^{b^{z}_n}X^{b^{x}_n},
\end{equation}
where $\bm{b} = (b^{z}_1, ...,b^{z}_n, b^{x}_1, ...,b^{x}_n)^{T}\equiv (\bm{b}^{z},\bm{b}^{x})^T$ is a $2n$-dimensional binary vector. From $i=1$ to $i=2^n$,  Algorithm~\ref{alg:circuit_construction} constructs the stabilizer matrix $M=(\bm{g}_{i,1},...,\bm{g}_{i,n})$ for a given Pauli group $G_{i}=\{ P_{i,j}\}_{j=1}^{2^n-1}$ so that $\{\bm{g}_{i,j}\}_{i}^{n}$ forms a basis of $\{\phi^{-1}(P_{i,j})\}_{j=1}^{2^n-1}$.

\begin{algorithm}\label{alg:circuit_construction}
\caption{Circuit constructions for $\{U_{i}\}_{i=1}^{2^n}$ in Theorem~\ref{thm 2}}\label{alg:produce_basis_main}
\SetKwInOut{KwIn}{Input}
\SetKwInOut{KwOut}{Output}
\KwIn{A set $\{G_{i}\}_{i=1}^{2^n+1}$ of Pauli groups satisfying the conditions (i)--(iii)}
\KwOut{A set of quantum circuits $C = \{U_{i}\}_{i=1}^{2^n}$ for implementing Theorem~\ref{thm 2}}
Set $C \leftarrow \{ \}$\;
 \For{\rm $i=1$ to $2^n$}{
   Set $U_{i} \leftarrow H^{\otimes n}$\;
   Set $M\leftarrow(\bm{g}_{i,1},...,\bm{g}_{i,j}) \in F_{2}^{2n \times n}$ where $\{\bm{g}_{i,j}\}_{j=1}^{n}$ is a basis of $\phi^{-1}(G_{i})$ defined in Eq.~(\ref{eq:definition of phi_main}) \;
   
  Performing column-wise Gaussian elimination on $M$ to transform it, such that the lower half of $M$ becomes $I_{n \times n}$ \;

  \For{\rm $k=1$ to $n$}{
  \If{\rm the $(k,k)$-th element of $M$ is 1}{
  Applying $S^{\dagger}$ gate on the $k$-th qubit: $U_{i} \leftarrow S_{k}^{\dagger} U_{i}$\;
  }
  }
  \For{\rm $l=1$ to $n-1$}{
  \For{\rm $m=l+1$ to $n$}{
  \If{\rm the $(l,m)$-th element of $M$ is $1$}{
  Applying $CZ$ gate on the $l$-th and the $m$-th qubits: $U_{i} \leftarrow CZ_{l,m} U_{i}$\;
  }
  }
  }
  Optimizing the positions of $CZ$ gates to minimize the depth of $U_{i}$\;
  Appending a circuit $U_{i}$ to a set $C$: $C \leftarrow C \cup U_{i}$
  }
\Return{$C = \{U_{i}\}_{i=1}^{2^n}$}
\end{algorithm}

Here we provide a numerical experiment, for the purpose of evaluating the maximum number of gates required for implementing the circuits (thus, we omitted the step of the depth optimization in Algorithm~\ref{alg:produce_basis_main}). 
 The experimental setup is as follows. 
 For each number of qubit $n=1,\ldots,12$, we first generate the set $\{G_{i}\}_{i=1}^{2^n}$ of $n$-qubit Pauli groups \footnote{To generate the set $\{U_{i}\}_{i=1}^{2^n}$, we have used the publicly available package $\mathtt{psfam.py}$ in Ref.~\cite{reggio2023fast}}, and employ Algorithm~\ref{alg:produce_basis_main} to convert the set $\{G_{i}\}_{i=1}^{2^n}$ into the set of quantum circuits $\{U_{i}\}_{i=1}^{2^n}$. Then, we calculate the maximum number of $CZ$ ($N_{CZ,max}$) and $S^{\dagger}$ ($N_{S^{\dagger},{\rm max}}$) as well as that of all gates contained in $U_{i}$ ($N_{\rm all,max}$) \footnote{Note that it is clear from the construction of Algorithm~\ref{alg:produce_basis_main} that each quantum circuit in $\{U_{i}\}_{i=1}^{2^n}$ contains $n$ Hadamard gates; hence we did not count this gate.}. 
 These processes were repeated from $n=1$ to $n=12$. 
 The results of this experiment are presented in Fig.~\ref{fig:num_of_gates}. 
 Also, the circuit diagram for $\{U_{i}\}_{i=1}^{2^n}$ are provided in Appendix~\ref{sec:circuits} up to 4-qubit.

\begin{figure*}[htb]
 \centering
 \begin{tabular}{ccc}
 \includegraphics[scale=0.56]{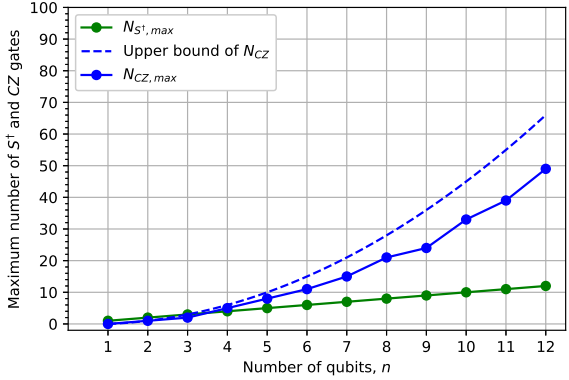}
 &&
 \includegraphics[scale=0.56]{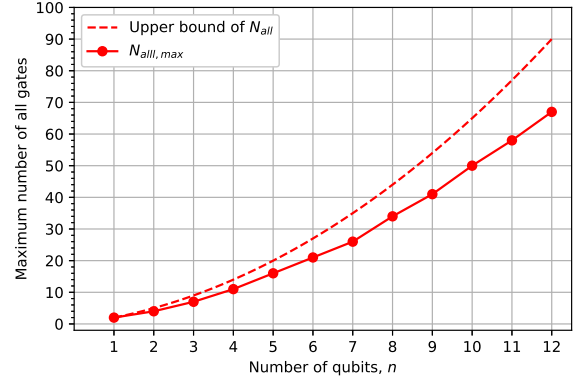}
 \\
(a) The maximum number of $S^{\dagger}$ and $CZ$ gates &&(b) The maximum number of all gates \\
 \end{tabular}
 \caption{The maximum number of $CZ$ and $S^{\dagger}$ as well as that of all gates contained in $U_{i}$, generated by the method of Ref.~\cite{reggio2023fast} and Algorithm~\ref{alg:produce_basis_main}. 
 The plots in Fig.~\ref{fig:num_of_gates}(a) displays the maximum number of $S^{\dagger}$ gates ($N_{S^{\dagger},{\rm max}}$, the green plots) and the maximum number of $CZ$ gates ($N_{CZ,{\rm max}}$, the blue plots) contained in the circuit $U_{i}$ for $i=1,\ldots,2^n$, plotted versus the number of qubits $n$. 
 The dashed blue curve represents the theoretical upper bound of the number of $CZ$ gates in each circuit as shown in Eq.~(\ref{eq:bound}) of Lemma~\ref{apdx:maximal_depth_proof}. 
 Fig.~\ref{fig:num_of_gates}(b) shows that the maximum number of total counts of $H, S^{\dagger}$ and $CZ$ gates ($N_{\rm all,max}$, the red plots) in each circuit and that the dashed red curve provides the theoretical upper bound of the total gates, derived from Eq.~(\ref{eq:bound}) of Lemma~\ref{apdx:maximal_depth_proof}.
 }
 \label{fig:num_of_gates}
\end{figure*}


From Fig.~\ref{fig:num_of_gates}(a), it can be observed that the maximum number of $S^{\dagger}$ gates, $N_{S^{\dagger},{\rm max}}$, increases as 
$N_{ S^{\dagger},{\rm max}}=n$ with $n$ the number of qubits. 
Also, Fig.~\ref{fig:num_of_gates}(a)(b) shows that both the maximum number of $CZ$ gates, $N_{CZ, {\rm max}}$, and that of all gates, $N_{\rm total,max}$, increase quadratically with respect to the number of qubits, yet below the theoretical upper bounds, implying an easier implementation of our method than we expected from the theory. 

Lastly, let us see the circuit depth of a random Clifford circuit $V$, which is employed in the ${\rm Id}^{\otimes n}$ decomposition given in \cite{Lowe2023fastquantumcircuit}. 
The circuit $V$ has at most $O(n^2)$ elementary gates, and its maximal depth is $O(n\log n)$ on a device with fully connected qubits \cite{9605330}. 
While the circuit depth and the number of elementary gates of the circuit $V$ is comparable to that of $U_{i}$ constructed by Algorithm~\ref{alg:circuit_construction}, the random Clifford circuit $V$ additionally requires $O(n^2)$ time complexity to sample a single circuit~\cite{9605330}.

\subsubsection{Implementation of Theorem~\ref{thm 2}}

As stated in Sec.~\ref{sec:preliminaries}, to implement the quasiprobability decomposition in Theorem~\ref{thm 2}, we use  
the Monte-Carlo sampling. 
Thus, it is convenient to rewrite the decomposition~\eqref{n-qubit cut in main text} as the following form:
\begin{equation}\label{eq:quasi-form_1}
    {\rm Id}^{\otimes n}(\bullet)
    = \gamma \left\{ \sum_{i=1}^{2^n} p_{i} (+1) \mathcal{E}_{i}^{(1)}(\bullet) + q (-1) \mathcal{E}^{(2)}(\bullet) \right\},
\end{equation}
where $\gamma:=(2^{n+1}-1)$, $p_{i}:=1/(2^{n+1}-1)$, and $q:=(2^{n}-1)/(2^{n+1}-1)$ are probabilities. 
This form contains $(2^n+1)$ m-p channels, and they can be classified into two types of channels: $\mathcal{E}_{i}^{(1)} (i=1,...,2^n)$ and $\mathcal{E}^{(2)}$. 
The first type of channels $\mathcal{E}_{i}^{(1)}$ are defined as
\begin{equation}
    \mathcal{E}_{i}^{(1)}(\bullet) := \sum_{\bm{j}\in\{0,1\}^{ n}} \mathrm{Tr}\left[ U_{i} \ket{\bm{j}}\bra{\bm{j}}U_{i}^{\dagger}(\bullet) \right]U_{i}\ket{\bm{j}}\bra{\bm{j}}U_{i}^{\dagger}.
\end{equation}
In practice, this channel can be realized by measurement on an entering state with the POVM $\{ U_{i} \ket{\bm{b}}\bra{\bm{b}}U_{i}^{\dagger} \} ~(\bm{b}\in\{0,1\}^{n})$ and preparation of a quantum state $U_{i} \ket{\bm{j}}$, which depends on the measurement result $\bm{j}\in\{0,1\}^{n}$.
The second channel is defined as
\begin{equation}
    \mathcal{E}^{(2)}(\bullet) := \sum_{\bm{j}\in\{0,1\}^{ n}} \mathrm{Tr}\left[ \ket{\bm{j}}\bra{\bm{j}}(\bullet) \right] \rho_{\bm{j}}.
\end{equation}
This can be implemented by measuring an entering state with $\{\ket{\bm{b}}\bra{\bm{b}}\} ~(\bm{b}\in\{0,1\}^{n})$ and preparing 
a new input state randomly selected from the set $\{\ket{0}\bra{0}, \ket{1}\bra{1}\}^{\otimes n}$ excluding the quantum state $\ket{\bm{j}}\bra{\bm{j}}$ with equal probability where $\bm{j}\in \{0,1\}^{n}$ is the measurement result.

\section{Discussion and Conclusions}
\label{sec:conclusion}

We propose new decompositions of the identity channels for simulating a large quantum circuit with limited quantum resources in the quantum circuit-cutting framework.
The core idea is to apply the grouping technique for mutually commuting observables on the original decomposition~\cite{PhysRevLett.125.150504} and eliminate the redundancy in measurement and state preparation operations required for the wire cutting. 
Then, we obtain a new decomposition of the parallel $n~(\geq 1)$-qubit identity channel comprised of the m-p channels, and these channels in our decomposition contain LOCC that naturally arise from the grouping. 
Here, the m-p channels defined in Eq.~\eqref{measure-and-prepare channel} (which is slightly extended from the original definition~\cite{doi:10.1142/S0129055X03001709}) also contain principle channels used in the previous works on parallel wire cutting~\cite{PhysRevLett.125.150504,Lowe2023fastquantumcircuit,brenner2023optimal}.

Notably, the proposed decomposition achieves both of the lower bounds in the sampling overhead $\gamma$ and the number of m-p channels $m$; in particular, the lower bound on $m$ for the ancilla-free parallel wire cutting (Theorem~\ref{thm:number_mp_channels}) is first derived in this paper. 
The sampling overhead in our decomposition for the parallel $n$ wires is $(2^{n+1}-1)^2$, which achieves the lower bound given in Ref.~\cite{brenner2023optimal} that allows any LOCC. 
Moreover, our method completely eliminates the need for ancilla qubits in the previous optimal method~\cite{brenner2023optimal} that uses $n$ ancilla qubits to introduce quantum teleportations. 
As for the number of channels, our decomposition contains $2^n+1$ m-p channels, which is strictly smaller than any existing methods. 
This is practically important, because the smaller $m$ implies the shorter classical processing time for quantum circuit compilation. 
Furthermore, quantum circuits required for our decomposition are efficiently constructed with at most $(n+2)$ depth on a device with fully connected qubits; actually, we explicitly show the circuits up to 12 qubits. 
Overall, our method is a doubly optimal LOCC-based, ancilla-free decomposition for the $n$-qubit identity channel in terms of both sampling overhead and the number of quantum circuits.
It should also be emphasized that our decomposition improves the (worst-case) sampling overhead for an arbitrary $k$-wire cutting from $16^k$ by the existing best decomposition to $9^k$ at the cost of classical communication but needs no ancilla qubits.

In the derivation of the proposed parallel $n$-wire cutting, the grouping technique naturally yields the MUBs in the m-p channels.
MUBs is an important concept for quantum state estimation; specifically, the measurements based on MUBs provide the minimal and optimal (in the sense of statistical errors) method to estimate a density matrix~\cite{WOOTTERS1989363}.
Such double optimality on MUBs seems to be closely related to the features of our method, i.e., the double optimality on $\gamma$ and $m$, and therefore the wire-cutting methods may have an underlying structure similar to quantum state estimation.

Considering the cut problem for multiple wires including non-parallel ones in a quantum circuit, we first partition the wires into the sets of parallel wires and then apply our optimal decomposition method for each of the sets.
However, this procedure does not have the optimal sampling overhead in the sense of {\it global}, though each set of parallel wires is decomposed by our decomposition with {\it locally} optimal cost.
Since the optimality of sampling overhead $(2^{n+1}-1)^2$ derived in~\cite{brenner2023optimal} holds for a non-parallel $n$-wire cutting, it is an open problem whether the globally optimal cutting can be accomplished without any additional ancilla qubits.

{\bf Note added}:
After the first preprint version of this work was presented on arXiv, \textcite{pednault2023alternative} also developed an $n$-qubit parallel wire-cutting method without ancilla qubits, which achieves the same sampling overhead as ours.
Furthermore, the number of channels in the decomposition presented in \cite{pednault2023alternative} achieves the lower bound in the Corollary~\ref{corollary:smallest_m} up to $n(\leq 2)$-qubit.
Also, Ref.~\cite{10025537} proposed a wire-cutting scheme that uses parameterized unitary bases in each cut with the intuition of MUBs to reduce the sampling overhead.


{\bf Acknowledgements:} 
We thank Suguru Endo and Hidetaka Manabe for the helpful discussions.
K.W. was supported by JST SPRING, Grant Number JPMJSP2123.
This work was supported by MEXT Quantum Leap Flagship Program Grants No. JPMXS0118067285 and No. JPMXS0120319794. 

\bibliography{bib}

\appendix
\newpage
\onecolumngrid

\begin{figure*}[htb]
\centering
\begin{center}
 \includegraphics[width=170mm]{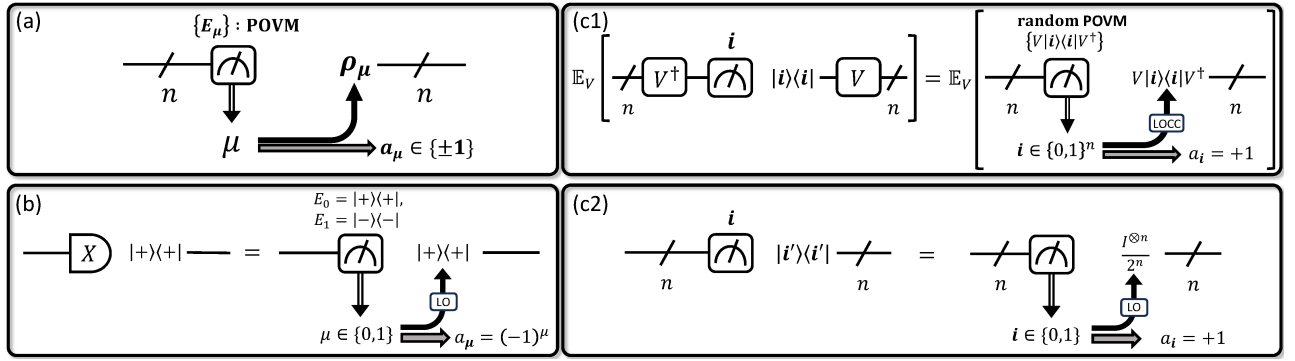}    
\end{center}
\caption{Schematic illustrations of an $n$-qubit m-p channel and its examples.
(a) an $n$-qubit m-p channel defined in Eq.~(\ref{measure-and-prepare channel}).
(b) The description of $\mathrm{Tr}[X(\bullet)]\ket{+}\bra{+}$ contained in the original quantum circuit cutting formulation~\cite{PhysRevLett.125.150504} by the 1-qubit m-p channel. 
(c1)(c2) The description of the circuit cutting scheme with randomized measurements~\cite{Lowe2023fastquantumcircuit} by the $n$-qubit m-p channels.}
\label{apdx:description of previous studies}
\end{figure*}
%
\section{The description of existing quantum circuit cutting by the m-p channels.}\label{sec:correspondence}



\subsection{Quantum circuit cutting introduced in~\textcite{PhysRevLett.125.150504}}\label{apdxsub:harrow_decom}
The decomposition of the single-qubit identity channel ${\rm Id}(\bullet)$ given in Eq.~(\ref{quantum circuit cutting}) can be represented with the use of the 1-qubit m-p channels defined in Eq.~(\ref{measure-and-prepare channel}):
\begin{equation}\label{apdx:harrow_decomp}
    {\rm Id}(\bullet) = \sum_{i=1}^{8} c_{i} \mathcal{E}_{i}(\bullet),~~~
    \mathcal{E}_{i}(\bullet) = \sum_{\mu\in\{0,1\}} a_{i{\mu}} \mathrm{Tr}\left[ E_{i{\mu}}(\bullet) \right] \rho_{i{\mu}},
\end{equation}
where the set $\{c_{i},(a_{i\mu}, E_{i\mu}, \rho_{i\mu})_{\mu=0,1} \}_{i=1}^{8}$ follows the Table~\ref{table:apdx_pengcut}.
For instance, the channel $\mathcal{E}_{3}(\bullet):=\mathrm{Tr}[X(\bullet)]\ket{+}\bra{+}$, depicted in Fig.~\ref{apdx:description of previous studies}(b), can be implemented by measuring the input state with POVM $\{E_{\mu}\}_{\mu=0,1}$ where $E_{0}=\ket{+}\bra{+}, E_{1}=\ket{-}\bra{-}$, newly inputting the state $\rho_{\mu=0}=\rho_{\mu=1}=\ket{+}\bra{+}$, and multiplying the output with coefficients $a_{\mu}=(-1)^\mu$.
Note that any channels in this decomposition do not require classical communication between the measured system and the state-input system; that is, the decomposition consists of only local operations (LO).

This decomposition is applicable to arbitrary $n$-wire cutting.
In particular, using this decomposition, the $n$-qubit identity channel $\mathrm{Id}^{\otimes n}(\bullet)$ can be represented by $n$-qubit m-p channels as follows.
\begin{equation}\label{apdx:harrow_decomp_n}
    \mathrm{Id}^{\otimes n}(\bullet) = \sum_{\bm{i}\in\{1,2,\cdots,8\}^n} c_{\bm{i}}\, \mathcal{E}_{\bm{i}}(\bullet),~~~
    \mathcal{E}_{\bm{i}}(\bullet) = \sum_{\bm{\mu}\in\{0,1\}^{n}} a_{\bm{i}\bm{\mu}} \mathrm{Tr}\left[ E_{\bm{i}\bm{\mu}}(\bullet) \right] \rho_{\bm{i}\bm{\mu}},
\end{equation}
where $\bm{i}=(i_{1},i_{2},...,i_{n})\in\{1,...,8\}^n$ and $c_{\bm{i}}=\prod_{l=1}^{n}c_{i_{l}} \in \{\pm \frac{1}{2^n}\}$ denote a label and a real weight of the channel $\mathcal{E}_{\bm{i}}$, respectively.
The $n$-qubit m-p channel $\mathcal{E}_{\bm{i}}$ consists of coefficients $a_{\bm{i}\bm{\mu}}=\prod_{l=1}^{n}a_{\mu_{l}}\in \{\pm 1\}$ for the measurement results $\bm{\mu}=(\mu_{1},\mu_{2},...,\mu_{n})\in\{0,1\}^n$, POVM $\{E_{\bm{i}\bm{\mu}}=\bigotimes_{l=1}^{n} 
E_{i_{l}\mu_{l}}\}_{\bm{\mu}}$, and the tensor product state of $1$-qubit input states $\rho_{\bm{i}\bm{\mu}}=\bigotimes_{l=1}^{n} \rho_{i_{l}\mu_{l}}$.
From Eq.~(\ref{apdx:harrow_decomp_n}), the sampling overhead is calculated as $(\sum_{\bm{i}} |c_{\bm{i}}| )^2 = 16^n$ and the number of channels is $8^n$.

\begin{table*}[ht]
\renewcommand{\arraystretch}{1.4}
\begin{tabular}{|ccc|ccc|ccccc ccc ccc|ccccc ccc ccc|}
 \hline
 &\multirow{2}{*}{$i$}&&& \multirow{2}{*}{$c_{i}$} && &&&&&& $\mu=0$ &&&& & &&&&&& $\mu=1$ &&&& \\
 \cline{7-28}
 & &&& &&&&&$a_{i\mu}$&&&$E_{i\mu}$&&&$\rho_{i\mu}$ &&&&&$a_{i\mu}$&&&$E_{i\mu}$&&&$\rho_{i\mu}$ &\\ 
 \hline\hline
 & 1 &&&$+1/2$&&&&&$+1$&&&$\ket{0}\bra{0}$&&&$\ket{0}\bra{0}$ &&&&&$+1$&&&$\ket{1}\bra{1}$&&&$\ket{0}\bra{0}$ &\\
 \hline
 & 2 &&&$+1/2$&&&&&$+1$&&&$\ket{0}\bra{0}$&&&$\ket{1}\bra{1}$ &&&&& $+1$ &&& $\ket{1}\bra{1}$&&& $\ket{1}\bra{1}$ &\\
 \hline
 & 3 &&&$+1/2$&&&&&$+1$&&&$\ket{+}\bra{+}$&&&$\ket{+}\bra{+}$ &&&&&$-1$&&&$\ket{-}\bra{-}$&&&$\ket{+}\bra{+}$&\\
 \hline
 & 4 &&&$-1/2$&&&&&$+1$&&&$\ket{+}\bra{+}$&&&$\ket{-}\bra{-}$ &&&&&$-1$&&&$\ket{-}\bra{-}$&&&$\ket{-}\bra{-}$ &\\
 \hline
 & 5 &&&$+1/2$&&&&&$+1$&&&$\ket{+i}\bra{+i}$&&&$\ket{+i}\bra{+i}$ &&&&& $-1$ &&& $\ket{-i}\bra{-i}$&&& $\ket{+i}\bra{+i}$ &\\
 \hline
 & 6 &&&$-1/2$&&&&&$+1$&&&$\ket{+i}\bra{+i}$&&&$\ket{-i}\bra{-i}$ &&&&&$-1$&&&$\ket{-i}\bra{-i}$&&&$\ket{-i}\bra{-i}$&\\
 \hline
 & 7 &&&$+1/2$&&&&&$+1$&&&$\ket{0}\bra{0}$&&&$\ket{0}\bra{0}$ &&&&&$-1$&&&$\ket{1}\bra{1}$&&&$\ket{0}\bra{0}$ &\\
 \hline
 & 8 &&&$-1/2$&&&&&$+1$&&&$\ket{0}\bra{0}$&&&$\ket{1}\bra{1}$ &&&&& $-1$ &&& $\ket{1}\bra{1}$&&& $\ket{1}\bra{1}$ &\\
 \hline
\end{tabular}
\caption{The elements $\{c_{i},(a_{i\mu}, E_{i\mu}, \rho_{i\mu})_{\mu=0,1} \}_{i=1}^{8}$ of the decomposition in Eq.~(\ref{apdx:harrow_decomp}).
The coefficient $a_{ki}$, the POVM element $E_{{k}i}$, and the new input state $\rho_{{k}i}$ are in the $k$th m-p channel $\mathcal{E}_{k}(\bullet)$, and the coefficient $c_{k}$ denotes a weight for each $\mathcal{E}_{k}(\bullet)$.}
\label{table:apdx_pengcut}
\end{table*}

\subsection{Quantum circuit cutting with randomized measurements in~\textcite{Lowe2023fastquantumcircuit}}\label{apdxsub:randomizedcutting}

The decomposition of the $n$-qubit identity channel ${\rm Id}^{\otimes n}$ given in~\cite{Lowe2023fastquantumcircuit} can be expressed with the principle $n$-qubit m-p channels as follows:
\begin{align}\label{apdx:decomp_random}
    {\rm Id}^{\otimes n}(\bullet) &= (2^n+1) \mathbb{E}_{V}\left[ \sum_{\bm{i}\in\{0,1\}^{n}} \mathrm{Tr}\left[ V\ket{\bm{i}}\bra{\bm{i}}V^{\dagger} (\bullet) \right] V\ket{\bm{i}}\bra{\bm{i}}V^{\dagger}\right] - 2^n \sum_{\bm{i}\in\{0,1\}^{ n}} \mathrm{Tr}\left[ \ket{\bm{i}}\bra{\bm{i}} (\bullet) \right] \frac{I^{\otimes n}}{2^n}\notag\\
    &=\mathbb{E}_{V}\sum_{\bm{i}\in\{0,1\}^{ n}} \mathrm{Tr}\left[ V\ket{\bm{i}}\bra{\bm{i}}V^{\dagger} (\bullet) \right]{\rho_{\rm cs}},~~~{\rho_{\rm cs}}:=(2^n+1)V\ket{\bm{{i}}}\bra{\bm{{i}}}V^{\dagger}-I^{\otimes n},
\end{align}
where $\{\ket{\bm{i}}\}$ is the $n$-qubit computational base, and $\mathbb{E}_{V}\left[\bullet\right]$ denotes the expectation over random unitary operators $V$ forming a unitary 2-design.
Note that when we enter a quantum state $\sigma$ into ${\rm Id}^{\otimes n}$ channel, the decomposition can be understood with the notion of classical shadow for $\sigma$, especially the classical shadow associated with random $n$-qubit Clifford measurements~\cite{Huang2020}.
That is, Eq.~(\ref{apdx:decomp_random}) indicates exactly that an infinite number of classical shadows (or the expectation of $\rho_{\rm cs}$) reproduces the underlying target state $\sigma$ completely.
The first channel of the decomposition in the upper line of Eq.~(\ref{apdx:decomp_random}) can be written as 
\begin{align}\label{apdx:rand_decom_1chan}
    (2^n+1)\sum_{V\in\mathcal{V}} p(V) \sum_{\bm{i}\in\{0,1\}^{n}} \mathrm{Tr}\left[ V\ket{\bm{i}}\bra{\bm{i}}V^{\dagger} (\bullet) \right] V\ket{\bm{i}}\bra{\bm{i}}V^{\dagger},
\end{align}
where $\mathcal{V}$ is a finite set of unitary operators and $p(\bullet)$ is a probability function over $\mathcal{V}$.
For a valid decomposition, it is sufficient that the set $\mathcal{V}$ with $p(\bullet)$ forms a unitary 2-design in $2^n$ dimension.
Clearly, Eq.~(\ref{apdx:rand_decom_1chan}) is a sum of m-p channels composed of a random POVM in the form of $\{E_{\bm{i}}\}=\{V\ket{\bm{i}}\bra{\bm{i}}V^{\dagger}\}$, a coefficient $a_{\bm{i}}=1$, and a new input state $\rho_{\bm{i}} = V\ket{\bm{i}}\bra{\bm{i}}V^{\dagger}$.
Also, the second channel is a m-p channel with the computational basis measurement and the maximally-mixed state.
These two channels are schematically illustrated in Fig.~\ref{apdx:description of previous studies}(c1)(c2).
From Eqs.~(\ref{apdx:decomp_random}) and (\ref{apdx:rand_decom_1chan}), the sampling overhead is calculated as 
\begin{align}
    \left(\sum_{V\in\mathcal{V}} \left|(2^n+1)p(V)\right|+ \left|-2^n\right| \right)^2 = (2^{n+1}+1)^2,
\end{align}
and the number of channels is $|\mathcal{V}|+1$ where $|\mathcal{V}|$ denotes the number of elements in $\mathcal{V}$. 
In Sec.~\ref{sec:min_of_mp_channels}, we detail the lower bound of $|\mathcal{V}|$.

\subsection{Quantum circuit cutting with quantum teleportation in~\textcite{brenner2023optimal}}\label{apdxsub:teleportationcutting}

\begin{figure*}[htb]
\centering
\begin{center}
 \includegraphics[width=170mm]{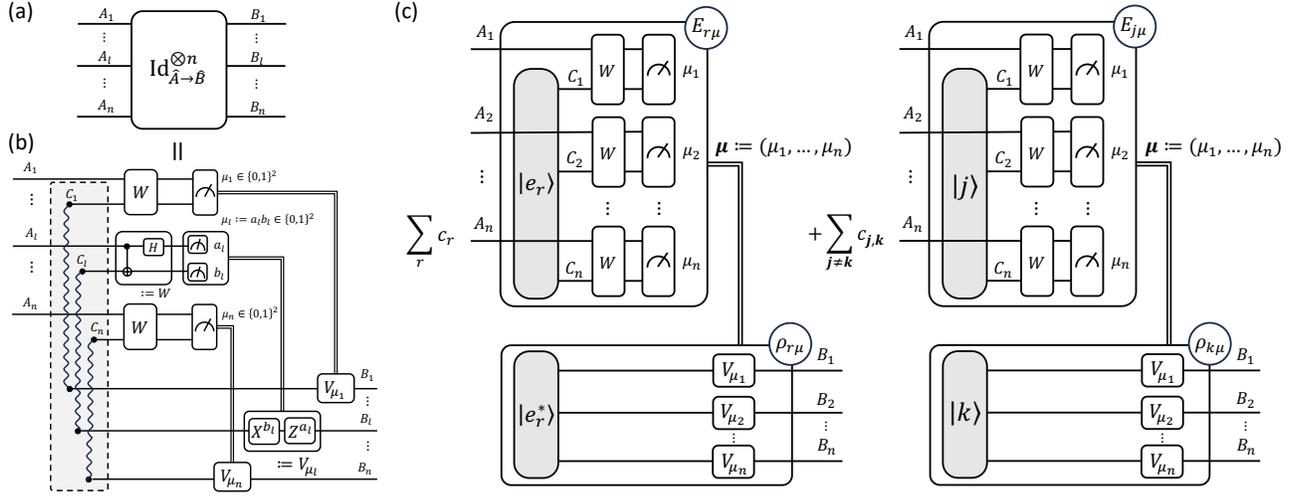}    
\end{center}
\caption{Schematic illustrations of the basic idea of the $n$-wire cutting proposed by \textcite{brenner2023optimal}. Here, we consider the parallel $n$-wire cutting, i.e., the decomposition of $\rm{Id}^{\otimes n}$. The $n$-qubit identity channel $\mathrm{Id}_{A\rightarrow B}^{\otimes n}$ in (a) can be regarded as $n$ quantum teleportation protocols in (b). By applying the quasiprobability decomposition for $n$ Bell pairs (the wavy lines) in (b), based on the decomposition proposed by \textcite{PhysRevA.59.141}, the decomposition of $\mathrm{Id}^{\otimes n}$ is obtained as in (c).
}
\label{apdx:teleportation_based_cutting}
\end{figure*}

\textcite{brenner2023optimal} proposed the $n$-wire cutting method based on quantum teleportation protocol, which is proven to have optimal sampling overhead. 
Now, we consider the application of their proposal for cutting the $n$-qubit identity channel $\mathrm{Id}^{\otimes n}_{\hat{A}\rightarrow \hat{B}}:=\bigotimes_{l=1}^{n} \mathrm{Id}_{A_{l}\rightarrow B_{l}}$, where $\hat{A}:=A_{1}\otimes A_{2} \otimes ... \otimes A_{n}$ and $\hat{B}:=B_{1}\otimes B_{2} \otimes ... \otimes B_{n}$ denote the $n$-qubit input and output systems, respectively (Fig.~\ref{apdx:teleportation_based_cutting}(a)). 
By replacing each single wire $\mathrm{Id}_{A_{l}\rightarrow B_{l}}$ with the quantum teleportation protocol for $l=1,...,n$, we obtain the quantum circuit shown in the bottom of Fig.~\ref{apdx:teleportation_based_cutting}(a).
In each teleportation protocol, with a pre-shared Bell pair between the receiver system $B_{l}$ and an additional ancilla system $C_{l}$, a quantum state on $A_{l}$ is transferred into $B_{l}$ by local operations and classical communication.
That is, the sender first performs unitary gates $W:= H_{A_l} {\rm CX}_{A_l,C_l}$ and the computational basis measurement on $A_{l}\otimes C_{l}$ with the measurement results $\mu_l=(a_l,b_l)\in \{0,1\}^2$.
Then, the receiver recovers the original state on $B_{l}$ by operating the unitary gate $V_{\mu_{l}}:=Z^{a_l}X^{b_l}$ based on the received result $\mu_l=(a_l,b_l)$ via classical communication.
Note that in the above procedure, the system $\hat{A}$ and $\hat{B}$ have quantum connections through the Bell pairs.

To ``cut'' the quantum connections between $\hat{A}$ and $\hat{B}$, the authors of Ref.~\cite{brenner2023optimal} employ the quasiprobabilistic state preparation of $n$ Bell pairs, based on the following decomposition proposed by~\textcite{PhysRevA.59.141}:
\begin{align}
    &\bigotimes_{l=1}^n \left(|\Psi_{\rm Bell}\rangle \langle \Psi_{\rm Bell}|_{C_{l},B_l}\right)= \frac{1}{{2^n}} \sum_{{j},{k}=1}^{2^n} \ket{{j}}\bra{{k}}_{\hat{C}} \otimes \ket{{j}}\bra{{k}}_{\hat{B}} \notag\\
    &= 2^n \left( \sum_{r=1}^{2^{2^n}-1} \frac{1}{2^{2^n}-1} \ket{e_{r}}\bra{e_{r}}_{\hat{C}}\otimes \ket{e_{r}^{*}}\bra{e_{r}^{*}}_{\hat{B}} \right)
    - (2^n-1) \left( \sum_{{j}\neq {k}} \frac{1}{2^n(2^n-1)} \ket{{j}}\bra{{j}}_{\hat{C}} \otimes \ket{{k}}\bra{{k}}_{\hat{B}} \right),
\end{align}
where $|\Psi_{\rm Bell}\rangle_{C_l,B_l}:=({|0\rangle_{C_l}|0\rangle_{B_l}+|1\rangle_{C_l}|1\rangle_{B_l}})/{\sqrt{2}}$, $\hat{C}:=C_{1}\otimes C_{2}\otimes ...\otimes C_{n}$ is the $n$-qubit ancilla system, and $\{\ket{j}\}_{j=1}^{2^n}$ denotes the $n$-qubit computational base with integer labels.
The $n$-qubit quantum states $\ket{e_{r}},\ket{e^{*}_{r}}$ are defined by
\begin{equation}
    \ket{e_{r}}=\sum_{j=1}^{2^n} \frac{1}{\sqrt{2^n}} \exp\left({i2\pi r}\frac{2^{j-1}-1}{2^{2^n}-1}\right) \ket{j},~~~\ket{e_{r}^{*}}=\sum_{j=1}^{2^n} \frac{1}{\sqrt{2^n}} \exp\left({-i2\pi r}\frac{2^{j-1}-1}{2^{2^n}-1}\right) \ket{j}.
\end{equation}
Combining this quasiprobabilistic state preparation method with the teleportation protocols in Fig.~\ref{apdx:teleportation_based_cutting}(a), the ${\rm Id}^{\otimes n}$ decomposition in Ref.~\cite{brenner2023optimal} can be represented as:
\begin{align}
    {\rm Id}^{\otimes n}(\bullet) 
    &= \sum_{r=1}^{2^{2^n}-1} c_{r} \,\mathcal{E}_{r}(\bullet) + \sum_{j\neq k} c_{j,k} \,\mathcal{E}_{j,k}(\bullet),\label{apdx:brenner_decom1}
\end{align}
where $c_{r}:=2^n/(2^{2^n}-1)$ and $c_{j,k}:=-1/2^n$ are coefficients of the channels $\mathcal{E}_{r}(\bullet)$ and $\mathcal{E}_{j,k}(\bullet)$, respectively, and these channels are $n$-qubit m-p channels defined by
\begin{equation}\label{apdx:eq_brenner_channels}
    \mathcal{E}_{r}(\bullet):= \sum_{\mu\in\{0,1\}^{2n}} \mathrm{Tr}\left[ E_{r,\mu} (\bullet) \right] \rho_{r,\mu},~~~ \mathcal{E}_{j,k}(\bullet):=\sum_{\mu\in\{0,1\}^{2n}} \mathrm{Tr}\left[ E_{j,\mu} (\bullet) \right] \rho_{k,\mu}.
\end{equation}
The channel $\mathcal{E}_{r}(\bullet)$ are composed of the POVM measurement $\{E_{r,\mu}:=\braket{e_{r}|\cdot \bigotimes_l ({W}^{\dagger}|\mu_l}\braket{\mu_l|{W})\cdot |e_{r}}\}_{\mu}$ and the preparation of the $n$-qubit state $\rho_{r,\mu}:=V_{\mu}\ket{e_{r}^*}\bra{e_{r}^*}V_{\mu}^{\dagger}$, where $V_{\mu}:=\bigotimes_l V_{\mu_l}$ and $\mu:=(\mu_{1},...,\mu_{n})\in\{0,1\}^{2n}$ are the measurement results.
As for the channel $\mathcal{E}_{j,k}(\bullet)$, it contains the POVM measurement 
$\{E_{j,\mu}:=\braket{j|\cdot \bigotimes_l ({W}^{\dagger}|\mu_l}\braket{\mu_l|{W})\cdot|j}\}_{\mu}$ and the preparation of the $n$-qubit state $\rho_{k,\mu}:=V_\mu \ket{k}\bra{k} V_{\mu}^\dagger$.
The implementation of the channels in Eq.~(\ref{apdx:eq_brenner_channels}) is schematically illustrated in Fig.~\ref{apdx:teleportation_based_cutting}(c).
From this formulation, the sampling overhead and the number of m-p channels of this decomposition are $(2^{n+1}-1)^2$ and $2^{2^n}+4^n-2^n-1$, respectively.




\section{Number of m-p channels for quantum channel decomposition}\label{sec:proof of lower bound_basis}
\subsection{Transfer matrix}
This subsection briefly reviews the \textit{transfer matrix} of a quantum channel~\cite{Nielsen2021gatesettomography,Mitarai_2021}.
Taking a basis set of the Hilbert space $\mathcal{H}=\mathbb{C}^{d}$ describing an $n$-qubit quantum system ($d=2^n$), (bounded) linear operators such as unitary operators act on $\mathcal{H}$ can be represented by $d\times d$ matrices.
In the same way, denoting a $d^2$-dimensional vector space comprised of the (bounded) linear operators as $\mathcal{B}(\mathcal{H})$, linear transformations on $\mathcal{B}(\mathcal{H})$ can also be represented as $d^2\times d^2$ matrices by choosing a basis for $\mathcal{B}(\mathcal{H})$.
This matrix representation of a linear transformation $\Lambda$ on $\mathcal{B}(\mathcal{H})$ is often called the \textit{transfer matrix} of $\Lambda$.
Thus, we can use the transfer matrices for the description of quantum channels, which are linear transformations with completely positive and trace-preserving constraints.

The vector space $\mathcal{B}(\mathcal{H})$, called Hilbert-Schmidt space, equipped with an inner product $\langle A, B\rangle:=\mathrm{Tr}\left[A^\dagger B\right]$ is Hilbert space as well as $\mathcal{H}$.
Then, we can write an element $A$ of $\mathcal{B}(\mathcal{H})$ and its dual as a \textit{superket} (a column vector) $|A\rangle\!\rangle$ and a \textit{superbra} (a row vector) $\langle\!\langle A|$, respectively, so that $\langle\!\langle A|B\rangle\!\rangle=\mathrm{Tr}\left[A^\dagger B\right]$ holds.
Here, the terminology comes from that operators on $\mathcal{B}(\mathcal{H})$ are called \textit{superoperators}.
Since normalized Pauli operators $\sigma_{0}=I/\sqrt{2}$, $\sigma_{1}=X/\sqrt{2}$, $\sigma_{2}=Y/\sqrt{2}$, and $\sigma_{3}=Z/\sqrt{2}$ are orthonormal under the inner product, it is useful to take an orthonormal basis of $\mathcal{B}(\mathcal{H})$ in the form of $\sigma_{\vec{i}}:=\sigma_{i_{1}}\otimes ... \otimes \sigma_{i_{n}},~(\vec{i}=(i_1,\cdots,i_n))$ with $i_{k}\in\{0,1,2,3\}$ being the index associated with the $k$-th qubit.
Using the notation, for instance, we can write the expectation of an observable $O$ with respect to an $n$-qubit quantum state $\rho$ evolved under a quantum channel $\Lambda$ as follows. 
An $n$-qubit quantum state $\rho$ can be represented as
\begin{equation}
    |\rho \rangle\!\rangle = \sum_{{\vec{i}}} \langle\!\langle \sigma_{\vec{i}}|\rho \rangle\!\rangle |\sigma_{\vec{i}} \rangle\!\rangle= \sum_{{\vec{i}}} \rho_{\sigma_{\vec{i}}} |\sigma_{\vec{i}} \rangle\!\rangle,
\end{equation}
where $\rho_{\sigma_{\vec{i}}} = \langle\!\langle \sigma_{\vec{i}}|\rho \rangle\!\rangle$ is a real coefficient.
Also, an $n$-qubit Hermitian operator $O$ can be expressed in the same manner.
\begin{equation}
    \langle\!\langle O|=\sum_{{\vec{i}}} \langle\!\langle O|\sigma_{\vec{i}}\rangle\!\rangle\langle\!\langle \sigma_{\vec{i}}|
    = \sum_{{\vec{i}}} O_{\sigma_{\vec{i}}} \langle\!\langle \sigma_{\vec{i}}|,
\end{equation}
where $O_{\sigma_{\vec{i}}} = \langle\!\langle O|\sigma_{\vec{i}}\rangle\!\rangle$. 
Lastly, the transfer matrix of a quantum channel $\Lambda(\bullet)$ with respect to the basis set $\{|\sigma_{\vec{i}}\rangle\!\rangle\}$ is denoted as
\begin{equation}
    S(\Lambda) = \sum_{{\vec{k}},{\vec{l}}} \langle\!\langle \sigma_{\vec{k}}\,|S(\Lambda)|\sigma_{\vec{l}}\rangle\!\rangle|\sigma_{\vec{k}} \rangle\!\rangle\langle\!\langle \sigma_{\vec{l}}|= \sum_{{\vec{k}},{\vec{l}}} S(\Lambda)_{\sigma_{\vec{k}},\sigma_{\vec{l}}} |\sigma_{\vec{k}} \rangle\!\rangle\langle\!\langle \sigma_{\vec{l}}|,
\end{equation}
where $S(\Lambda)_{\sigma_{\vec{k}},\sigma_{\vec{l}}}=\langle\!\langle \sigma_{\vec{k}}\,|S(\Lambda)|\sigma_{\vec{l}}\rangle\!\rangle$ can be calculated as $\mathrm{Tr}\left[\sigma_{\vec{k}}^{\dagger} \, \Lambda(\sigma_{\vec{l}}) \right]$. 
Combining these elements $|\rho \rangle\!\rangle$, $\langle\!\langle O|$, and $S(\Lambda)$, we can write the expectation of the observable $O$ for the quantum state $\Lambda(\rho)$ as 
\begin{equation}
    \mathrm{Tr}\left[ O\, \Lambda(\rho) \right]=\langle\!\langle O\,|S(\Lambda)|\rho\rangle\!\rangle.
\end{equation}
Note that the example is based on the Pauli basis on $\mathcal{B}(\mathcal{H})$, and the transfer matrix on this basis is also called \textit{Pauli transfer matrix (PTM)}.

\subsection{Proof of Theorem~\ref{thm:number_mp_channels}}
\newtheorem*{T0}{Theorem~\ref{thm:number_mp_channels}}
\begin{T0}
    Suppose that an $n$-qubit quantum channel $\Gamma(\bullet)$ can be decomposed by $m$ (extended) m-p channels $\{\mathcal{E}_i(\bullet)\}_{i=1}^m$ 
    defined in Eq.~(\ref{measure-and-prepare channel}) and some weights $c_i\in \mathbb{R}$, in the form of Eq.~(\ref{eq:quasiprobability}).
    If the channels $\{\mathcal{E}_i(\bullet)\}_{i=1}^m$ do not use any ancilla qubits to implement each POVM, then $m$ 
    is lower bounded as
    \begin{align}
        \frac{{\rm Rank}\left(\Gamma\right)-1}{2^n-1}\leq m,
    \end{align}
    where ${\rm Rank}\left(\Gamma\right)$ is the rank of the transfer matrix of $\Gamma(\bullet)$.
\end{T0}
\begin{proof}

An $n$-qubit m-p channel $\mathcal{E}_i(\bullet)$ defined in Eq.~(\ref{measure-and-prepare channel}) can be represented as
\begin{align}
  S\left(\mathcal{E}_i\right) = \sum_{\mu=1}^{d_i} a_{i\mu} |\rho_{i\mu}\rangle\!\rangle\langle\!\langle E_{i\mu}|,~~~a_{i\mu}\in\{+1,-1\},
\end{align}
where $\rho_{i\mu}$ is an $n$-qubit density operator, and $d_i$ is the number of POVM elements $\{E_{i\mu}\}_{\mu=1}^{d_i}$.
By the assumption that no ancilla qubits are allowed, $d_i$ is upper bounded as $d_i\leq 2^n$.
Here, the POVM elements satisfy $\sum_{\mu} E_{i\mu}=I^{\otimes n}$, and this constraint yields
\begin{align}
    S\left(\mathcal{E}_i\right) &= a_{i1} |\rho_{i1}\rangle\!\rangle\langle\!\langle E_{i1}|+\sum_{\mu=2}^{d_i} a_{i\mu} |\rho_{i\mu}\rangle\!\rangle\langle\!\langle E_{i\mu}|\notag\\
    &=a_{i1} |\rho_{i1}\rangle\!\rangle\left(\langle\!\langle I^{\otimes n}|-\sum_{\mu=2}^{d_i} \langle\!\langle E_{i\mu}|
    \right)+\sum_{\mu=2}^{d_i} a_{i\mu} |\rho_{i\mu}\rangle\!\rangle\langle\!\langle E_{i\mu}|\notag\\
    &=a_{i1} |\rho_{i1}\rangle\!\rangle\langle\!\langle I^{\otimes n}|
    -\sum_{\mu=2}^{d_i} a_{i1} |\rho_{i1}\rangle\!\rangle\langle\!\langle E_{i\mu}|
    +\sum_{\mu=2}^{d_i} a_{i\mu} |\rho_{i\mu}\rangle\!\rangle\langle\!\langle E_{i\mu}|\notag\\
    &=|a_{i1} \rho_{i1}\rangle\!\rangle\langle\!\langle I^{\otimes n}|
    +\sum_{\mu=2}^{d_i}  |a_{i\mu}\rho_{i\mu}-a_{i1}\rho_{i1}\rangle\!\rangle\langle\!\langle E_{i\mu}|.
\end{align}
Thus, we obtain
\begin{align}
    S(\Gamma)&=S\left(\sum_{i=1}^m c_i\mathcal{E}_i\right)=\sum_{i=1}^m c_i S(\mathcal{E}_i)\notag\\
    &=\sum_{i=1}^m c_i \left(|a_{i1} \rho_{i1}\rangle\!\rangle\langle\!\langle I^{\otimes n}|
    +\sum_{\mu=2}^{d_i}  |a_{i\mu}\rho_{i\mu}-a_{i1}\rho_{i1}\rangle\!\rangle\langle\!\langle E_{i\mu}|\right)\notag\\
    &=|\Sigma_{i=1}^m c_i a_{i1} \rho_{i1}\rangle\!\rangle\langle\!\langle I^{\otimes n}|
    +\sum_{i=1}^m\sum_{\mu=2}^{d_i}  |c_i(a_{i\mu}\rho_{i\mu}-a_{i1}\rho_{i1})\rangle\!\rangle\langle\!\langle E_{i\mu}|.
\end{align}
Because the rank of a linear transformation $|A\rangle\!\rangle\langle\!\langle B|$ is one for any $A,B\in\mathcal{B}(\mathcal{H})$, the subadditivity of the rank leads to 
\begin{align}
    \mathrm{Rank}\left(S(\Gamma)\right)\leq 1+\sum_{i=1}^m{(d_i-1)}\leq 1+m(2^n-1).
\end{align}
\end{proof}

\subsection{Minimum number of m-p channels for ${\rm Id}^{\otimes n}$ decomposition}\label{sec:min_of_mp_channels}
The $n$-qubit identity channel ${\rm Id}^{\otimes n}$ corresponds to the $4^n\times 4^n$ identity matrix, which is full-rank, in the transfer matrix representation.
Therefore, Theorem~\ref{thm:number_mp_channels} leads that the number $m_{{\rm Id}^{\otimes n}}$ of m-p channels in ${\rm Id}^{\otimes n}$ decomposition is lower bounded as
\begin{align}\label{apdx:lower_bound_m}
    2^n+1 \leq m_{{\rm Id}^{\otimes n}},
\end{align}
when we are not allowed to use any ancilla qubits.
Importantly, the lower bound of $m_{{\rm Id}^{\otimes n}}$ can be achieved together with the optimal sampling overhead $(\gamma^{(\rm mp)}_n)^2=(2^{n+1}-1)^2$ for any number of qubits, as proved in Appendix~\ref{sec:proof of thm2}.
Also, the minimum number of $m_{{\rm Id}^{\otimes n}}$ is significantly small in comparison with the ${\rm Id}^{\otimes n}$ decomposition with randomized measurement~\cite{Lowe2023fastquantumcircuit}, which also uses m-p channels without ancilla qubits.
To show this, the following lemma is useful:
\begin{lemma}[\cite{10.1063/1.2716992,Roy2009}]\label{apdx:Lemma_2_design}
    A unitary 2-design in dimension $d$ has no fewer than $d^4-2d^2+2$ elements.
\end{lemma}
\noindent
This lemma immediately leads that the decomposition given in~\cite{Lowe2023fastquantumcircuit} (or Eq.~(\ref{apdx:decomp_random})) has at least $2^{4n}-2\cdot2^{2n}+3$ m-p channels, compared to the minimum number $2^n+1$.
Recently, Pednault~\cite{pednault2023alternative} shows that a set $\mathcal{V}$ of unitary operators for Eq.~(\ref{apdx:rand_decom_1chan}) does not have to be a unitary 2-design and provides unitary designs with more compact size for the decomposition.
Note that the size of unitary designs shown in~\cite{pednault2023alternative} matches the lower bound in Eq.~(\ref{apdx:lower_bound_m}) only up to 2-qubit.

\section{Proof of Eq.~(\ref{eq:lower_bound})}\label{sec:proof of lower bound}
A single-qubit m-p channel defined in Eq.~(\ref{measure-and-prepare channel}) can be written as:
\begin{equation}
    S(\mathcal{E}) = \sum_{\mu} a_{\mu} |\rho_{\mu} \rangle\!\rangle \langle\!\langle E_{\mu} |,
\end{equation}
where $\rho_{\mu}$ is a single-qubit quantum state with the bloch vector $\bm{r}_\mu:=(r_{\mu}^{x}, r_{\mu}^{y},r_{\mu}^{z})$, and its PTM representation is given by $|\rho_{i}\rangle\!\rangle=(1,r_{\mu}^{x},r_{\mu}^{y},r_{\mu}^{z})^{T}/\sqrt{2}$.
$E_{\mu}=e_{\mu}^{\mathbbm{1}}I+e_{\mu}^{x}X+e_{\mu}^{y}Y+e_{\mu}^{z}Z$ (where each $e_{\mu}^{(\cdot)}$ is a real coefficient) is a POVM element, and the corresponding matrix is $\langle\!\langle E_{\mu}|=\sqrt{2}(e_{\mu}^{\mathbbm{1}},e_{\mu}^{x},e_{\mu}^{y},e_{\mu}^{z})$.
Also, we can write the transfer matrix of the single-qubit identity channel $S({\rm Id})$ as 
\begin{equation}
    S({\rm Id}) = \begin{pmatrix} 1 & 0 & 0 & 0 \\ 0 & 1 & 0 & 0 \\ 0 & 0 & 1 & 0 \\ 0 & 0 & 0 & 1 \end{pmatrix}.
\end{equation}
Then, using these representations, the single-qubit identity channel can be decomposed by a following set of coefficients and single-qubit m-p channels:
\begin{align}
    \left\{c_{i},~S(\mathcal{E}_{i})=\sum_{{\mu}} a_{i{\mu}}|\rho_{i{\mu}}\rangle\!\rangle\langle\!\langle E_{i{\mu}}|\right\}_{i},
\end{align}
if and only if the elements of the set satisfy
\begin{equation}\label{necessary and sufficient condition}
    \begin{pmatrix} 1 & 0 & 0 & 0 \\ 0 & 1 & 0 & 0 \\ 0 & 0 & 1 & 0 \\ 0 & 0 & 0 & 1 \end{pmatrix}=\sum_{i} c_{i} \sum_{\mu} a_{{i}\mu} \begin{pmatrix} e_{{i}\mu}^{\mathbbm{1}} & e_{{i}\mu}^{x} & e_{{i}\mu}^{y} & e_{{i}\mu}^{z} \\ r_{{i}\mu}^{x}e_{{i}\mu}^{\mathbbm{1}} & r_{{i}\mu}^{x}e_{{i}\mu}^{x} & r_{{i}\mu}^{x}e_{{i}\mu}^{y} & r_{{i}\mu}^{x}e_{{i}\mu}^{z} \\ r_{{i}\mu}^{y}e_{{i}\mu}^{\mathbbm{1}} & r_{{i}\mu}^{y}e_{{i}\mu}^{y} & r_{{i}\mu}^{y}e_{{i}\mu}^{y} & r_{{i}\mu}^{y}e_{{i}\mu}^{z} \\ r_{{i}\mu}^{z}e_{{i}\mu}^{\mathbbm{1}} & r_{{i}\mu}^{z}e_{{i}\mu}^{x} & r_{{i}\mu}^{z}e_{{i}\mu}^{y} & r_{{i}\mu}^{z}e_{{i}\mu}^{z} \end{pmatrix}.
\end{equation}
Focusing on the lower right $3\times 3$ matrix in Eq.~(\ref{necessary and sufficient condition}), we obtain a necessary condition that a set $\{c_{i}, \mathcal{E}_{i}\}$ forms a valid decomposition as follows:
\begin{equation}\label{necessary condition}
    \begin{pmatrix}  1 & 0 & 0 \\ 0 & 1 & 0 \\ 0 & 0 & 1 \end{pmatrix}=\sum_{i} c_{i} \sum_{\mu} a_{{i}\mu} \begin{pmatrix} r_{{i}\mu}^{x}e_{{i}\mu}^{x} & r_{{i}\mu}^{x}e_{{i}\mu}^{y} & r_{{i}\mu}^{x}e_{{i}\mu}^{z} \\ r_{{i}\mu}^{y}e_{{i}\mu}^{y} & r_{{i}\mu}^{y}e_{{i}\mu}^{y} & r_{{i}\mu}^{y}e_{{i}\mu}^{z} \\ r_{{i}\mu}^{z}e_{{i}\mu}^{x} & r_{{i}\mu}^{z}e_{{i}\mu}^{y} & r_{{i}\mu}^{z}e_{{i}\mu}^{z} \end{pmatrix}.
\end{equation}
Now, evaluating the trace norm $\| \bullet\|_{1}$ for the both side of Eq.~(\ref{necessary condition}), and we get the following relation:
\begin{equation}\label{norm eneqn1}
\begin{split}
    3 = \left|\left| \sum_{i} c_{i} \sum_{\mu} a_{{i}\mu} \begin{pmatrix} r_{i\mu}^{x}e_{i\mu}^{x} & r_{i\mu}^{x}e_{i\mu}^{y} & r_{i\mu}^{x}e_{i\mu}^{z} \\ r_{i\mu}^{y}e_{{i}\mu}^{y} & r_{i\mu}^{y}e_{i\mu}^{y} & r_{i\mu}^{y}e_{i\mu}^{z} \\ r_{i\mu}^{z}e_{i\mu}^{x} & r_{i\mu}^{z}e_{i\mu}^{y} & r_{i\mu}^{z}e_{i\mu}^{z} \end{pmatrix} \right|\right|_1 &\leq \sum_{i} |c_{i}| \sum_{\mu} |a_{i\mu}| \left|\left| \begin{pmatrix} r_{i\mu}^{x}e_{i\mu}^{x} & r_{i\mu}^{x}e_{i\mu}^{y} & r_{i\mu}^{x}e_{i\mu}^{z} \\ r_{i\mu}^{y}e_{i\mu}^{y} & r_{i\mu}^{y}e_{i\mu}^{y} & r_{i\mu}^{y}e_{i\mu}^{z} \\ r_{i\mu}^{z}e_{i\mu}^{x} & r_{i\mu}^{z}e_{i\mu}^{y} & r_{i\mu}^{z}e_{i\mu}^{z} \end{pmatrix} \right|\right|_1\\
    &= \sum_{i} |c_{i}| \sum_{\mu} \left|\left| \begin{pmatrix} r_{i\mu}^{x}e_{i\mu}^{x} & r_{i\mu}^{x}e_{i\mu}^{y} & r_{i\mu}^{x}e_{i\mu}^{z} \\ r_{i\mu}^{y}e_{i\mu}^{y} & r_{{i}\mu}^{y}e_{{i}\mu}^{y} & r_{i\mu}^{y}e_{{i}\mu}^{z} \\ r_{i\mu}^{z}e_{i\mu}^{x} & r_{i\mu}^{z}e_{i\mu}^{y} & r_{i\mu}^{z}e_{i\mu}^{z} \end{pmatrix} \right|\right|_1,\\
\end{split}
\end{equation}
where the first inequality comes from the triangle inequality and the homogeneity of the trace norm, and the second equality uses $|a_{i\mu}|=1$.
From the definition of the trace norm, we have
\begin{equation}\label{norm eneqn2}
\begin{split}
    \left|\left| \begin{pmatrix} r_{i\mu}^{x}e_{i\mu}^{x} & r_{i\mu}^{x}e_{i\mu}^{y} & r_{i\mu}^{x}e_{i\mu}^{z} \\ r_{i\mu}^{y}e_{i\mu}^{y} & r_{{i}\mu}^{y}e_{{i}\mu}^{y} & r_{i\mu}^{y}e_{{i}\mu}^{z} \\ r_{i\mu}^{z}e_{i\mu}^{x} & r_{i\mu}^{z}e_{i\mu}^{y} & r_{i\mu}^{z}e_{i\mu}^{z} \end{pmatrix} \right|\right|_1 
    &= \sqrt{(r_{{i}\mu}^{x})^2 + (r_{{i}\mu}^{y})^2 + (r_{{i}\mu}^{z})^2}\sqrt{(e_{{i}\mu}^{x})^2 + (e_{{i}\mu}^{y})^2 + (e_{{i}\mu}^{z})^2}\\
    &\leq \sqrt{(e_{{i}\mu}^{x})^2 + (e_{{i}\mu}^{y})^2 + (e_{{i}\mu}^{z})^2},
\end{split}
\end{equation}
where the inequality comes from the fact that the magnitude of the bloch vector is less than 1, i.e., $(r_{{i}\mu}^{x})^2 + (r_{{i}\mu}^{y})^2 + (r_{{i}\mu}^{z})^2 \leq 1$. 
Here, the eigenvalues $\mu_{{i}\mu,+}$, $\mu_{{i}\mu,-}$ of a POVM element corresponding to $\langle\!\langle E_{{i}\mu}|=\sqrt{2}(e_{{i}\mu}^{\mathbbm{1}},e_{{i}mu}^{x},e_{{i}\mu}^{y},e_{{i}mu}^{z})$ are calculated as
\begin{equation}\nonumber
    \mu_{{i}\mu,\pm}=e_{{i}\mu}^{\mathbbm{1}} \pm \sqrt{(e_{{i}\mu}^{x})^2 + (e_{{i}\mu}^{y})^2 + (e_{{i}\mu}^{z})^2}.
\end{equation}
Since $E_{{i}\mu}$ is a positive semi-definite operator, we have
\begin{equation}\label{norm eneqn4}
    \sqrt{(e_{{i}\mu}^{x})^2 + (e_{{i}\mu}^{y})^2 + (e_{{i}\mu}^{z})^2} \leq e_{{i}\mu}^{\mathbbm{1}}.
\end{equation}
Combining Eqs.~(\ref{norm eneqn1})-(\ref{norm eneqn4}), we obtain
\begin{equation}\label{norm eneqn5}
    3 \leq \sum_{i} |c_{i}| \sum_{\mu} e_{{i}\mu}^{\mathbbm{1}}.
\end{equation}
Since the operator set $\{E_{{i}\mu}\}_{\mu}$ is a POVM, it satisfies $\sum_{\mu} E_{{i}\mu} = I$, and the trace of this equation leads to
\begin{equation}\label{norm eneqn6}
    \sum_{\mu} e_{{i}\mu}^{\mathbbm{1}} = 1.
\end{equation}
Consequently, we derive
\begin{equation}\label{norm eneqn7}
    \gamma_{1}^{\mathrm{(mp)}} = \sum_{i} |c_{i}| \ge 3.
\end{equation}
Also, we can conclude that $\gamma_{1,\mathrm{min}}^{\mathrm{(mp)}} \ge 3$.

\section{Proof of Theorem~\ref{thm 1}}\label{sec:proof of thm1}

\newtheorem*{T1}{Theorem~\ref{thm 1}}
\begin{T1}
The single-qubit identity channel ${\rm Id}(\bullet)$ can be decomposed as
\begin{equation}\label{1 cut}
    {\rm Id}(\bullet)=\sum_{i=1,2} \sum_{j\in\{0,1\}}\mathrm{Tr}\left[ U_{i} \ket{j}\bra{j}U_{i}^{\dagger}(\bullet) \right]U_{i} \ket{j}\bra{j}U_{i}^{\dagger} -\sum_{j\in\{0,1\}} \mathrm{Tr}\left[ \ket{j}\bra{j}(\bullet) \right]X \ket{j}\bra{j}X,
\end{equation}
where $U_1=H$ and $U_2=SH$, with $H$ the Hadamard gate and $S$ the phase gate. 
This decomposition achieves both of the lower bounds in Eqs.~(\ref{eq:lower_bound_of_m}) and (\ref{eq:lower_bound}); that is, 
\begin{equation}
    m~\mbox{of Eq.~(\ref{1 cut})}=3,~\mbox{and}~~\gamma^{\mathrm{(mp)}}_1~\mbox{of Eq.~(\ref{1 cut})}=3.
\end{equation}
\end{T1}
\begin{proof}

The decomposition of a single-qubit identity channel ${\rm Id}(\bullet)$ proposed by~\textcite{PhysRevLett.125.150504} is expressed as
\begin{equation}\label{pengcut}
    {\rm Id}(\bullet)=\frac{1}{2}\sum_{i=0}^{3}  \mathrm{Tr} \left[ O_{i} (\bullet) \right] O_{i},
\end{equation}
where $O_{0}=I$, $O_{1}=X$, $O_{2}=Y$, and $O_{3}=Z$.
Distributing the 0th term with the identity $I$ into the other terms, we can transform Eq.~(\ref{pengcut}) into
\begin{equation}\label{C4}
    {\rm Id}(\bullet)=\frac{1}{2}\left\{ \mathrm{Tr}\left[ X (\bullet) \right] X + \mathrm{Tr}\left[ I (\bullet) \right] I \right\} + \frac{1}{2}\left\{ \mathrm{Tr}\left[ Y (\bullet) \right] Y + \mathrm{Tr}\left[ I (\bullet) \right] I \right\} - \frac{1}{2}\left\{ \mathrm{Tr}\left[ I (\bullet) \right] I - \mathrm{Tr}\left[ Z (\bullet) \right] Z \right\}.
\end{equation}
Here, by a simple calculation, each term in Eq.~(\ref{C4}) can be rewritten as follows.
\begin{align}
    \frac{1}{2}\left\{ \mathrm{Tr}\left[ X (\bullet) \right] X + \mathrm{Tr}\left[ I (\bullet) \right] I \right\} 
    &= \sum_{j\in\{0,1\}} \mathrm{Tr}\left[ H \ket{j}\bra{j} H (\bullet) \right] H \ket{j}\bra{j} H, \\
    \frac{1}{2}\left\{ \mathrm{Tr}\left[ Y (\bullet) \right] Y + \mathrm{Tr}\left[ I (\bullet) \right] I \right\} 
    &= \sum_{j\in\{0,1\}} \mathrm{Tr}\left[ SH \ket{j}\bra{j} HS^{\dagger} (\bullet) \right] SH \ket{j}\bra{j} HS^{\dagger},\\
    \frac{1}{2}\left\{ \mathrm{Tr}\left[ I (\bullet) \right] I - \mathrm{Tr}\left[ Z (\bullet) \right] Z \right\}
    &= \sum_{j\in\{0,1\}} \mathrm{Tr}\left[ \ket{j}\bra{j}(\bullet) \right]X \ket{j}\bra{j}X.
\end{align}
Therefore, we obtain the decomposition of a single-qubit identity channel in Eq.~(\ref{1 cut}).


Next, we show that our decomposition in Eq.~(\ref{1 cut}) achieve both of the lower bounds in Eqs.~(\ref{eq:lower_bound_of_m}) and (\ref{eq:lower_bound}). 
Following the definition of the m-p channel in Eq.(\ref{measure-and-prepare channel}), the decomposition in Eq.~(\ref{1 cut}) can be represented as
\begin{equation}\label{B8}
    {\rm Id}(\bullet) = \sum_{i=1}^{3} c_{i} \mathcal{E}_{i}(\bullet),~~~\mathcal{E}_{i}(\bullet)=\sum_{\mu=0}^1 a_{i\mu}\mathrm{Tr}\left[ E_{i\mu}(\bullet) \right] \rho_{i\mu},
\end{equation}
where a set $\{c_{i},(a_{i\mu},E_{i\mu},\rho_{i\mu})_{\mu=0,1}\}_{i=1,2,3}$ follows Table~\ref{table:1-cut}.
By the definition of $\gamma_{1}^{\mathrm{(mp)}}$ in Sec.~\ref{mp_channel}, $\gamma_{1}^{\mathrm{(mp)}}$ of this decomposition is clearly 3. Furthermore, this decomposition consists of only 3 m-p channels, i.e., $m=3$. 
Hence, this decomposition simultaneously achieves the lower bounds of $m$ and $\gamma_{1}^{\mathrm{(mp)}}$ as presented in Eqs.~(\ref{eq:lower_bound_of_m}) and (\ref{eq:lower_bound}). 

\begin{table*}[ht]
\renewcommand{\arraystretch}{1.4}
\begin{tabular}{|ccc|ccc|ccccc ccc ccc|ccccc ccc ccc|}
 \hline
 &\multirow{2}{*}{$i$}&&& \multirow{2}{*}{$c_{i}$} && &&&&&& $\mu=0$ &&&& & &&&&&& $\mu=1$ &&&& \\
 \cline{7-28}
 & &&& &&&&&$a_{i\mu}$&&&$E_{i\mu}$&&&$\rho_{i\mu}$ &&&&&$a_{i\mu}$&&&$E_{i\mu}$&&&$\rho_{i\mu}$ &\\ 
 \hline\hline
 &1&&&$+1$&&&&&$+1$&&&$\ket{+}\bra{+}$&&&$\ket{+}\bra{+}$ &&&&&$+1$&&&$\ket{-}\bra{-}$&&&$\ket{-}\bra{-}$ &\\
 \hline
 &2&&&$+1$&&&&&$+1$&&&$\ket{+i}\bra{+i}$&&&$\ket{+i}\bra{+i}$ &&&&& $+1$ &&& $\ket{-i}\bra{-i}$&&& $\ket{-i}\bra{-i}$ &\\
 \hline
 &3&&&$-1$&&&&&$+1$&&&$\ket{0}\bra{0}$&&&$\ket{1}\bra{1}$ &&&&&$+1$&&&$\ket{1}\bra{1}$&&&$\ket{0}\bra{0}$&\\
 \hline
\end{tabular}
\caption{The elements $\{c_{i},(a_{i\mu}, E_{i\mu}, \rho_{i\mu})_{\mu=0,1} \}_{i=1,2,3}$ of the decomposition in Eq.~(\ref{B8}).
The coefficient $a_{i\mu}$, the POVM element $E_{i\mu}$, and the new input state $\rho_{i\mu}$ are in the $k$-th m-p channel $\mathcal{E}_{i}(\bullet)$, and the coefficient $c_{i}$ denotes a weight for each $\mathcal{E}_{i}(\bullet)$.}
\label{table:1-cut}
\end{table*}

\end{proof}

\section{Proof of Theorem~\ref{thm 2}}\label{sec:proof of thm2}
\newtheorem*{T2}{Theorem~\ref{thm 2}}
\begin{T2}
The $n$-qubit identity channel ${\rm Id}^{\otimes n}(\bullet)$ can be decomposed as
\begin{align}\label{n-qubit cut}
    {\rm Id}^{\otimes n}(\bullet) =\sum_{i=1}^{2^{n}} \sum_{\bm{j}\in\{0,1\}^{n}} \mathrm{Tr}\left[ U_{i} \ket{\bm{j}}\bra{\bm{j}}U_{i}^{\dagger}(\bullet) \right]U_{i} \ket{\bm{j}}\bra{\bm{j}}U_{i}^{\dagger}
   -(2^{n}-1)\sum_{\bm{j}\in\{0,1\}^{ n}} \mathrm{Tr}\left[ \ket{\bm{j}}\bra{\bm{j}}(\bullet) \right] \rho_{\bm{j}},
\end{align}
where $\ket{\bm{j}}$ denotes the $n$-qubit computational basis. 
The new input state $\rho_{\bm{j}}$, which depends on the measurement result $\bm{j}$, is defined as
\begin{equation}
    \rho_{\bm{j}}=\sum_{\bm{k}\in\{0,1\}^{n}}\frac{1}{2^n-1}(1-\delta_{\bm{j},\bm{k}})\ket{\bm{k}}\bra{\bm{k}}.
\end{equation}
where $\delta_{\bm{j},\bm{k}}$ is the Kronecker delta. 
Also, $\{U_i\}_{i=1}^{2^n}\cup \{I^{\otimes n}\}$ denotes a set of unitary operators that transform the computational base into the $2^n+1$ MUBs. 
In particular, each unitary $U_i$ can be implemented by a Clifford circuit with the maximal depth of $n+2$ on a device with fully-connected qubits. 
\end{T2}

\begin{proof}
Using Eq.~(\ref{pengcut}), the $n$-qubit identity channel ${\rm Id}^{\otimes n} := {\rm Id}\otimes \cdots \otimes {\rm Id}$ can be written as 
\begin{equation}\label{B4}
    {\rm Id}^{\otimes n}(\bullet)=\frac{1}{2^{n}}\sum_{P\in\{I,X,Y,Z\}^{\otimes n}} \mathrm{Tr} \left[ P (\bullet) \right] P.
\end{equation}
We first partition the $4^{n}-1$ $n$-qubit Pauli strings $\{I,X,Y,Z\}^{\otimes n} \setminus I^{\otimes n}$ into $2^{n}+1$ disjoint sets, each consisting of $2^n-1$ mutually commuting Pauli strings.
As shown in~\cite{PhysRevA.65.032320}, we can perform the above partitioning because of the existence of $2^n+1$ mutually unbiased bases in the $n$-qubit system.
Then we write the disjoint set as $G_{i} = \{ P_{ij} \}_{j=1}^{2^{n}-1},~(i=1,2,\cdots,2^{n}+1)$, and especially we can take $G_{2^n+1}=\{I,Z\}^{\otimes n}\backslash\{I^{\otimes n}\}$~\cite{Bandyopadhyay2002,seyfarth2019cyclic}.
Thus, Eq.~(\ref{B4}) can be rewritten as
\begin{align}\label{B5}
    {\rm Id}^{\otimes n}(\bullet)=\frac{1}{2^n}\mathrm{Tr}\left[I^{\otimes n} (\bullet)\right]I^{\otimes n}+\frac{1}{2^{n}}\sum_{i=1}^{2^n+1} \sum_{P_{ij}\in G_{i}} \mathrm{Tr}\left[P_{ij}(\bullet)\right]P_{ij}.
\end{align}
The disjoint sets $\{G_i\}_{i=1}^{2^n+1}$ have $2^n+1$ distinct eigenbasis sets, which are exactly MUBs.
For any $G_i$, we can explicitly construct a Clifford circuit $U_i$ that transforms the computational base into one of the MUBs for $G_i$, meaning that $G_i$ is diagonalized by $U_i$ as
\begin{equation}\label{B6}
    U_i^\dagger G_i U_i = \{U_i^\dagger P_{ij} U_i:P_{ij}\in G_i\}=\{p_i(P^{(z)})P^{(z)}:P^{(z)}\in \{I,Z\}^{\otimes n}\backslash \{I^{\otimes n}\}\},~~~p_{i}(P^{(z)})\in\{+1,-1\}.
\end{equation}
Here, the construction of $U_i$ is provided in Appendix~\ref{apdx:mubs_pre} (Lemma~\ref{apdx:Lemma_diag}).
That is, $U_i^\dagger G_i U_i$ is equivalent to the set $\{I,Z\}^{\otimes n}\backslash\{I^{\otimes n}\}$ of diagonal matrices regardless of the index $i$ (up to some phases $p_{i}(P^{(z)})\in\{+1,-1\}$).
Combining Eqs.~(\ref{B5}) and~(\ref{B6}), we obtain 
\begin{align}\label{B9}
    {\rm Id}^{\otimes n}(\bullet) &=\frac{1}{2^n}\mathrm{Tr}\left[I^{\otimes n} (\bullet)\right]I^{\otimes n}+\frac{1}{2^{n}}\sum_{i=1}^{2^n+1} \sum_{P_{ij}\in G_{i}} \mathrm{Tr}\left[P_{ij}(\bullet)\right]P_{ij} \nonumber\\
    &=\frac{1}{2^n}\mathrm{Tr}\left[I^{\otimes n} (\bullet)\right]I^{\otimes n}+\frac{1}{2^{n}}\sum_{i=1}^{2^n} \sum_{P_{ij}\in G_{i}} \mathrm{Tr}\left[P_{ij}(\bullet)\right]P_{ij}
    +\frac{1}{2^{n}}\sum_{P^{(z)}\in \{I,Z\}^{\otimes n}\setminus I^{\otimes n}} \mathrm{Tr}\left[P^{(z)}(\bullet)\right]P^{(z)} \nonumber\\
    &=\frac{1}{2^n}\sum_{i=1}^{2^n}\left\{ \mathrm{Tr}\left[I^{\otimes n}(\bullet)\right]I^{\otimes n} +\!\!\!\!\! \sum_{P_{ij}\in G_{i}} \mathrm{Tr}\left[ P_{ij} (\bullet) \right] P_{ij} \right\}-\frac{1}{2^n}\left\{ 2^n \mathrm{Tr}\left[ I^{\otimes n} (\bullet) \right]I^{\otimes n}-\!\!\!\!\!\!\!\!\!\! \sum_{P^{(z)}\in\{I,Z\}^{\otimes n}} \!\!\!\!\!\!\!\!\!\! \mathrm{Tr}\left[ P^{(z)}(\bullet) \right]P^{(z)} \right\}\nonumber\\
    &=\frac{1}{2^n}\sum_{i=1}^{2^n} \sum_{P^{(z)}\in\{I,Z\}^{\otimes n}}\!\!\!\!\!\!\!\! \mathrm{Tr}\left[ U_{i}P^{(z)}U^{\dagger}_{i} (\bullet) \right]U_{i}P^{(z)}U^{\dagger}_{i}-\frac{1}{2^{n}}\left\{ 2^n \mathrm{Tr}\left[ I^{\otimes n} (\bullet) \right] I^{\otimes n} -\!\!\!\!\!\!\!\! \sum_{P^{(z)}\in \{I,Z\}^{\otimes n}}\!\!\!\!\!\!\!\! \mathrm{Tr}\left[ P^{(z)} (\bullet) \right] P^{(z)} \right\},
\end{align}
where we used $p_i(P^{(z)})^2=1$.
For the first term in Eq.~(\ref{B9}), we can write it as
\begin{eqnarray}\label{1st term}
    \lefteqn{ \sum_{P^{(z)}\in\{I,Z\}^{\otimes n}}\!\!\!\!\!\!\!\! \mathrm{Tr}\left[ U_{i}P^{(z)}U_{i}^{\dagger} (\bullet) \right]U_{i}P^{(z)}U_{i}^{\dagger}} \nonumber\\
    &=& \sum_{P^{(z)}\in\{I,Z\}^{\otimes n}} \mathrm{Tr}\left[  U_{i} \left( \sum_{\bm{k},\bm{k'} \in\{0,1\}^{ n}} \braket{\bm{k}|P^{(z)}|\bm{k'}}\ket{\bm{k}}\bra{\bm{k'}} \right) U_{i}^{\dagger} (\bullet) \right] U_{i} \left( \sum_{\bm{l},\bm{l'}\in\{0,1\}^{ n}} \braket{\bm{l}|P^{(z)}|\bm{l'}}\ket{\bm{l}}\bra{\bm{l'}} \right) U_{i}^{\dagger}\nonumber\\
    &=& \sum_{P^{(z)}\in\{I,Z\}^{\otimes n}} \sum_{\bm{k},\bm{l}\in\{0,1\}^{ n}} \braket{\bm{k}|P^{(z)}|\bm{k}}\braket{\bm{l}|P^{(z)}|\bm{l}} \mathrm{Tr}\left[ U_{i} \ket{\bm{k}}\bra{\bm{k}} U_{i}^{\dagger} (\bullet) \right] U_{i} \ket{\bm{l}}\bra{\bm{l}} U_{i}^{\dagger}\nonumber\\
    &=& \sum_{\bm{k},\bm{l}\in\{0,1\}^{n}} \left\{ \sum_{P_{1}^{(z)} \in \{I,Z\}} \sum_{P_{2}^{(z)} \in \{I,Z\}}\cdots \sum_{P_{n}^{(z)} \in \{I,Z\}} \prod_{m=1}^{n} \braket{k_{m}|P_{m}^{(z)}|k_{m}}\braket{l_{m}|P_{m}^{(z)}|l_{m}} \right\} \mathrm{Tr}\left[ U_{i} \ket{\bm{k}}\bra{\bm{k}} U_{i}^{\dagger} (\bullet) \right] U_{i} \ket{\bm{l}}\bra{\bm{l}} U_{i}^{\dagger}\nonumber\\
    &=& \sum_{\bm{k},\bm{l}\in\{0,1\}^{ n}}  \left\{ \prod_{m=1}^{n} \left(1+(-1)^{k_{m}+l_{m}}\right) \right\} \mathrm{Tr}\left[ U_{i} \ket{\bm{k}}\bra{\bm{k}} U_{i}^{\dagger} (\bullet) \right]U_{i} \ket{\bm{l}}\bra{\bm{l}} U_{i}^{\dagger},
\end{eqnarray}
where $\ket{\bm{k}}=\ket{k_{1}}\otimes \cdots\otimes \ket{k_{n}}$ denotes the $n$-qubit computational basis, and $P_{m}^{(z)}$ is a Pauli operator acting on the $m$-th qubit.
Since $P^{(z)}$ is a diagonal matrix with respect to the computational basis, we can omit the indices $\bm{k'}$ and $\bm{l'}$ in the first equality.
Noting that
\begin{eqnarray}
    \prod_{m=1}^{n} (1+(-1)^{k_{m}+l_{m}}) &= \left\{ \begin{array}{ll} 2^n,& (\bm{k} = \bm{l})\\ 0, & (\bm{k} \neq \bm{l}) \end{array} \right.= 2^n\delta_{\bm{k},\bm{l}},
\end{eqnarray}
we have the following equation:
\begin{equation}\label{B11}
\begin{split}
    \frac{1}{2^n}\sum_{i=1}^{2^n} \sum_{P^{(z)}\in\{I,Z\}^{\otimes n}}\!\!\!\!\!\!\!\! \mathrm{Tr}\left[ U_{i}P^{(z)}U_{i}^{\dagger} (\bullet) \right]U_{i}P^{(z)}U_{i}^{\dagger}
    &= \frac{1}{2^n}\sum_{i=1}^{2^n} \sum_{\bm{k},\bm{l}\in\{0,1\}^{ n}}
    2^n\delta_{\bm{k},\bm{l}}\mathrm{Tr}\left[ U_{i} \ket{\bm{k}}\bra{\bm{k}} U_{i}^{\dagger} (\bullet) \right]  U_{i} \ket{\bm{l}}\bra{\bm{l}} U_{i}^{\dagger}\\
    &= \sum_{i=1}^{2^{n}} \sum_{\bm{k}\in\{0,1\}^{n}} \mathrm{Tr}\left[ U_{i} \ket{\bm{k}}\bra{\bm{k}}U^{\dagger}_{i}(\bullet) \right]U_{i} \ket{\bm{k}}\bra{\bm{k}}U_{i}^{\dagger}.
\end{split}
\end{equation}

For the second term in Eq.~(\ref{B9}), we can transform the two components as
\begin{equation}
    2^n \mathrm{Tr}\left[ I^{\otimes n} (\bullet) \right] I^{\otimes n} = 2^n \sum_{\bm{k},\bm{l}\in\{0,1\}^{ n}} \mathrm{Tr}\left[ \ket{\bm{k}}\bra{\bm{k}} (\bullet) \right] \ket{\bm{l}}\bra{\bm{l}},
\end{equation}
and
\begin{equation}\label{B13}
    \sum_{P^{(z)}\in\{I,Z\}^{\otimes n}} \mathrm{Tr}\left[ P^{(z)}(\bullet) \right] P^{(z)} = 2^n \sum_{\bm{k}\in\{0,1\}^{n}} \mathrm{Tr}\left[\ket{\bm{k}}\bra{\bm{k}} (\bullet)\right] \ket{\bm{k}}\bra{\bm{k}},
\end{equation}
where we used a special case of Eq.~(\ref{1st term}) with $U_{i}=I^{\otimes n}$.
Hence, we have
\begin{equation}
\begin{split}\label{B14}
    \frac{1}{2^{n}}\left\{ 2^n \mathrm{Tr}\left[ I^{\otimes n} (\bullet) \right] I^{\otimes n} -\!\!\!\!\!\!\!\! \sum_{P^{(z)}\in \{I,Z\}^{\otimes n}}\!\!\!\!\!\!\!\! \mathrm{Tr}\left[ P^{(z)} (\bullet) \right] P^{(z)} \right\}
    &= \sum_{\bm{k}\in\{0,1\}^{n}} \mathrm{Tr}\left[ \ket{\bm{k}}\bra{\bm{k}} (\bullet) \right] \left(-\ket{\bm{k}}\bra{\bm{k}}+\sum_{\bm{l}\in\{0,1\}^{n}}\ket{\bm{l}}\bra{\bm{l}}\right)\\
    &= (2^{n}-1)\sum_{\bm{k}\in\{0,1\}^{n}} \mathrm{Tr}\left[ \ket{\bm{k}}\bra{\bm{k}}(\bullet) \right] \sum_{\bm{l}\in\{0,1\}^{ n}}\frac{1-\delta_{\bm{k},\bm{l}}}{2^n-1}\ket{\bm{l}}\bra{\bm{l}}.
\end{split}
\end{equation}
Combining Eqs.~(\ref{B9}), (\ref{B11}) and (\ref{B14}), we establish the decomposition of Eq.~(\ref{n-qubit cut}).

The last part of Theorem~\ref{thm 2} on the maximal circuit depth of $U_i$ is the consequence of Lemma~\ref{apdx:maximal_depth_proof}, which is proved after introducing the binary bit representation of Pauli strings in Appendix~\ref{sec:construct_qc}.
Lemma~\ref{apdx:maximal_depth_proof} provides the procedure for constructing an (at most) $(n+2)$-depth Clifford circuit that diagonalizes $2^n-1$ mutually commuting Pauli strings (excluding $I^{\otimes n}$), if the Pauli strings are not in $\{I,Z\}^{\otimes n}\backslash\{I^{\otimes n}\}$.
Since the sets $\{G_i\}_{i=1}^{2^n+1}$ are disjoint, $G_i~(i=1,2,\cdots,2^n)$ does not have any elements of $\{I,Z\}^{\otimes n}\backslash\{I^{\otimes n}\}=G_{2^n+1}$.
Accordingly, the condition on Lemma~\ref{apdx:maximal_depth_proof} is satisfied for each $G_i$, and thus the Clifford circuit $U_i$ appered in Eq.~(\ref{B6}) has at most $(n+2)$-depth. 
This completes the proof of Theorem~\ref{thm 2}.

\end{proof}

\section{Construction of quantum circuits for MUBs}\label{sec:construct_qc}

\subsection{Preliminaries}\label{apdx:mubs_pre}
\subsubsection{Binary bit representation of $n$-qubit Pauli strings}\label{apdx:pre1}
Four elements of Pauli matrices, $I$, $X$, $Y$, and $Z$, can be represented by a matrix product of $X$ and $Z$ with some phase (e.g., $Y=-iZX$), which enables us to associate the Pauli matrices with the binary vectors~\cite{PhysRevLett.78.405,PhysRevA.70.052328,seyfarth2019cyclic,Crawford2021efficientquantum,9248636}: 
\begin{align*}
    I=Z^0 X^0 &\rightarrow (0,0), & X =Z^0 X^1 &\rightarrow (0,1), & Y =-iZ^1 X^1 &\rightarrow (1,1), & Z =Z^1 X^0 &\rightarrow (1,0),
\end{align*}
where the first and second elements of the binary vector above correspond to the exponents of $Z$ and $X$, respectively. 
For an $n$-qubit case, the $n$-qubit Pauli string is $n$ tensor product of the Pauli matrices, and then a $2n$-dimensional binary vector specifies a single $n$-qubit Pauli string.
To clarify this relation, let us define the following function $\phi:\{0,1\}^{2n} \rightarrow \{I,X,Y,Z\}^{\otimes n}$, which maps a $2n$-dimensional binary vector to a single Pauli string with coefficient being $+1$ as
\begin{equation}\label{eq:definition of phi}
    \phi(\bm{b}) :=(-i)^{(\bm{b}^{z})^T\bm{b}^{x}} Z^{b^{z}_1}X^{b^{x}_1} \otimes Z^{b^{z}_2}X^{b^{x}_2} \otimes ... \otimes Z^{b^{z}_n}X^{b^{x}_n},
\end{equation}
where $\bm{b} = (b^{z}_1, ...,b^{z}_n, b^{x}_1, ...,b^{x}_n)^{T}$ is a $2n$-dimensional binary vector, $\bm{b}^z=(b^{z}_1, ...,b^{z}_n)^{T}$, and $\bm{b}^x=(b^{x}_1, ...,b^{x}_n)^{T}$.
Here, $T$ denotes the transpose operation.
Note that the function $\phi$ is clearly bijective, i.e., we can take its inverse map $\phi^{-1}$.

\subsubsection{Maximally commuting family of Pauli strings}\label{apdx:pre2}
Hereafter, we call a set of Pauli strings $\{P_{1}, P_{2}, ..., P_{m} \}$ a \textit{commuting family} if any pair of Pauli strings in the set is commuting.
Note that the maximal size of a commuting family is $2^n-1$ (excluding ${I}^{\otimes n}$). For instance, each of the sets $\{G_i\}_{i=1}^{2^n+1}$ appeared in the proof of Theorem~\ref{thm 2} is a maximally commuting family (excluding ${I}^{\otimes n}$).
The following lemma enables us to identify ``generators'' in a maximally commuting family $G$ with a basis set of an $n$-dimensional vector space (over the binary field $\mathbb{F}_2$) defined by $\phi^{-1}(G)$.


\begin{lemma}\label{apdx:Lemma_basis}
    Let $G\subset \{I,X,Y,Z\}^{\otimes n}$ be a maximally commmuting family with $I^{\otimes n}$.
    Then, a set $\phi^{-1}(G)$ of binary vectors forms an $n$-dimensional vector space over the binary filed $\mathbb{F}_2$.
    (Here, the addition "$\oplus$" of vectors and the scalar multiplication follow the rules of mod-2 arithmetic element-wise.)
    Taking a basis set $\{\bm{g}_{k}\}_{k=1}^{n}$ in $\phi^{-1}(G)$, $G$ can be written as 
    \begin{align}
        G=\left\{\phi\left(\sum_{k=1}^n c_k\bm{g}_k\right):\bm{c}=(c_1,c_2,\cdots,c_n)\in \{0,1\}^n\right\}.
    \end{align}
    Moreover, for any $\bm{c}\in\{0,1\}^n$, there exists a scalar $p\in\{+1,-1\}$ such that
    \begin{align}
        \phi\left(\sum_{k=1}^n c_k \bm{g}_k\right)=p\,\phi\left(\bm{g}_1\right)^{c_1}\phi\left(\bm{g}_2\right)^{c_2}\cdots \phi\left(\bm{g}_n\right)^{c_n}.
    \end{align}
\end{lemma}
\begin{proof}
First, we show that $\phi^{-1}(G) :=\{\phi^{-1}(P) : P \in G \}$ is a vector space over $\mathbb{F}_2$ (or a subspace of $\{0,1\}^n$)~\cite{Bandyopadhyay2002,seyfarth2019cyclic}.
To this end, it suffices to show the following properties: (i) closure under scalar multiplication $\alpha \in \mathbb{F}_2,~ \bm{v} \in \phi^{-1}(G) \Rightarrow \alpha \bm{v} \in \phi^{-1}(G)$ and (ii) closure under addition $\bm{u}, \bm{v} \in \phi^{-1}(G) \Rightarrow \bm{u} \oplus \bm{v} \in \phi^{-1}(G)$.
By definition, $G$ contains $I^{\otimes n}$, so (i) is trivially satisfied.
For any $\bm{u}, \bm{v} \in \phi^{-1}(G)$, we obtain
\begin{equation}\label{eq:closure under addition}
\begin{split}
    \phi(\bm{u} \oplus \bm{v}) 
    &= (-i)^{(\bm{u}^{z}\oplus \bm{v}^{z})^{T}(\bm{u}^{x} \oplus \bm{v}^{x})} \bigotimes_{i=1}^{n} Z^{(u^{z}_i \oplus v^{z}_i)}X^{(u^{x}_i \oplus v^{x}_i)}\\
    &= (-i)^{(\bm{u}^{z})^T\bm{v}^{x} \oplus ({\bm{v}^{z}})^T\bm{u}^{x}} (-1)^{(\bm{v}^{z})^T\bm{u}^{x}} \left( (-i)^{(\bm{u}^{z})^T\bm{u}^{x}} \bigotimes_{i=1}^{n} Z^{u^{z}_i} X^{u^{x}_{i}}\right)\left((-i)^{(\bm{v}^{z})^T\bm{v}^{x}}\bigotimes_{i=1}^{n} Z^{v^{z}_i} X^{u^{x}_{i}}\right)\\
    &= (-1)^{(\bm{v}^{z})^T\bm{u}^{x}} \phi(\bm{u})\phi(\bm{v}),
\end{split}
\end{equation}
where $Z^{(u^{z}_i \oplus v^{z}_i)}X^{(u^{x}_i \oplus v^{x}_i)} = (-1)^{v_i^z u_i^x} (Z^{u^{z}_i} X^{u^{x}_{i}}) (Z^{v^{z}_i} X^{u^{x}_{i}}),~(i=1,...,n)$ is used in the second equality, and the last equality comes from the condition of commuting relations between $\phi(\bm{u})$ and $\phi(\bm{v})$, i.e., $(\bm{u}^{z})^T\bm{v}^{x}=(\bm{v}^{z})^T\bm{u}^{x}$.
From the Eq.~(\ref{eq:closure under addition}), it can be seen that for any $P\in G$, 
\begin{equation}
    [\phi(\bm{u} \oplus \bm{v}), P] = (-1)^{(\bm{v}^{z})^T\bm{u}^{x}} \left( [\phi(\bm{u}), P]\phi(\bm{v}) + \phi(\bm{u})[\phi(\bm{v}), P] \right) = 0,
\end{equation}
so $\phi(\bm{u}\oplus\bm{v})$ is in $G$.
From the discussion up to this point, $\phi^{-1}(G)$ is a vector space over $\mathbb{F}_2$.
Next, we see that dim~$\phi^{-1}(G)$ is $n$.
Since we take $G$ as a maximally commuting family with $I^{\otimes n}$, the cardinality of $G$ is $2^{n}$.
Recalling that the function $\phi$ is bijective, we have that the cardinality of $\phi^{-1}(G)$ is also $2^{n}$.
Suppose that dim$~\phi^{-1}(G) = k \neq n$.
By choosing a basis set $\{\bm{g}_{l}\}_{l=1}^{k}$ in $\phi^{-1}(G)$, any vector $\bm{v}$ in $\phi^{-1}(G)$ can be represented by $\bm{v} = \sum_{l=1}^{k} c_{l} \bm{g}_{l}$ with $c_{l} \in \{0,1\}~(l=1,...,k)$.
This implies that the cardinality of $\phi^{-1}(G)$ is $2^{k}$, which contradicts that $\phi^{-1}(G)$ has $2^n$ distinct elements.

Now, we can take $n$ linearly independent vectors in $\phi^{-1}(G)$:
\begin{equation}
    \phi^{-1}(G) = {\rm Span}\{\bm{g}_{1}, \bm{g}_{2},...,\bm{g}_{n}\}.
\end{equation}
Note that fixing a basis set, all elements of $\phi^{-1}(G)$ can be written as $\sum_{k=1}^{n} c_{k} \bm{g}_{k}$, where ${\bm c}=(c_{1},c_{2},...,c_{n}) \in \{0,1\}^{n}$ is uniquely determined for each element. Recalling that $\phi$ is bijective, there is a one-to-one correspondence between $\bm{c}\in \{0,1\}^{n}$ and $P \in G$. That is,
\begin{align}
    G=\left\{\phi\left(\sum_{k=1}^n c_k\bm{g}_k\right):\bm{c}=(c_1,c_2,\cdots,c_n)\in \{0,1\}^n\right\}.
\end{align}
Substituting $\bm{v}_{\bm c} = \sum_{k=1}^{n} c_{k} \bm{g}_{k}$ into Eq.~(\ref{eq:definition of phi}), $\phi(\bm{v}_{\bm c})$ can be expressed as
\begin{equation}\label{eq:phi_vc}
    \phi(\bm{v}_{\bm c}) = (-i)^{(\bm{v}_{\bm c}^{z})^{T} \bm{v}_{\bm c}^{x}} \bigotimes_{\mu =1}^{n} Z^{v_{{\bm c}\mu}^{z}} X^{v_{{\bm c}\mu}^{x}},
\end{equation}
where we defined $\bm{v}_{\bm c}^{z} = \sum_{k=1}^{n} c_{k} \bm{g}_{k}^{z}$ and $\bm{v}_{\bm c}^{x} = \sum_{k=1}^{n} c_{k} \bm{g}_{k}^{x}$.
For each qubit $\mu$, it holds that 
\begin{equation}\label{eq:interchanging order of matrix multiplication}
\begin{split}
    Z^{{v}_{\bm{c}\mu}^{z}} X^{{v}_{\bm{c}\mu}^{x}} &= \left(Z^{c_{1}\bm{g}_{1\mu}^{z}}~Z^{c_{2}\bm{g}_{2\mu}^{z}} \cdots Z^{c_{n}\bm{g}_{n\mu}^{z}}\right) \left(X^{c_{1}\bm{g}_{1\mu}^{x}}~ X^{c_{2}\bm{g}_{2\mu}^{x}} \cdots X^{c_{n}\bm{g}_{n\mu}^{x}}\right) \\
    &\propto \left(Z^{c_{1}\bm{g}_{1\mu}^{z}}~X^{c_{1}\bm{g}_{1\mu}^{x}}\right)\left(Z^{c_{2}\bm{g}_{2\mu}^{z}}~X^{c_{2}\bm{g}_{2\mu}^{x}}\right) \cdots \left(Z^{c_{n}\bm{g}_{n\mu}^{z}}~X^{c_{n}\bm{g}_{n\mu}^{x}}\right).
\end{split}
\end{equation}
In the second step, we used $\propto$ instead of the multiplication of a norm-one scalar arising from the swapping of Pauli matrices.
Thus, we obtain
\begin{equation}\label{eq:interchanging order of matrix multiplication, n}
\begin{split}
    \bigotimes_{\mu =1}^{n} Z^{{v}_{\bm{c}\mu}^{z}} X^{{v}_{\bm{c}\mu}^{x}} 
    &\propto \bigotimes_{\mu =1}^{n} \left\{Z^{\bm{g}_{1\mu}^{z}}~X^{\bm{g}_{1\mu}^{x}}\right\}^{c_{1}}\left\{Z^{\bm{g}_{2\mu}^{z}}~X^{\bm{g}_{2\mu}^{x}}\right\}^{c_{2}} \cdots \left\{Z^{\bm{g}_{n\mu}^{z}}~X^{\bm{g}_{n\mu}^{x}}\right\}^{c_{n}}\\
    &\propto \left\{ (-i)^{({\bm{g}_{1}^{z}})^{T}{\bm{g}_{1}^{x}}} \bigotimes_{\mu=1}^{n} Z^{\bm{g}_{1\mu}^{z}}~X^{\bm{g}_{1\mu}^{x}} \right\}^{c_{1}} \left\{ (-i)^{({\bm{g}_{2}^{z}})^{T}{\bm{g}_{2}^{x}}} \bigotimes_{\mu=1}^{n} Z^{\bm{g}_{2\mu}^{z}}~X^{\bm{g}_{2\mu}^{x}} \right\}^{c_{2}} \cdots \left\{ (-i)^{({\bm{g}_{n}^{z}})^{T}{\bm{g}_{n}^{x}}} \bigotimes_{\mu=1}^{n} Z^{\bm{g}_{n\mu}^{z}}~X^{\bm{g}_{n\mu}^{x}} \right\}^{c_{n}}\\
    &= \phi(\bm{g}_{1})^{c_{1}} \phi(\bm{g}_{2})^{c_{2}} \cdots \phi(\bm{g}_{n})^{c_{n}}
\end{split}
\end{equation}
From Eq.~(\ref{eq:phi_vc}) and Eq.~(\ref{eq:interchanging order of matrix multiplication, n}), 
\begin{equation}\label{eq:phi_vc_improved}
    \phi\left(\sum_{k=1}^n c_k \bm{g}_k\right)=p\,\phi\left(\bm{g}_1\right)^{c_1}\phi\left(\bm{g}_2\right)^{c_2}\cdots \phi\left(\bm{g}_n\right)^{c_n},
\end{equation}
where $p$ is a scalar (depending on $\{\bm{g}_k\}$ and $\bm{c}$ implicitly). Here, considering $\phi(\bm{g}_{k})~(k=1,...,n)$ and $\phi(\sum_{k=1}^n c_k \bm{g}_k)$ are the $n$-qubit Pauli strings, by squaring both sides of Eq.~(\ref{eq:phi_vc_improved}), we obtain $I^{\otimes n}=p^2 I^{\otimes n}$ because of the commutivity of the set $\{\phi({\bm{g}_{k}})\}_{k=1}^{n}$.
Thus, $p$ takes $+1$ or $-1$, and we complete the proof.
\end{proof}

\subsubsection{Action of Clifford circuits on $\{0,1\}^{2n}$}\label{apdx:pre3}

Here, we characterize the action of Clifford circuits on a $2n$-dimensional vector space $\{0,1\}^{2n}$ (over the binary field $\mathbb{F}_2$). 
Note that the Clifford circuit is defined as a unitary operator whose conjugate action is a transformation on Pauli strings with a phase $\{\pm1, \pm i\}$.
First, the definition of Clifford circuits and the bijective function $\phi$ induce the action of Clifford circuits on $\{0,1\}^{2n}$, as summarized in the following remark.
\begin{remark}\label{remark: the action of Clifford operator}
    Given $P \in \{I,X,Y,Z\}^{\otimes n}$ and an arbitrary Clifford circuit $V$, then there exist a (unique) phase $p \in \{+1, -1\}$ and $Q \in \{I,X,Y,Z\}^{\otimes n}$ such that $VPV^{\dagger} = pQ$.
    Thus, using the bijective function $\phi$, a Clifford circuit defines a map such that $\phi^{-1}(P)\mapsto \phi^{-1}(Q)$.
    More specifically, we can induce a following map on $\{0,1\}^{2n}$ from a Clifford circuit $V$:
    \begin{align}\label{eq: VGV^dagger}
        \tilde{V}:\{0,1\}^{2n} \rightarrow \{0,1\}^{2n}~;~\bm{v} \in \{0,1\}^{2n} ~\mapsto~ \tilde{V}\bm{v} = \phi^{-1} ([V \phi(\bm{v}) V^{\dagger}]) \in \{0,1\}^{2n}.
    \end{align} 
    Here, we defined a function $[e^{i\theta}\,P] := P$ for all $\theta \in \mathbb{R}$ and all $P \in \{I,X,Y,Z\}^{\otimes n}$, which drops a phase such that $[VPV^{\dagger}] = Q$.
    The above discussion is summarized as the following commutative diagram:
\[
\xymatrix{
\phi(\bm{v}) \in \{I,X,Y,Z\}^{\otimes n} \ar[r]^-{V} \ar@<0.5ex>[d]^-{\phi^{-1}}
& V G V^{\dagger} \ar@<0.5ex>[r]^-{ \lbrack \bullet \rbrack }
& \lbrack V \phi(\bm{v}) V^{\dagger} \ar@<0.5ex>[d]^-{\phi^{-1}} \rbrack \\
\bm{v} \in \{0,1\}^{2n} \ar@<0.5ex>[u]^-{\phi} \ar[rr]^-{\tilde{V}}
& 
& \tilde{V}\bm{v}\in\{0,1\}^{2n} \ar@<0.5ex>[u]^-{\phi}
}
\] 
\end{remark}

Here, one of the key properties inherent in the map $\tilde{V}$ is linearity on $\{0,1\}^{2n}$. To prove this, we firstly show that for all $\bm{u}, \bm{v} \in \{0,1\}^{2n}, \alpha,\beta \in \{0,1\}$, there exists a $\theta \in \mathbb{R}$ such that 
\begin{equation}\label{eq:eq_for_linearity_proof_1}
    \phi(\bm{u})^{\alpha}\phi(\bm{v})^{\beta} = e^{i\theta} \phi(\alpha \bm{u} \oplus \beta \bm{v}).
\end{equation}
From the discussion in the proof of Lemma~\ref{apdx:Lemma_basis}, we obtain $\phi(\bm{u} \oplus \bm{v})=p_{\bm uv} \phi(\bm{u})\phi(\bm{v})$ where $p_{\bm{uv}}$ is a constant factor. Since $2^n = \mathrm{Tr}~\phi^{\dagger}(\bm{u}\oplus \bm{v})\phi(\bm{u}\oplus \bm{v}) = |p_{\bm{uv}}|^2\, \mathrm{Tr}~\phi^{\dagger}(\bm{v})\phi^{\dagger}(\bm{u})\phi(\bm{u})\phi(\bm{v}) = |p_{\bm{uv}}|^2 2^n$, we obtain $p_{\bm{uv}}=e^{i\theta}$ with $\theta \in \mathbb{R}$. 
Thus, we establish Eq.~(\ref{eq:eq_for_linearity_proof_1}).
Using Eq.~(\ref{eq:eq_for_linearity_proof_1}), we have
\begin{align}
    \tilde{V}(\alpha \bm{u} \oplus \beta \bm{v}) 
    &=\phi^{-1}\left( [ V \phi(\alpha \bm{u} \oplus \beta \bm{v}) V^{\dagger} ]\right) \\
    &=\phi^{-1}\left([ V \phi(\bm{u})^{\alpha} \phi(\bm{v})^{\beta} V^{\dagger} ]\right)\\
    &=\phi^{-1}\left([ V \phi(\bm{u})^{\alpha} V^{\dagger} V \phi(\bm{v})^{\beta} V^{\dagger}]\right)\\
    &=\phi^{-1}\left([ \phi(\tilde{V}\bm{u})^{\alpha} \phi(\tilde{V}\bm{v})^{\beta} ]\right)\\
    &= \phi^{-1}\phi(\alpha \tilde{V} \bm{u} \oplus \beta \tilde{V} \bm{v} ) ,
\end{align}
which implies the linearity of the map $\tilde{V}$, that is, $\tilde{V}(\alpha \bm{u} \oplus \beta \bm{v}) = \alpha \tilde{V}\bm{u} \oplus \beta \tilde{V}\bm{v}$. 

Now, of particular interest to us here is the transformation of a maximally commuting family induced by the conjugate action of a Clifford operator: {\it how the conjugate action of a Clifford operator maps a maximally commuting family to a new set?}.
The following remark gives us the answer to this question.

\begin{remark}\label{apdx:remark2}
    Let $G \subset \{I,X,Y,Z\}^{\otimes n}$ be a maximally commuting family with $I^{\otimes n}$. 
    For an arbitrary Clifford circuit $V$, $[VGV^{\dagger}] \coloneqq \{[VPV^{\dagger}] : P \in G\}$ is also a maximally commuting family with $I^{\otimes n}$.
    Thus, the following relation holds for any base $\{\bm{g}'_{k}\}_{k=1}^{n}$ of $\phi^{-1}([VGV^{\dagger}])$:
    \begin{equation}\label{apdx:VGV_dg}
        VGV^{\dagger} = \left\{ ~p_{V}(\bm{c}';\{\bm{g}'_{k}\}) \, \phi\left(\sum_{k=1}^{n} c'_{k} \bm{g}'_{k} \right) : \bm{c}' \in \{0,1\}^{n} \right\},
    \end{equation}
    where $p_{V}(\bm{c}';\{\bm{g}'_{k}\}) \in \{+1,-1\}$. Moreover, let $\tilde{V}$ and $\{\bm{g}_{k}\}_{k=1}^{n}$ be the map on $\{0,1\}^{2n}$ associated with a Clifford circuit $V$ and a basis set in $\phi^{-1}(G)$, respectively, and then the set of binary vectors $\{\tilde{V}\bm{g}_{k}\}_{k=1}^{n}$ forms a basis set of $\phi^{-1}([VGV^{\dagger}])$.
\end{remark}
We start by showing that $[VGV^{\dagger}]$ is a maximally commuting family with $I^{\otimes n}$. 
To prove this, we show the following two propositions: (i) $P,Q \in \{I,X,Y,Z\}^{\otimes n} \land [P,Q]=0  \Rightarrow [[VPV^{\dagger}],[VQV^{\dagger}]] = 0$ and (ii) $P,Q \in \{I,X,Y,Z\}^{\otimes n} \land P \neq Q \Rightarrow [VPV^{\dagger}] \neq [VQV^{\dagger}]$.
From Remark~\ref{remark: the action of Clifford operator}, there exist $p,q \in \{+1,-1\}$ and $P',Q' \in \{I,X,Y,Z\}^{\otimes n}$ which satisfy $VPV^{\dagger} = pP'$ and $VQV^{\dagger} = qQ'$. Based on this fact and the assumption $PQ=QP$, we obtain 
\begin{equation}
    [VPV^{\dagger}][VQV^{\dagger}] = pqVPQV^{\dagger} = pqVQPV^{\dagger} = [VQV^{\dagger}][VPV^{\dagger}],
\end{equation}
meaning that the proposition (i) holds. 
Next, to prove proposition (ii), suppose $[VPV^{\dagger}] = [VQV^{\dagger}]$ for the sake of contradiction. Then, we have
\begin{equation}
2^n = \mathrm{Tr}( [VPV^{\dagger}][VQV^{\dagger}] ) = pq \mathrm{Tr}'( VPV^{\dagger} VQV^{\dagger}) = pq \mathrm{Tr}(PQ).
\end{equation}
Since $P$ and $Q$ are $n$-qubit Pauli strings such that $P \neq Q$, we derive $pq \mathrm{Tr}(PQ) = 0$, which is contradiction.
Thus, it must be $[VPV^{\dagger}] \neq [VQV^{\dagger}]$, which establishes the proposition (ii).
From the two propositions, if $G$ is a maximally commuting family with $I^{\otimes n}$, $[VGV^{\dagger}]$ also forms a maximally commuting family. 
Additionally, from Lemma~\ref{apdx:Lemma_basis} and the fact that $[VGV^{\dagger}]$ is a maximally commuting family, we derive
\begin{equation}
    [VGV^{\dagger}] = \left\{\phi\left(\sum_{k=1}^n c_k\bm{g}'_k\right):\bm{c}=(c_1,c_2,\cdots,c_n)\in \{0,1\}^n\right\},
\end{equation}
where $\{\bm{g}'_{k}\}_{k=1}^{n}$ is a basis set of $\phi^{-1}([VGV^{\dagger}])$.
For any $[VPV^\dagger]\in [VGV^{\dagger}]$, there exists a phase $p_V(P) \in \{+1,-1\}$ such that $[VPV^\dagger]=p_V(P)VPV^\dagger$.
Thus, specifying $P\in G$ as $(\bm{c},\{\bm{g}'_{k}\})$ (and rewriting $p_V(P)$ into $p_V(\bm{c},\{\bm{g}'_{k}\})$), we arrive at Eq.~(\ref{apdx:VGV_dg}).

Next, we show the set $\{\tilde{V}\bm{g}_{k}\}_{k=1}^{n}$ is a basis set in $\phi^{-1}([VGV^{\dagger}])$. Here we consider the vector $\bm{c} \in \{0,1\}^{n}$ which satisfy $\sum_{k=1}^{n} c_{k} \tilde{V}\bm{g}_{k} = \bm{0}$. Since $\tilde{V}$ is a linear transformation, we obtain
\begin{equation}
    \tilde{V}\left( \sum_{k=1}^{n} c_{k} \bm{g}_{k} \right) = \bm{0}.
\end{equation}
Noting that $\tilde{V}$ is an injective map, we obtain $\sum_{k=1}^{n} c_{k} \bm{g}_{k}=\bm{0}$. Based on this and the assumption that $\{\bm{g}_{k}\}_{k=1}^{n}$ is a basis set of $\phi^{-1}(G)$, $\bm{c}=\bm{0}$ is a unique solution.
Therefore, the set $\{\tilde{V}\bm{g}_{k}\}_{k=1}^{n}$ forms a basis set of $\phi^{-1}([VGV^{\dagger}])$.

At the end of this section, three Clifford operators $V=H$ (the Hadamard gate), $S$ (the phase gate), and $CZ$ (the controlled-Z) gate play an important role.
Here, we provide a detailed description of the action of $\tilde{V} (\bullet) = \phi^{-1}[ V\phi(\bullet)V^{\dagger}]$ associated with the three Clifford operators in preparation for the following subsection.
Let $H_i$ and $S_i$ be a Hadamard gate and a phase gate acting on the $i$-th qubit, respectively, and let $CZ_{i,j}$ be a CZ gate acting on the $i$-th and the $j$-th qubits. Also, let us assume that a Pauli string is expressed by the vector $\bm{b}$ as in Eq.~(\ref{eq:definition of phi}). 
Then, the Pauli string $\phi(\bm{b})$ is transformed by the conjugate action of the three operators $H_i,S_i,CZ_{i,j}$ as follows:
\begin{align}
    H_{i} \phi(\bm{b}) H_{i}^{\dagger} &= (-i)^{(\bm{b}^{z})^T\bm{b}^{x}} \left\{ (-1)^{b_{i}^{z} b_{i}^{x}} Z^{b_{i}^{x}}X^{b_{i}^{z}}  \right\} \bigotimes_{k \neq i} Z^{b_{k}^{z}}X^{b_{k}^{x}}, \\
    S_{i} \phi(\bm{b}) S_{i}^{\dagger} &= (-i)^{(\bm{b}^{z})^T\bm{b}^{x}} \left\{ (-i)^{b_{i}^{x}} Z^{b_{i}^{z} \oplus b_{i}^{x} }X^{b_{i}^{x}} \right\}
    \bigotimes_{k \neq i} Z^{b_{k}^{z}}X^{b_{k}^{x}}, \\
    CZ_{i,j} \phi(\bm{b}) CZ_{i,j}^{\dagger} &= (-i)^{(\bm{b}^{z})^T\bm{b}^{x}} \left\{ (-1)^{b_{i}^{x} b_{j}^{x}} Z^{b_{i}^{z} \oplus b_{j}^{x} }X^{b_{i}^{x}} \otimes Z^{b_{j}^{z} \oplus b_{i}^{x} }X^{b_{j}^{x}} \right\} \bigotimes_{k \neq i,j} Z^{b_{k}^{z}}X^{b_{k}^{x}}. 
\end{align}
Recalling that the function $[\bullet]$ transforms the input Pauli operator with a phase into one without a phase and the function $\phi^{-1}$ maps the Pauli string into the binary vector, the action of $\tilde{V} (\bullet) = \phi^{-1}[ V\phi(\bullet)V^{\dagger}]$ on the vector $\bm{b}=(b_{1}^{z},...,b_{i}^{z},...,b_{j}^{z},..., b_{n}^{z},b_{1}^{x},...,b_{i}^{x},...,b_{j}^{x},..., b_{n}^{x})^{T}$ with $V = H_{i},\, S_{i},\, CZ_{i,j} ~(1 \leq i \leq n )$ can be summarized as follows:
\begin{itemize}
    \item$\tilde{H}_{i}$: Swapping $b_{i}^{z}$ and $b_{i}^{x}$ (i.e., $b_{i}^{z} \leftarrow b_{i}^{x}$, $b_{i}^{x} \leftarrow b_{i}^{z}$)
    \item$\tilde{S}_{i}$: Adding $b_{i}^{x}$ to $b_{i}^{z}$ (i.e., $b_{i}^{z} \leftarrow b_{i}^{z} \oplus b_{i}^{x}$)
    \item$\widetilde{CZ}_{i,j}$: Adding $b_{j}^{x}$ to $b_{i}^{z}$ and $b_{i}^{x}$ to $b_{j}^{z}$ (i.e., $b_{i}^{z} \leftarrow b_{i}^{z} \oplus b_{j}^{x}$,~ $b_{j}^{z} \leftarrow b_{j}^{z} \oplus b_{i}^{x}$),
\end{itemize}
which reproduces the previous result~\cite{PhysRevA.70.052328,Crawford2021efficientquantum,9248636}.

\subsubsection{Diagonalization of maximally commuting family}
Based on the two remarks in Sec.~\ref{apdx:pre3}, we formally describe the procedure to obtain a quantum circuit $U$ that diagonalizes a given maximally commuting family, which is related to the proof of Theorem~\ref{thm 2} (e.g., Eq.~(\ref{B6}) in the proof).
Note that the construction strategy for $U$ provided in the proof of the following lemma is essentially equivalent to the existing method based on stabilizer formalism as in Ref.~\cite{PhysRevA.70.052328,Crawford2021efficientquantum,9248636}.

\begin{lemma}\label{apdx:Lemma_diag}
    Let $G\subset \{I,X,Y,Z\}^{\otimes n}$ be a maximally commuting family with $I^{\otimes n}$. There exists a Clifford circuit $U$ that diagonalizes all elements of $G$. 
    That is, for a phase function $p:\{I,Z\}^{\otimes n}\to \{+1,-1\}$, the following equality holds:
    \begin{align}\label{eq:diag_G}
        UGU^\dagger = \left\{p(D)D:D\in\{I,Z\}^{\otimes n}\right\}.
    \end{align}
\end{lemma}

\begin{proof}
We prove Lemma~\ref{apdx:Lemma_diag} by explicitly constructing a Clifford circuit $U$ which diagonalizes all elements of $G$.
Let $\{\bm{g}_{i}\}_{i=1}^{n}$ be a basis set of $\phi^{-1}(G)$ where $\bm{g}_{i} = (g_{i,1}^{z},...,g_{i,n}^{z}, g_{i,1}^{x},...,g_{i,n}^{x})^{T}$, and let $M_{0}$ be a $2n\times n$ matrix given by
\begin{equation}\label{eq:M0}
    M_{0} = (\bm{g}_{1},...,\bm{g}_{n}) 
    =\begin{pmatrix} \bm{g}_{1}^{z} & \cdots & \bm{g}_{n}^{z}  \\ \hline \bm{g}_{1}^{x} & \cdots & \bm{g}_{n}^{x} \end{pmatrix},
\end{equation}
where $\bm{g}_{i}^{z} = (g_{i,1}^{z},...,g_{i,n}^{z})^{T}$ and $\bm{g}_{i}^{x} = (g_{i,1}^{x},...,g_{i,n}^{x})^{T}$. We refer to the upper half of $M_{0}$ as the $Z$-matrix of $M_{0}$ and refer to the lower half of $M_{0}$ as the $X$-matrix of $M_{0}$.

In the first step, we make the $X$-matrix of $M_{0}$ full-rank by applying $2n \times 2n$ matrix $\tilde{V}_{1} = \prod_{i} \tilde{H_{i}}$ to $M_0$, which corresponds to the (conjugate) operation of $V_{1}=\otimes_{i} H_{i}$ to the Pauli strings $\phi(\bm{g}_{j})~(j=1,2,...,n)$ (ignoring the phase):
\begin{equation}\label{eq:M0_to_M1}
    M_{0}
    \rightarrow 
    M_{1} := (\tilde{V}_{1}\bm{g}_{1},...,\tilde{V}_{1}\bm{g}_{n}) 
    =
    \begin{pmatrix}A \\ \hline B \end{pmatrix},
\end{equation}
where the $X$-matrix of $M_1$ denoted by $B$ is a full-rank matrix. 
Note that this operation is always possible, as guaranteed by Ref.~\cite[Lemma~6]{PhysRevA.70.052328}.

In the second step, we perform the Gaussian elimination in the columns of $M_1$ such that the matrix $B$ in Eq.~(\ref{eq:M0_to_M1}) turns into the $n\times n$ identity matrix $I_{n \times n}$:
\begin{equation}\label{eq:M1_to_M2}
    M_{1}
    \rightarrow
    M_{2} := (\bm{g}'_{1},...,\bm{g}'_{n}) = \begin{pmatrix} C \\ \hline I_{n\times n} \end{pmatrix}.
\end{equation}
Note that the $Z$-matrix of $M_{2}$ denoted by $C$ is proven to be a symmetric matrix in Ref.~\cite[Lemma 4.3]{Bandyopadhyay2002}.
The procedure in the second step corresponds to changing the choice of base in $\phi^{-1}([ V_{1}GV_{1}^{\dagger}])$, that is,
\begin{equation}
    \phi^{-1}([ V_{1}GV_{1}^{\dagger} ]) = \mathrm{Span}\{\tilde{V}_{1}\bm{g}_{1},...,\tilde{V}_{1}\bm{g}_{n}\} 
    = \mathrm{Span}\{\bm{g}'_{1},...,\bm{g}'_{n}\}.
\end{equation}

In the third step, by appropriately applying the $S$ and $CZ$ gates to $M_2$, we transform the $Z$-matrix of $M_{2}$ into the zero matrix $0_{n\times n}$ where all elements are zero:
\begin{equation}\label{eq:M2_to_M3}
    M_{2}
    \rightarrow
    M_{3} := (\bm{x}_{1},...,\bm{x}_{n}) = \begin{pmatrix} 0_{n\times n} \\ \hline I_{n\times n} \end{pmatrix},
\end{equation}
where $\bm{x}_{i}~(i=1,...,n)$ is a $2n$-dimensional vector such that the $(i+n)$-th element is one and all other elements are zero.
To construct the transformation Eq.~(\ref{eq:M2_to_M3}), we first describe the action of $\tilde{S}_{i}$ and $\widetilde{CZ}_{i,j}$ on the $2n\times n$ matrix $(C^{T}|I_{n \times n})^{T}$, in which the $(i,j)$-th entry of the $Z$-matrix is denoted by $c_{i,j}^{z}$, as follows:
\begin{itemize}
    \item$\tilde{S}_{i}$: Adding $1$ to $c_{i,i}^{z}$ (i.e., $c_{i,j}^{z} \leftarrow c_{i,j}^{z} \oplus 1$)
    \item$\widetilde{CZ}_{i,j}$: Adding 1 to $c_{i,j}^{z}$ and $c_{j,i}^{z}$ (i.e., $c_{i,j}^{z} \leftarrow c_{i,j}^{z} \oplus 1$,~ $c_{j,i}^{z} \leftarrow c_{j,i}^{z} \oplus 1$ ).
\end{itemize}
Note that the operations $\tilde{S}_{i}$ and $\widetilde{CZ}_{i,j}$ do not affect the elements of the $ X$-matrix when the target matrix is $M_2$. Therefore, by applying $\tilde{S}_{i}$ for all $i$ satisfying $c_{i,i}^{z} = 1$ and $\widetilde{CZ}_{j,k}$ for all $j,k (j < k) $ satisfying $c_{j,k}^{z}=1$ to $M_{2}$, we can eliminate the all elements of $C$ and obtain $M_{3}$.
For convenience, we denote the operation used in this step by $\tilde{V}_{2}$.

In the fourth step, we swap the $X$-matrix and the $Z$-matrix of $M_{3}$ by applying $\tilde{V}_{3}\coloneqq \prod_{i=1}^{n} \tilde{H}_{i}$ to $M_{3}$ and obtain 
\begin{equation}\label{eq:M3_to_M4}
    M_{3}
    \rightarrow
    M_{4} := (\bm{z}_{1},...,\bm{z}_{n}) = \begin{pmatrix} I_{n\times n} \\ \hline 0_{n\times n} \end{pmatrix},
\end{equation}
where $\bm{z}_{i}~(i=1,...,n)$ is a $2n$-dimensional vector such that the $i$-th element is one and all other elements are zero. 

The transformations from the first step to the fourth step can be summarized in the following diagram:
\[
\xymatrix{
G \ar@<0.5ex>[d]^-{\phi^{-1}} \ar[r]^-{[V_{1} (\bullet) V_{1}^{\dagger}]}
& [ V_{1}GV_{1}^{\dagger} ] \ar[r]^-{[V_{2} (\bullet) V_{2}^{\dagger}]} \ar@<0.5ex>[d]^-{\phi^{-1}}
& [ V_{2} V_{1}GV_{1}^{\dagger} V_{2}^{\dagger}] \ar[r]^-{[V_{3} (\bullet) V_{3}^{\dagger}]} \ar@<0.5ex>[dd]^-{\phi^{-1}}
& [V_{3} V_{2} V_{1}GV_{1}^{\dagger} V_{2}^{\dagger} V_{3}^{\dagger}] \ar@<0.5ex>[dd]^-{\phi^{-1}}
\\
\mathrm{Span}\{\bm{g}_{1},...,\bm{g}_{n}\} \ar@<0.5ex>[u]^-{\phi} \ar[r]^-{\tilde{V}_{1}}_-{(\mathrm{step1})}
& \mathrm{Span}\{\tilde{V}_{1}\bm{g}_{1},...,\tilde{V}_{1}\bm{g}_{n}\} \ar@<0.5ex>[u]^-{\phi}
&
&
\\
& \mathrm{Span}\{\bm{g}'_{1},...,\bm{g}'_{n}\} \ar@{=}[u]_{(\mathrm{step2})} \ar[r]^-{\tilde{V}_{2}}_{(\mathrm{step3})}
& \mathrm{Span}\{\bm{x}_{1},...,\bm{x}_{n}\} \ar[r]^-{\tilde{V}_{3}}_{(\mathrm{step4})} \ar@<0.5ex>[uu]^-{\phi}
& \mathrm{Span}\{\bm{z}_{1},...,\bm{z}_{n}\} \ar@<0.5ex>[uu]^-{\phi}
}
\]
To obtain the above diagram, we utilize the following property: 
{\it Given a $P\in\{I,X,Y,Z\}^{\otimes n}$ and Clifford circuits $U$ and $V$, then $[V[UPU^{\dagger}]V^{\dagger}] = [VUPU^{\dagger}V^{\dagger}] $ holds}. 
This property allows us to simplify the diagram as $[V_{2}[V_{1}PV_{1}^{\dagger}]V_{2}^{\dagger}] = [V_{2}V_{1}PV_{1}^{\dagger}V_{2}^{\dagger}]$. 
From Remark~\ref{apdx:remark2}, we have
\begin{equation}
    V_{123} G V_{123}^{\dagger} = \left\{ p_{V_{123}}(\bm{c};\{\bm{z}_{k}\}_{k=1}^{n})\, \phi\left(\sum_{k=1}^{n} c_{k} \bm{z}_{k}\right) : \bm{c} \in \{0,1\}^{n} \right\},
\end{equation}
where $V_{123} = V_{3}V_{2}V_{1}$ and $p_{V_{123}}(\bm{c};\{\bm{z}_{k}\}_{k=1}^{n}) \in \{+1,-1\}$. 
Additionally, Lemma~\ref{apdx:Lemma_basis} says that for any $\bm{c}$, there exists $p\in\{+1,-1\}$ such that
\begin{equation}
    \phi\left(\sum_{k=1}^{n} c_{k} \bm{z}_{k}\right) = p\, \phi(\bm{z}_{1})^{c_{1}} \phi(\bm{z}_{2})^{c_{2}} \cdots \phi(\bm{z}_{n})^{c_{n}}.
\end{equation}
Note that the scalar $p$ depends on $\bm{c}$ and $\{\bm{z}_k\}_{k=1}^{n}$. Thus, we obtain 
\begin{align}
    V_{123} G V_{123}^{\dagger} 
    &= \biggl\{ p_{V_{123}}(\bm{c};\{\bm{z}_{k}\}_{k=1}^{n})\, p\, \phi(\bm{z}_{1})^{c_{1}} \phi(\bm{z}_{2})^{c_{2}} \cdots \phi(\bm{z}_{n})^{c_{n}} : \bm{c} \in \{0,1\}^{n} \biggr\} \\
    &= \biggl\{ p_{V_{123}}(\bm{c};\{\bm{z}_{k}\}_{k=1}^{n})\, p\, Z^{c_{1}}_{1}Z^{c_{2}}_{2} \cdots Z^{c_{n}}_{n} : \bm{c} \in \{0,1\}^{n} \biggl\} \\
    &= \biggl\{ p(D)D : D\in\{I,Z\}^{\otimes n} \biggl\},
\end{align}
where we write the phase of each element as $p(D)\in\{+1,-1\}$ for simplicity. This establishes Lemma~\ref{apdx:Lemma_diag}.
\end{proof}

\subsection{Details of the Clifford circuit in Lemma~\ref{apdx:Lemma_diag}}\label{apdx:construct_MUBs_qc}
Here, we clarify the components of the Clifford circuit constructed by the procedure in Lemma~\ref{apdx:Lemma_diag}.
The following lemma completes the proof of Theorem~\ref{thm 2}.
\begin{lemma}\label{apdx:maximal_depth_proof}
    Let $G$ be a maximally commuting family with $I^{\otimes n}$. If $G\setminus I^{\otimes n}$ has no overlap with $\{I,Z\}^{\otimes n}\setminus I^{\otimes n}$, then, we can explicitly construct a Clifford circuit $U$ that diagonalizes $G$ with at most $(n+2)$-depth on fully connected topologies.
    Moreover, the Clifford circuit contains the following elementary gates:
    \begin{align}\label{eq:bound}
        N_{H} = n, ~~ N_{S} \leq n, ~~ N_{CZ} \leq \frac{n(n-1)}{2}
    \end{align}
    where $N_{H}$,$N_{S}$, and $N_{CZ}$ denote the total number of $H$, $S$, and $CZ$ gates in the Clifford circuit, respectively.
\end{lemma}
\begin{proof}
We start by showing that if $G\setminus I^{\otimes n}$ has no overlap with $\{I,Z\}^{\otimes n}\setminus I^{\otimes n}$, then it holds that $\bm{u}^{x} \neq \bm{v}^{x}$ for any $\bm{u}, \bm{v} \in \phi^{-1}(G)$ such that $\bm{u} \neq \bm{v}$.
This is true because, suppose $\bm{u}^x = \bm{v}^x$, the fact: $\phi(\bm{u}\oplus\bm{v})\in\{I,Z\}^{\otimes n}\setminus{I^{\otimes n}}$ contradicts that $G$ has no overlap with $\{I,Z\}^{\otimes n}\setminus{I^{\otimes n}}$.
Accordingly, every $\bm{u}\in \phi^{-1}(G)$ has unique $\bm{u}^{x}$, which result in $2^n$ distinct vectors in $\{0,1\}^{n}$, and then we can take a basis set $\{\bm{g}_k'\}_{k=1}^n$ in $\phi^{-1}(G)$ satisfying
\begin{equation}\label{apdx:auto_full_rank}
    (\bm{g}'_{1},...,\bm{g}'_{n}) = \begin{pmatrix} C \\ \hline I_{n\times n} \end{pmatrix},
\end{equation}
where $C$ is an $n\times n$ symmetric matrix.
Now, we join the third step in the construction of the Clifford circuit for $G$ (more specifically, Eq.~(\ref{eq:M2_to_M3}) in the proof of Lemma~\ref{apdx:Lemma_diag}), denoting the $2n\times n$ matrix Eq.~(\ref{apdx:auto_full_rank}) by $M_2$.
In the third step, we transform the $Z$-matrix of $M_{2}$ into the zero matrix $0_{n\times n}$ as shown in Eq.~(\ref{eq:M2_to_M3}), by applying the operator $V_{2}$ consisting of $S$ and $CZ$ gates. 
Here, it is evident that the circuit depth of $V_{2}$ is maximum when $V_{2}$ contains the maximum number of $S$ and $CZ$ gates as
\begin{equation}\label{apdx:worst_V2}
    V_{2,{\rm max}} = \prod_{1 \leq j<k \leq n} CZ_{j,k} \prod_{1 \leq i \leq n} S_{i} 
\end{equation}
where $S_{i}$ denotes a phase gate acting on the $i$-th qubit, and $CZ_{j,k}$ denotes a $CZ$ gate acting on the $j$-th and the $k$-th qubits.
Since the adjacent $CZ$ gates are commuting with each other, 
the optimal configuration of full-connected $CZ$ gates to minimize the circuit depth on an $n$-qubit fully connected topology can be obtained by solving the \textit{edge-coloring problem} for the complete graph $K_{n}$ with $n$ vertices.
In graph theory, \textit{coloring} edges of a given graph is to assign colors to each edge of the graph so that no two adjacent edges share the same color.
Also, the \textit{edge-coloring problem} is to find the minimum number of colors, which fill all edges in the graph, and the minimum number is called {\it the chromatic index}, denoted by $\chi'$.
Now, we associate $n$ qubits on a fully connected topology with the vertices of the complete graph $K_{n}$ and consider the $CZ_{j,k}$ gate as the edge between the $j$-th and the $k$-th vertices.
Note that each edge in the complete graph corresponds to one of the full-connected $CZ$ gates.
Edges with the same color in the complete graph $K_n$ corresponds to the CZ gates that can be driven simultaneously (or at only 1-depth) on a fully connected device.
Therefore, we can identify the circuit-depth of full-connected CZ gates with the chromatic index of the complete graph $K_n$.
For the complete graph $K_{n}$, it is widely known that
\begin{equation}
    \chi'(K_{n}) = 
    \begin{cases}
        n & n: odd \\
        n-1 & n: even,
    \end{cases}
\end{equation}
so we conclude that the circuit depth of $V_{2}$ is upper bounded by $n+1$.


Adding the Clifford gate $V_3$ in the fourth step, which clearly has 1-depth, we finally obtain the Clifford circuit $V_{3}V_{2}$ to diagonalize $G$.
The preceding discussions lead that the circuit depth of $V_{3}V_{2}$ can be upper bounded by $n+2$.
In addition, Eq.~(\ref{apdx:worst_V2}) and $V_3=\otimes_{i=1}^n H_i$ yield
\begin{align}
    N_{H} = n, ~~ N_{S} \leq n, ~~ N_{CZ} \leq \frac{n(n-1)}{2},
\end{align}
where $N_{H}$, $N_{S}$, and $N_{CZ}$ denote the total number of $H$, $S$, and $CZ$ gates in the Clifford circuit, respectively.
\end{proof}

\subsection{A Method for obtaining quantum circuits $\{U_{i}\}_{i=1}^{2^n}$ in Theorem~\ref{thm 2}}\label{apdx:obtain_circuits}
In this section, we will introduce the procedure to construct a set $\{U_{i}\}_{i=1}^{2^n}$ of quantum circuits for implementing Theorem~\ref{thm 2}.
To generate the set $\{U_{i}\}_{i=1}^{2^n} \cup I^{\otimes n}$, it is necessary to first prepare a set $G_{i}~(i=1,2,\cdots,2^n+1)$ of Pauli strings  (excluding the identity) satisfying the following conditions:
\begin{itemize}
\item[(i)] $G_i$ and $G_j$ are disjoint for all $i\neq j$.
\item[(ii)] All elements of $G_i$ are $n$-qubit Pauli strings in $\{I,X,Y,Z\}^{\otimes n} \setminus I^{\otimes n}$, and the cardinality of $G_i$ is $2^n-1$.
\item[(iii)] $G_{2^n+1}=\{I,Z\}^{\otimes n} \setminus I^{\otimes n}$.
\end{itemize}
These conditions imply that Pauli strings $\{I,X,Y,Z\}^{\otimes n}\backslash \{I^{\otimes n}\}$ are grouped into the set $\{G_i\}_{i=1}^{2^n+1}$ as
\begin{align}
\bigcup_{i=1}^{2^n+1} G_i=\{I,X,Y,Z\}^{\otimes n}\backslash\{I^{\otimes n}\}.
\end{align}
The existence of the set fulfilling the conditions (i)--(iii) is guaranteed by the existing construction~\cite{Bandyopadhyay2002, seyfarth2019cyclic, reggio2023fast}.
Especially, Ref.~\cite{reggio2023fast} provides a publicly available package $\mathtt{psfam.py}$ to output a set satisfying conditions (i)--(iii), and we utilize their code for the following constructions.

Next, we provide the way of constructing the quantum circuits $\{U_{i}\}_{i=1}^{2^n}$ for Theorem~\ref{thm 2} from a given set $\{G_{i}\}_{i=1}^{2^n+1}$ satisfying (i)--(iii).
The basic strategy follows the procedure in the proof of Lemma~\ref{apdx:maximal_depth_proof}.
Note that we should adopt the Hermitian conjugation of the output Clifford circuit from the procedure in Lemma~\ref{apdx:maximal_depth_proof} as $U_i$  in order to match the notation in the proof of Theorem~\ref{thm 2} (See Eqs.~(\ref{B6}) and (\ref{eq:diag_G})).
Then, the detailed procedure for generating quantum circuits $\{U_{i}\}_{i=1}^{2^n}$ is summarized as Algorithm~\ref{alg:produce_basis_main}. The results of executing this algorithm for ranging from $n=1$ to $n=12$ can be found in the main text.

\section{Examples of the sets of quantum circuits in Theorem~\ref{thm 2}}\label{sec:circuits}

In this section, we provide examples of the sets of quantum circuits $\{U_{i}\}_{i=1}^{2^n}$ enabling the implementation of Theorem~\ref{thm 2} for the case when $n=1,2,3,4$. The sets $\{U_{i}\}_{i=1}^{2^n}$ of quantum circuits for $n=1,2,3,4$ are presented in Table.~\ref{tab:n_1}, Table.~\ref{tab:n_2}, Table.~\ref{tab:n_3}, and Table.~\ref{tab:n_4}, respectively. Each table includes $2^{n}$ quantum circuits along with the generators of the corresponding maximally commuting family (for detail on ``generator'', refer to Appendix~\ref{apdx:pre2}). These sets of circuits $\{U_{i}\}_{i=1}^{2^n}$ are derived by applying Algorithm~\ref{alg:produce_basis_main} to the disjoint sets $\{G_{i}\}_{i=1}^{2^n+1}$ of maximally commuting families with $G_{2^n+1}=\{I,Z\}^{\otimes n}\setminus I^{\otimes n}$. The detailed procedure for constructing the circuits is elaborated in Appendix~\ref{apdx:obtain_circuits}. 

\newpage

\begin{table}[ht]
\renewcommand{\arraystretch}{1.3}
\centering
\begin{tabular}{|p{4.5em}||p{11em}||p{11em}|}
\hline
&&\\
\begin{tabular}{c}
\bf{\footnotesize Quantum} \\ \bf{\footnotesize Circuit}
\end{tabular}
&
\multicolumn{1}{c||}{
\Qcircuit @C=1em @R=0.7em  {
& \gate{H} & \qw
}
}
&
\multicolumn{1}{c|}{
\Qcircuit @C=1em @R=0.7em  {
& \gate{H} & \gate{S^{\dagger}} & \qw
}
}
\\
&&
\\
\hline
\multicolumn{1}{|c||}{\footnotesize $\bm{U_{i}}$}&\multicolumn{1}{c||}{$U_{1}$}&\multicolumn{1}{c|}{$U_{2}$}\\
\multicolumn{1}{|c||}{\footnotesize \bf{Generator}}&\multicolumn{1}{c||}{$X$}&\multicolumn{1}{c|}{$Y$}\\
\hline
\end{tabular}
\caption{One example of the sets of quantum circuits $\{U_{i}\}^{2}_{i=1}$ that enables the implementation of Theorem~\ref{thm 2} when $n=1$.}
\label{tab:n_1}
\end{table}


\begin{table}[ht]
\renewcommand{\arraystretch}{1.3}
\centering
\begin{tabular}{|p{4.5em}||p{11em}||p{11em}||p{11em}||p{11em}|}
\hline
&&&&\\
\begin{tabular}{c}
\bf{\footnotesize Quantum} \\ \bf{\footnotesize Circuit}
\end{tabular}
&
\multicolumn{1}{c||}{
\Qcircuit @C=1em @R=0.7em  {
& \gate{H} & \qw \\
& \gate{H} & \qw \\ \\
}
}
&
\multicolumn{1}{c||}{
\Qcircuit @C=1em @R=0.7em  {
& \gate{H} & \gate{S^{\dagger}} & \ctrl{1} & \qw \\
& \gate{H} & \qw                & \ctrl{0} & \qw \\ \\
}
}
&
\multicolumn{1}{c||}{
\Qcircuit @C=1em @R=0.7em  {
& \gate{H} & \qw                & \ctrl{1} & \qw \\
& \gate{H} & \gate{S^{\dagger}} & \ctrl{0} & \qw \\ \\
}
}
&
\multicolumn{1}{c|}{
\Qcircuit @C=1em @R=0.7em  {
& \gate{H} & \gate{S^{\dagger}} & \qw \\
& \gate{H} & \gate{S^{\dagger}} & \qw \\ \\
}
}
\\
&&&&
\\
\hline
\multicolumn{1}{|c||}{\footnotesize $\bm{U_{i}}$}&\multicolumn{1}{c||}{$U_{1}$}&\multicolumn{1}{c||}{$U_{2}$}&\multicolumn{1}{c||}{$U_{3}$}&\multicolumn{1}{c|}{$U_{4}$}\\
\multicolumn{1}{|c||}{\bf{\footnotesize Generator}}&\multicolumn{1}{c||}{$XI, IX$}&\multicolumn{1}{c||}{$YZ, ZX$}&\multicolumn{1}{c||}{$XZ, ZY$}&\multicolumn{1}{c|}{$YI, IY$}\\
\hline
\end{tabular}
\caption{One example of the sets of quantum circuits $\{U_{i}\}^{4}_{i=1}$ that enables the implementation of Theorem~\ref{thm 2} when $n=2$.}
\label{tab:n_2}
\end{table}


\begin{table}[ht]
\renewcommand{\arraystretch}{1.3}
\centering
\begin{tabular}{|p{4.5em}||p{11em}||p{11em}||p{11em}||p{11em}|}
\hline
&&&&\\
\begin{tabular}{c}
\bf{\footnotesize Quantum} \\ \bf{\footnotesize Circuit}
\end{tabular}
&
\multicolumn{1}{c||}{
\Qcircuit @C=1em @R=0.7em  {
& \gate{H} & \gate{S^{\dagger}} & \qw          & \ctrl{1} & \qw \\
& \gate{H} & \qw                & \ctrl{1}     & \control \qw & \qw \\ 
& \gate{H} & \qw                & \control \qw & \qw & \qw
}
}
&
\multicolumn{1}{c||}{
\Qcircuit @C=1em @R=0.7em  {
& \gate{H} & \qw                & \ctrl{2}     & \qw          & \qw \\
& \gate{H} & \gate{S^{\dagger}} & \qw          & \ctrl{1}     & \qw \\ 
& \gate{H} & \qw                & \control \qw & \control \qw & \qw
}
}
&
\multicolumn{1}{c||}{
\Qcircuit @C=1em @R=0.7em  {
& \gate{H} & \gate{S^{\dagger}} & \ctrl{1}      & \ctrl{2}      & \qw \\
& \gate{H} & \gate{S^{\dagger}} & \control \qw  & \qw           & \qw \\ 
& \gate{H} & \qw                & \qw           & \control \qw  & \qw
}
}
&
\multicolumn{1}{c|}{
\Qcircuit @C=1em @R=0.7em  {
& \gate{H} & \qw                & \ctrl{1}     & \ctrl{2}     & \qw \\
& \gate{H} & \qw                & \control \qw & \qw          & \qw \\ 
& \gate{H} & \gate{S^{\dagger}} & \qw          & \control \qw & \qw
}
}
\\
&&&&
\\
\hline
\multicolumn{1}{|c||}{\footnotesize $\bm{U_{i}}$}&\multicolumn{1}{c||}{$U_{1}$}&\multicolumn{1}{c||}{$U_{2}$}&\multicolumn{1}{c||}{$U_{3}$}& \multicolumn{1}{c|}{$U_{4}$}\\
\multicolumn{1}{|c||}{\footnotesize \bf{Generator}}&\multicolumn{1}{c||}{$YZI,ZXZ,IZX$}&\multicolumn{1}{c||}{$XIZ,IYZ,ZZX$}&\multicolumn{1}{c||}{$YZZ,ZYI,ZIX$}&\multicolumn{1}{c|}{$XZZ,ZXI,ZIY$}\\
\hline
\hline
&&&&\\
\begin{tabular}{c}
\bf{\footnotesize Quantum} \\ \bf{\footnotesize Circuit}
\end{tabular}
&
\multicolumn{1}{c||}{
\Qcircuit @C=1em @R=0.7em  {
& \gate{H} & \gate{S^{\dagger}} & \ctrl{2}     & \qw          & \qw \\
& \gate{H} & \qw                & \qw          & \ctrl{1}     & \qw \\ 
& \gate{H} & \gate{S^{\dagger}} & \control \qw & \control \qw & \qw
}
}
&
\multicolumn{1}{c||}{
\Qcircuit @C=1em @R=0.7em  {
& \gate{H} & \qw                & \qw          & \ctrl{1}     & \qw \\
& \gate{H} & \gate{S^{\dagger}} & \ctrl{1}     & \control \qw & \qw \\ 
& \gate{H} & \gate{S^{\dagger}} & \control \qw & \qw          & \qw
}
}
&
\multicolumn{1}{c||}{
\Qcircuit @C=1em @R=0.7em  {
& \gate{H} & \gate{S^{\dagger}} & \qw \\
& \gate{H} & \gate{S^{\dagger}} & \qw \\ 
& \gate{H} & \gate{S^{\dagger}} & \qw
}
}
&
\multicolumn{1}{c|}{
\Qcircuit @C=1em @R=0.7em  {
& \gate{H} & \qw \\
& \gate{H} & \qw \\ 
& \gate{H} & \qw 
}
}
\\
&&&&
\\
\hline
\multicolumn{1}{|c||}{\footnotesize $\bm{U_{i}}$}&\multicolumn{1}{c||}{$U_{5}$}&\multicolumn{1}{c||}{$U_{6}$}&\multicolumn{1}{c||}{$U_{7}$}&\multicolumn{1}{c|}{$U_{8}$}\\
\multicolumn{1}{|c||}{\footnotesize \bf{Generator}}&\multicolumn{1}{c||}{$YIZ,IXZ,ZZY$}&\multicolumn{1}{c||}{$XZI,ZYZ,IZY$}&\multicolumn{1}{c||}{$YII,IYI,IIY$}&\multicolumn{1}{c|}{$XII,IXI,IIX$}\\
\hline
\end{tabular}
\caption{One example of the sets of quantum circuits $\{U_{i}\}^{8}_{i=1}$ that enables the implementation of Theorem~\ref{thm 2} when $n=3$.}
\label{tab:n_3}
\end{table}


\begin{table}[ht]
\renewcommand{\arraystretch}{1.3}
\centering
\begin{tabular}{|p{4.5em}||p{11em}||p{11em}||p{11em}||p{11em}|}
\hline
&&&&\\
\begin{tabular}{c}
\bf{\footnotesize Quantum} \\ \bf{\footnotesize Circuit}
\end{tabular}
&
\multicolumn{1}{c||}{
\Qcircuit @C=0.8em @R=0.7em  {
& \gate{H} & \gate{S^{\dagger}} & \ctrl{1}     & \ctrl{2}     & \qw          & \qw \\
& \gate{H} & \qw                & \control \qw & \qw          & \ctrl{1}     & \qw \\ 
& \gate{H} & \qw                & \ctrl{1}     & \control \qw & \control \qw & \qw \\
& \gate{H} & \qw                & \control \qw & \qw          & \qw          & \qw
}
}
&
\multicolumn{1}{c||}{
\Qcircuit @C=0.8em @R=0.7em  {
& \gate{H} & \qw                & \qw          & \ctrl{3}     & \qw          & \qw \\
& \gate{H} & \gate{S^{\dagger}} & \ctrl{1}     & \qw          & \ctrl{2}     & \qw \\ 
& \gate{H} & \qw                & \control \qw & \qw          & \qw          & \qw \\
& \gate{H} & \qw                & \qw          & \control \qw & \control \qw & \qw
}
}
&
\multicolumn{1}{c||}{
\Qcircuit @C=0.8em @R=0.7em  {
& \gate{H} & \gate{S^{\dagger}} & \ctrl{1}     & \ctrl{2}     & \qw          & \ctrl{3}     & \qw \\
& \gate{H} & \gate{S^{\dagger}} & \control \qw & \qw          & \ctrl{2}     & \qw          & \qw \\ 
& \gate{H} & \qw                & \ctrl{1}     & \control \qw & \qw          & \qw          & \qw \\
& \gate{H} & \qw                & \control \qw & \qw          & \control \qw & \control \qw & \qw
}
}
&
\multicolumn{1}{c|}{
\Qcircuit @C=0.8em @R=0.7em  {
& \gate{H} & \qw                & \ctrl{2}     & \qw          & \ctrl{3}     & \qw \\
& \gate{H} & \qw                & \qw          & \ctrl{1}     & \qw          & \qw \\ 
& \gate{H} & \gate{S^{\dagger}} & \control \qw & \control \qw & \qw          & \qw \\
& \gate{H} & \qw                & \qw          & \qw          & \control \qw & \qw
}
}
\\
&&&&
\\
\hline
\multicolumn{1}{|c||}{\footnotesize $\bm{U_{i}}$}&\multicolumn{1}{c||}{$U_{1}$}&\multicolumn{1}{c||}{$U_{2}$}&\multicolumn{1}{c||}{$U_{3}$}& \multicolumn{1}{c|}{$U_{4}$}\\
\multicolumn{1}{|c||}{\footnotesize \bf{Generator}}&\multicolumn{1}{c||}{\footnotesize $YZZI,ZXZI,ZZXZ,IIZX$}&\multicolumn{1}{c||}{\footnotesize $XIIZ,IYZZ,IZXI,ZZIX$}&\multicolumn{1}{c||}{\footnotesize $YZZZ,ZYIZ,ZIXZ,ZZZX$}&\multicolumn{1}{c|}{\footnotesize $XIZZ,IXZI,ZZYI,ZIIX$}\\
\hline
\hline
&&&&\\
\begin{tabular}{c}
\bf{\footnotesize Quantum} \\ \bf{\footnotesize Circuit}
\end{tabular}
&
\multicolumn{1}{c||}{
\Qcircuit @C=1em @R=0.7em  {
& \gate{H} & \gate{S^{\dagger}} & \ctrl{1}     & \ctrl{3}     & \qw \\
& \gate{H} & \qw                & \control \qw & \qw          & \qw \\ 
& \gate{H} & \gate{S^{\dagger}} & \ctrl{1}     & \qw          & \qw \\
& \gate{H} & \qw                & \control \qw & \control \qw & \qw 
}
}
&
\multicolumn{1}{c||}{
\Qcircuit @C=1em @R=0.7em  {
& \gate{H} & \qw                & \ctrl{2}     & \qw          & \qw \\
& \gate{H} & \gate{S^{\dagger}} & \qw          & \ctrl{2}     & \qw \\ 
& \gate{H} & \gate{S^{\dagger}} & \control \qw & \qw & \qw    & \qw \\
& \gate{H} & \qw                & \qw          & \control \qw & \qw
}
}
&
\multicolumn{1}{c||}{
\Qcircuit @C=1em @R=0.7em  {
& \gate{H} & \gate{S^{\dagger}} & \ctrl{1}     & \qw          & \qw          & \qw \\
& \gate{H} & \gate{S^{\dagger}} & \control \qw & \ctrl{1}     & \ctrl{2}     & \qw \\ 
& \gate{H} & \gate{S^{\dagger}} & \ctrl{1}     & \control \qw & \qw          & \qw \\
& \gate{H} & \qw                & \control \qw & \qw          & \control \qw & \qw
}
}
&
\multicolumn{1}{c|}{
\Qcircuit @C=1em @R=0.7em  {
& \gate{H} & \qw                & \ctrl{1}     & \qw          & \qw          & \qw \\
& \gate{H} & \qw                & \control \qw & \ctrl{1}     & \ctrl{2}     & \qw \\ 
& \gate{H} & \qw                & \ctrl{1}     & \control \qw & \qw          & \qw \\
& \gate{H} & \gate{S^{\dagger}} & \control \qw & \qw          & \control \qw & \qw
}
}
\\
&&&&
\\
\hline
\multicolumn{1}{|c||}{\footnotesize $\bm{U_{i}}$}&\multicolumn{1}{c||}{$U_{5}$}&\multicolumn{1}{c||}{$U_{6}$}&\multicolumn{1}{c||}{$U_{7}$}&\multicolumn{1}{c|}{$U_{8}$}\\
\multicolumn{1}{|c||}{\footnotesize \bf{Generator}}&\multicolumn{1}{c||}{\footnotesize $YZIZ,ZXII,IIYZ,ZIZX$}&\multicolumn{1}{c||}{\footnotesize $XIZI,IYIZ,ZIYI,IZIX$}&\multicolumn{1}{c||}{\footnotesize $YZII,ZYZZ,IZYZ,IZZX$}&\multicolumn{1}{c|}{\footnotesize $XZII,ZXZZ,IZXZ,IZZY$}\\
\hline
\hline
&&&&\\
\begin{tabular}{c}
\bf{\footnotesize Quantum} \\ \bf{\footnotesize Circuit}
\end{tabular}
&
\multicolumn{1}{c||}{
\Qcircuit @C=1em @R=0.7em  {
& \gate{H} & \gate{S^{\dagger}} & \ctrl{2}     & \qw          & \qw \\
& \gate{H} & \qw                & \qw          & \ctrl{2}     & \qw \\ 
& \gate{H} & \qw                & \control \qw & \qw & \qw    & \qw \\
& \gate{H} & \gate{S^{\dagger}} & \qw          & \control \qw & \qw
}
}
&
\multicolumn{1}{c||}{
\Qcircuit @C=1em @R=0.7em  {
& \gate{H} & \qw                & \ctrl{1}     & \ctrl{3}     & \qw \\
& \gate{H} & \gate{S^{\dagger}} & \control \qw & \qw          & \qw \\ 
& \gate{H} & \qw                & \ctrl{1}     & \qw          & \qw \\
& \gate{H} & \gate{S^{\dagger}} & \control \qw & \control \qw & \qw 
}
}
&
\multicolumn{1}{c||}{
\Qcircuit @C=1em @R=0.7em  {
& \gate{H} & \gate{S^{\dagger}} & \ctrl{2}     & \qw          & \ctrl{3}     & \qw \\
& \gate{H} & \gate{S^{\dagger}} & \qw          & \ctrl{1}     & \qw          & \qw \\ 
& \gate{H} & \qw & \control \qw & \control \qw & \qw          & \qw \\
& \gate{H} & \gate{S^{\dagger}} & \qw          & \qw          & \control \qw & \qw
}
}
&
\multicolumn{1}{c|}{
\Qcircuit @C=1em @R=0.7em  {
& \gate{H} & \qw                & \ctrl{1}     & \ctrl{2}     & \qw          & \ctrl{3}     & \qw \\
& \gate{H} & \qw                & \control \qw & \qw          & \ctrl{2}     & \qw          & \qw \\ 
& \gate{H} & \gate{S^{\dagger}} & \ctrl{1}     & \control \qw & \qw          & \qw          & \qw \\
& \gate{H} & \gate{S^{\dagger}} & \control \qw & \qw          & \control \qw & \control \qw & \qw
}
}
\\
&&&&
\\
\hline
\multicolumn{1}{|c||}{\footnotesize $\bm{U_{i}}$}&\multicolumn{1}{c||}{$U_{9}$}&\multicolumn{1}{c||}{$U_{10}$}&\multicolumn{1}{c||}{$U_{11}$}&\multicolumn{1}{c|}{$U_{12}$}\\
\multicolumn{1}{|c||}{\footnotesize \bf{Generator}}&\multicolumn{1}{c||}{\footnotesize $YIZI,IXIZ,ZIXI,IZIY$}&\multicolumn{1}{c||}{\footnotesize $XZIZ,ZYII,IIXZ,ZIZY$}&\multicolumn{1}{c||}{\footnotesize $YIZZ,IYZI,ZZXI,ZIIY$}&\multicolumn{1}{c|}{\footnotesize $XZZZ,ZXIZ,ZIYZ,ZZZY$}\\
\hline
\hline
&&&&\\
\begin{tabular}{c}
\bf{\footnotesize Quantum} \\ \bf{\footnotesize Circuit}
\end{tabular}
&
\multicolumn{1}{c||}{
\Qcircuit @C=1em @R=0.7em  {
& \gate{H} & \gate{S^{\dagger}} & \qw          & \ctrl{3}     & \qw          & \qw \\
& \gate{H} & \qw                & \ctrl{1}     & \qw          & \ctrl{2}     & \qw \\ 
& \gate{H} & \gate{S^{\dagger}} & \control \qw & \qw          & \qw          & \qw \\
& \gate{H} & \gate{S^{\dagger}} & \qw          & \control \qw & \control \qw & \qw
}
}
&
\multicolumn{1}{c||}{
\Qcircuit @C=1em @R=0.7em  {
& \gate{H} & \qw                & \ctrl{1}     & \ctrl{2}     & \qw          & \qw \\
& \gate{H} & \gate{S^{\dagger}} & \control \qw & \qw          & \ctrl{1}     & \qw \\ 
& \gate{H} & \gate{S^{\dagger}} & \ctrl{1}     & \control \qw & \control \qw & \qw \\
& \gate{H} & \gate{S^{\dagger}} & \control \qw & \qw          & \qw          & \qw
}
}
&
\multicolumn{1}{c||}{
\Qcircuit @C=1em @R=0.7em  {
& \gate{H} & \gate{S^{\dagger}} & \qw \\
& \gate{H} & \gate{S^{\dagger}} & \qw \\ 
& \gate{H} & \gate{S^{\dagger}} & \qw \\
& \gate{H} & \gate{S^{\dagger}} & \qw
}
}
&
\multicolumn{1}{c|}{
\Qcircuit @C=1em @R=0.7em  {
& \gate{H} & \qw \\
& \gate{H} & \qw \\ 
& \gate{H} & \qw \\
& \gate{H} & \qw
}
}
\\
&&&&
\\
\hline
\multicolumn{1}{|c||}{\footnotesize $\bm{U_{i}}$}&\multicolumn{1}{c||}{$U_{13}$}&\multicolumn{1}{c||}{$U_{14}$}&\multicolumn{1}{c||}{$U_{15}$}&\multicolumn{1}{c|}{$U_{16}$}\\
\multicolumn{1}{|c||}{\footnotesize \bf{Generator}}&\multicolumn{1}{c||}{\footnotesize $YIIZ,IXZZ,IZYI,ZZIY$}&\multicolumn{1}{c||}{\footnotesize $XZZI,ZYZI,ZZYZ,IIZY$}&\multicolumn{1}{c||}{\footnotesize $YIII,IYII,IIYI,IIIY$}&\multicolumn{1}{c|}{\footnotesize $XIII,IXII,IIXI,IIIX$}\\
\hline
\end{tabular}
\caption{One example of the sets of quantum circuits $\{U_{i}\}^{16}_{i=1}$ that enables the implementation of Theorem~\ref{thm 2} when $n=4$.}
\label{tab:n_4}
\end{table}

\end{document}